\documentclass[prx, reprint, amsmath,amssymb,showpacs,floatfix,longbibliography, aps, twocolumn, superscriptaddress,nofootinbib]{revtex4-2}
\usepackage{times} 
\usepackage{graphicx}
\usepackage{needspace}
\usepackage{dcolumn}
\usepackage{bm,bbm}
\usepackage{color}
\usepackage[dvipsnames]{xcolor}
\usepackage{hyperref}
\usepackage{enumitem}
\usepackage{cancel}
\usepackage{braket}
\usepackage[normalem]{ulem}
\hypersetup{colorlinks=true,citecolor=blue,linkcolor=blue, urlcolor=blue}
\hypersetup{linktocpage}
\newcolumntype{M}[1]{>{\centering\arraybackslash}m{#1}}
\newcolumntype{N}{@{}m{0pt}@{}}
\usepackage{environ}

\bibpunct{[}{]}{,}{n}{}{}

\usepackage{amsfonts,amssymb,amsmath}
\usepackage[T1]{fontenc}
\usepackage{tikz}
\usetikzlibrary{arrows, snakes, backgrounds, arrows.meta, positioning, calc, decorations.markings}
\tikzset{
  op/.style={inner sep=1pt, fill=white},
  every picture/.style={line width=0.5pt}
}
\newcommand{\Ledge}{1.2} %
\usepackage{verbatim}

\newcommand{\ZZ}{{\mathbb Z}}
\newcommand{\RR}{{\mathbb R}}

\newcommand{\GG}{{\mathbb G}}

\NewEnviron{eqs}{%
\begin{equation}\begin{split}
    \BODY
\end{split}\end{equation}
}

\makeatletter
\def\l@subsubsection#1#2{}
\makeatother

\begin{document}

\author{Meng Sun}
\thanks{These authors contributed equally to this work.}
\affiliation{International Center for Quantum Materials, School of Physics, Peking University, Beijing 100871, China}

\author{Zongyuan Wang}
\thanks{These authors contributed equally to this work.}
\affiliation{International Center for Quantum Materials, School of Physics, Peking University, Beijing 100871, China}

\author{Nathanan Tantivasadakarn}
\email[E-mail: ]{nathanan.tantivasadakarn@stonybrook.edu}
\affiliation{C. N. Yang Institute for Theoretical Physics, Stony Brook University, Stony Brook, NY 11794, USA}

\author{Yu-An Chen}
\email[E-mail: ]{yuanchen@pku.edu.cn}
\affiliation{International Center for Quantum Materials, School of Physics, Peking University, Beijing 100871, China}

\title{Majorana-Pauli stabilizer codes and duality webs of fermionic topological phases}

\date{October 1, 2026}

\begin{abstract}
Stabilizer codes provide exact lattice realizations of bosonic topological orders. In contrast, systematic stabilizer descriptions of intrinsically fermionic topological phases remain much less developed. In this work, we introduce \emph{Majorana-Pauli stabilizer codes}, a class of exactly solvable fermionic lattice models whose stabilizers are built from both generalized Pauli operators and Majorana operators.  As a main example, we construct an exactly solvable stabilizer realization of the fermionic toric code: an intrinsically fermionic $\mathbb Z_2$ topological order in $(2{+}1)$ dimensions, using $\mathbb Z_8$ Pauli operators coupled to Majorana modes. Within this stabilizer framework, the anyons, string operators, fusion rules, and braiding statistics all follow naturally from the stabilizer algebra. More broadly, we show that the fermionic toric code belongs to a duality web generated by anyon condensation and by gauging or ungauging bosonic or fermionic symmetries. This web connects bosonic topological orders, symmetry-enriched topological {(SET)} phases, and both bosonic and fermionic symmetry-protected topological {(SPT)} phases, all within a common stabilizer description. {We extend this construction to all Abelian fermionic topological orders with gapped boundaries and to fermionic SPT phases whose fully gauged theories are Abelian.} Going beyond Majorana operators, we introduce fermionic versions of the clock and shift operators and use them to construct an exact bosonization map for $\mathbb Z_D^F$ symmetries for $D$ even. Using this, we realize a stabilizer model for a nontrivial $\mathbb Z_8^F$ fermionic SPT phase with no free-fermion analog. Altogether, these results extend the stabilizer-code paradigm to a broad class of intrinsically fermionic phases bridging fermionic quantum many-body physics to quantum error correction.

\end{abstract}

\maketitle
\tableofcontents

\section{Introduction}

Exactly solvable lattice models play a central role in the study of quantum phases of matter~\cite{AKLT1987,KITAEV20032,Levin2005stringnet,Kitaev2006anyon,ChenLiuWen2011,ChenGuLiuWen,LevinGu2012Z2SPT,ChenLuVishwanath2014,Heinrich16,Cheng17,WalkerWang2012,GDS,FHH19,Hybridfracton1,Hybridfracton2,Ellison22,Shirley22,Ellison2023subsystem,Chamon2005Fractonmodel,Haah2011FractonCodes,FractonDebut2016,Ma2017FractonCoupledLayer,Prem2019CageNetFractons,Wilbur2019FoliatedFractonOrder,Aasen2020DefectNetworkFractons,Vadali2024CompositeGauging,feng2026paulistabilizerformalismtopological,haah_invertible_2023,Sun2026QCAandISA}. They provide microscopic Hamiltonians whose physical properties can be analyzed directly at the lattice level. Among the most successful examples are stabilizer models~\cite{gottesman1997stabilizer}, which offer a particularly transparent and tractable description of quantum phases.
In contrast to commuting-projector models, which provide an exactly solvable description primarily of the ground-state sector, stabilizer models give direct algebraic control over the full spectrum: their eigenstates are organized by stabilizer eigenvalues, and are also mapped into each other by simple Pauli operators.
Kitaev's toric code provides a paradigmatic example of this framework, realizing a $\mathbb Z_2$ topological order using Pauli operators on qubits and serving as a prominent model of topological quantum memory~\cite{dennis_topological_2002}.
The stabilizer framework naturally extends to qudit systems. 
In two dimensions, translation-invariant Pauli stabilizer models over prime-dimensional qudits realize stacks of $\mathbb Z_p$ toric codes~\cite{BDCP12,bombin2014structure,Haah2018a}, while more general constructions using composite-dimensional qudits realize all Abelian bosonic topological phases with a gapped boundary~\cite{Ellison22}.
These developments establish stabilizer models not only as a cornerstone of quantum error correction but also as a powerful microscopic framework for understanding a large class of bosonic topological phases.

On the other hand, many topological phases can only be realized in the presence of physical fermions. Such phases are called \emph{intrinsically fermionic} topological phases, which {include} those with intrinsic topological order, as well as fermionic symmetry-protected topological phases \cite{GW14,GuWangWen14,GaiottoKapustin2016,W16,BGK17,WLG17,AasenLakeWalker2019,WangGu2020Construction}. Although explicit commuting-projector Hamiltonians for such phases have been constructed \cite{GW14,GuWangWen14,BGK17,ChengTantivasadakarnWang2018,WangGu2020Construction,ChenWang18,TarantinoFidkowski2016,Wareetal2016,SonAlicea18,SonAlicea19,ChenEllisonTantivasadakarn21,Sullivan20}, a simple stabilizer description has been lacking.
{
Within the broader class of commuting-projector models, stabilizer
Hamiltonians provide particularly direct algebraic control over
the spectrum through stabilizer eigenvalues~\cite{EF19, TV18, Eisert2021nonpauli}.}
This raises a natural question: can intrinsically fermionic topological phases be realized with a similar ``Pauli" stabilizer framework, akin to the bosonic case?
{
Such a framework should incorporate physical fermions while retaining
the algebraic transparency of Pauli stabilizer models, allowing anyons,
anomalous symmetries, condensation, gauging, and ungauging to be
analyzed directly on the lattice while also supporting applications
to quantum error correction.
}

{One candidate is the Majorana stabilizer formalism} \cite{BravyiTerhalLeemhuis2010}, which uses the anticommutation relations of fermions instead of the commutation relations of bosons. Indeed, Majorana stabilizer codes can be used to describe many topological and fracton phases~\cite{VijayHsiehFu15,VijayHaahFu2015}. However, such anticommutation relations only produce $\pm$ signs, which is insufficient to reproduce the fractional statistics of intrinsically fermionic topological orders. In this paper, we therefore employ a hybrid stabilizer structure to explicitly realize these phases. In particular, we show that by combining generalized Pauli operators and Majorana operators, we are able to realize all Abelian fermionic topological phases in two spatial dimensions that admit a gapped boundary as stabilizer error correcting codes.

The stabilizer models in this work are obtained by starting from a Pauli stabilizer model realizing a (twisted) quantum double model constructed in Ref.~\cite{Ellison22}. {We bind an emergent fermion $f$ in that model to a physical fermion $\psi$ and condense the composite boson $f\psi$}\footnote{The conventional term ``fermion condensation'' of an emergent fermion $f$ refers to condensation of the bosonic bound state $f\psi$, where $\psi$ is a physical fermion. We retain this terminology when discussing the resulting intrinsic fermionic topological order, with only fermion parity $\ZZ_2^F$ retained.}. This condensation procedure can be realized while preserving the stabilizer structure. Thus, the resulting code is also a stabilizer code. Furthermore, we can establish a web of dualities relating intrinsically fermionic topological phases by gauging various symmetries. For example, by further ungauging suitable charge sectors, we obtain stabilizer models realizing fermionic symmetry-enriched topological (SET) phases, as well as fermionic symmetry-protected topological (SPT) phases.

{Throughout this work, we use ``condensation'' when only the resulting intrinsic topological order is at issue. When the operation also retains the resulting dual 0-form symmetry, we call it ``ungauging''; for a bosonic charge sector, this is gauging the 1-form symmetry generated by the indicated anyons while keeping its dual 0-form symmetry. In fermionic constructions, the emergent charges are bound to local bosonic or fermionic charges to recover the full fermionic symmetry.Condensation may also preserve a pre-existing symmetry without retaining an additional dual symmetry.}

We also extend the construction beyond ordinary Majorana operators. For example, in charge-$D$ superconductors for an even number $D$, the conserved symmetry is only fermion number modulo $D$, denoted $\ZZ_D^F$. In this case, the local symmetry generator has an order larger than two, so the ordinary Majorana stabilizer framework is not sufficient to capture the full symmetry action. To accommodate this, we introduce a further level of hybridization between Pauli and Majorana operators, which are analogues of the $\ZZ_D$ clock and shift operators for fermions.  These operators can be realized as an extension of Majorana operators by the usual clock and shift operators. With this, we are able to give a simple stabilizer representation for fermionic SPT phases, whose symmetry is $\ZZ_D^F$, some of which do not admit a free-fermion realization~\cite{W16}.

{Our construction provides Majorana-Pauli stabilizer realizations of all Abelian fermionic topological orders with gapped boundaries and of families of fermionic SPT phases whose fully gauged theories are Abelian.}
Majorana-Pauli stabilizer codes therefore provide a unified lattice framework for fermionic Abelian phases and their bosonic shadows, making their anyon content, symmetry enrichment, and duality relations accessible within a stabilizer code formalism.

{The duality webs in Figs.~\ref{fig:Z_8_commutative_diagram} and~\ref{fig:Z_16_commutative_diagram} outline the structure of this paper, summarizing the phases we construct and the condensation, gauging, and ungauging operations that connect them.
Sec.~\ref{sec:general_constructions} presents our general constructions and specifies the classes of phases they realize.}

\begin{figure*}[t]
\centering
\resizebox{0.95\linewidth}{!}{%
\begin{tikzpicture}[
    >=Latex,
    thick,
    box/.style={
        draw,
        rectangle,
        minimum width=3.7cm,
        minimum height=1.15cm,
        align=center,
        font=\large
    },
    lab/.style={font=\large}
]

\node[box] (Z8TC)    at (0,0) {$\mathbb{Z}_8$ toric code};
\node[box] (FermTC)  at (7.5,3.6) {fermionic \\ toric code \\ (Sec.~\ref{sec:fermionic_toric_code})};
\node[box] (Z2FSPT)  at (15,3.6) {$\mathbb{Z}_2 \times \mathbb{Z}_2^F$ SPT \\ ($\nu=2$) \\ (Sec.~\ref{sec:Z2Z2F_SPT})};

\node[box] (Kmat)    at (7.5,0)
{$U(1)_4 \times U(1)_{-4}$\\
Chern-Simons theory\\
(Sec.~\ref{sec:U1_4_U1_minus4_TQD})
};

\node[box] (Z2enTC)  at (15,0) {$\mathbb{Z}_2$-enriched \\ $\mathbb{Z}_2$ toric code \\ (Sec.~\ref{sec:Z2_enriched_TC})};

\node[box] (Z8SPT)   at (0,-3.6) {trivial $\mathbb{Z}_8$ SPT};
\node[box] (Z4SPT)   at (7.5,-3.6) {$\mathbb{Z}_4$ SPT \\ ($p=2$)};

\draw[->] ([xshift=-0.22cm, yshift=0.18cm]Z8SPT.north) -- node[left=2pt,lab] {gauge $\mathbb{Z}_8$} ([xshift=-0.22cm, yshift=-0.18cm]Z8TC.south);

\draw[<-] ([xshift=0.22cm, yshift=0.18cm]Z8SPT.north) -- node[right=2pt,lab,align=center] {ungauge \\ $\mathbb{Z}_8$ charge} ([xshift=0.22cm, yshift=-0.18cm]Z8TC.south);

\draw[->] ([xshift=0.22cm, yshift=-0.18cm]FermTC.south) -- node[right=0.5pt,lab] {gauge $\mathbb{Z}_2^F$} ([xshift=0.22cm, yshift=0.18cm]Kmat.north);

\draw[<-] ([xshift=-0.22cm, yshift=-0.18cm]FermTC.south) -- node[left=0.5pt,lab,align=center] {condense \\ composite boson {\footnotesize ${\begin{pmatrix} 0 \\ 2 \end{pmatrix}}$}$\times \psi$}([xshift=-0.22cm, yshift=0.18cm]Kmat.north);

\draw[->] ([xshift=0.22cm, yshift=-0.18cm]Z2FSPT.south) -- node[right=2pt,lab] {gauge $\mathbb{Z}_2^F$} ([xshift=0.22cm, yshift=0.18cm]Z2enTC.north);

\draw[<-] ([xshift=-0.22cm, yshift=-0.18cm]Z2FSPT.south) -- node[left=2pt,lab] {ungauge $f\psi$} ([xshift=-0.22cm, yshift=0.18cm]Z2enTC.north);

\draw[->] ([xshift=-0.22cm, yshift=0.18cm]Z4SPT.north) -- node[left=2pt,lab] {gauge $\mathbb{Z}_4$} ([xshift=-0.22cm, yshift=-0.18cm]Kmat.south);

\draw[<-] ([xshift=0.22cm, yshift=0.18cm]Z4SPT.north) -- node[right=0.5pt,lab,align=center] {ungauge \\ $\mathbb{Z}_4$ boson {\footnotesize $\begin{pmatrix} 1 \\ 1 \end{pmatrix}$}
} ([xshift=0.22cm, yshift=-0.18cm]Kmat.south);

\draw[->]
([xshift=0.18cm]Z8TC.east) --
node[below=3pt,lab] {condense $e^4m^4$}
([xshift=-0.18cm]Kmat.west);

\draw[->]
([xshift=-0.18cm, yshift=-0.22cm]Z2enTC.west) --
node[below=2pt,lab] {gauge $\mathbb{Z}_2$}
([xshift=0.18cm, yshift=-0.22cm]Kmat.east);

\draw[<-]
([xshift=-0.18cm, yshift=0.22cm]Z2enTC.west) --
node[above=0pt,lab,align=center] {ungauge \\ $\mathbb{Z}_2$ boson {\footnotesize $\begin{pmatrix} 2 \\ 2 \end{pmatrix}$}}
([xshift=0.18cm, yshift=0.22cm]Kmat.east);

\draw[->]
([yshift=0.22cm,xshift=0.18cm]Z8SPT.east) --
node[above=2pt,lab] {projection}
([yshift=0.22cm,xshift=-0.18cm]Z4SPT.west);

\draw[->]
([yshift=-0.22cm,xshift=-0.18cm]Z4SPT.west) --
node[below=2pt,lab] {symmetry extension}
([yshift=-0.22cm,xshift=0.18cm]Z8SPT.east);

\draw[->]
([yshift=0.22cm,xshift=0.18cm]FermTC.east) --
node[above=2pt,lab] {ungauge $d^2\psi$}
([yshift=0.22cm,xshift=-0.18cm]Z2FSPT.west);

\draw[->]
([yshift=-0.22cm,xshift=-0.18cm]Z2FSPT.west) --
node[below=2pt,lab] {gauge $\mathbb{Z}_2$}
([yshift=-0.22cm,xshift=0.18cm]FermTC.east);

\draw[->]
([xshift=0cm,yshift=0.3cm]Z8TC.north) --
node[above=6pt,sloped,lab] {condense $e^2m^6\psi$}
([xshift=-0.3cm,yshift=-0.cm]FermTC.west);

\draw[->]
([xshift=0.3cm,yshift=0.3cm]Z4SPT.east) --
node[above=2pt,sloped,lab] {gauge $\mathbb{Z}_2$}
([xshift=-0.3cm,yshift=-0.3cm]Z2enTC.south);
\draw[<-]
([xshift=0.4cm,yshift=-0.1cm]Z4SPT.east) --
node[below=2pt,sloped,lab] {ungauge $e$}
([xshift=-0.2cm,yshift=-0.7cm]Z2enTC.south);

\end{tikzpicture}}
\caption{
Duality web of the two-dimensional stabilizer models with $\mathbb{Z}_8$ Pauli and Majorana operators constructed in this work. Each box denotes a topological phase, and the arrows denote gauging, ungauging, or anyon condensation. Ungauging retains the resulting 0-form symmetry, whereas condensation specifies the intrinsic topological order; fermion condensation with only fermion parity retained is included in the latter convention. Here, $\psi$ denotes the transparent physical fermion. Starting from the $\mathbb{Z}_8$ toric code, condensing the bosonic bound state $e^2m^6\psi$ gives the fermionic toric code, while condensing $e^4m^4$ gives the bosonic $U(1)_4\times U(1)_{-4}$ Chern-Simons theory, equivalently the gauged $\mathbb{Z}_4$ SPT phase with $p=2$ in the $\mathbb{Z}_4$ classification. All column-vector anyon labels are written in the basis of the diagonal $K=\mathrm{diag}(4,-4)$.
Ungauging the order-two charge sector generated by $d^2\psi$ in the fermionic toric code restores a bosonic $\mathbb{Z}_2$ symmetry and produces a $\mathbb{Z}_2\times\mathbb{Z}_2^F$ fermionic SPT phase ($\nu = 2$ in the $\ZZ_8$ classification). Gauging its fermion parity gives the bosonic shadow, a $\mathbb{Z}_2$-enriched toric code in which the remaining $\mathbb{Z}_2$ symmetry acts projectively on both $e$ and $m$. The bottom row relates the $\mathbb{Z}_4$ SPT to the trivial $\mathbb{Z}_8$ SPT by symmetry extension~\cite{WWW19}. The commutativity of the diagram summarizes the central result of this work: the unified stabilizer framework realizes and relates bosonic/fermionic topological orders and SPT/SET phases through explicit condensation, gauging, and ungauging operations. Gauging $\mathbb{Z}_2^F$ is implemented by coupling the physical fermion to a $\mathbb{Z}_2$ gauge field~\cite{Kitaev2006anyon,CKR18}. The four purely bosonic phases, the trivial $\mathbb{Z}_8$ SPT phase, the $\mathbb{Z}_4$ SPT phase, the $\mathbb{Z}_8$ toric code, and the $U(1)_4\times U(1)_{-4}$ theory, admit Pauli-stabilizer realizations using the construction of Ref.~\cite{Ellison22}. }
\label{fig:Z_8_commutative_diagram}
\end{figure*}

\begin{figure*}[t]
\centering
\begin{tikzpicture}[
    scale=0.67,
    transform shape,
    >=Latex,
    thick,
    box/.style={
        draw,
        rectangle,
        minimum width=4.35cm,
        minimum height=1.15cm,
        align=center,
        font=\large
    },
    smallbox/.style={
        draw,
        rectangle,
        minimum width=3.9cm,
        minimum height=1.15cm,
        align=center,
        font=\large
    },
    lab/.style={font=\large}
]

\node[box] (Z16TC)    at (0,10) {$\mathbb{Z}_{16}$ toric code};

\node[box] (Kmat)    at (7.0,8)
{
$U(1)_8 \times U(1)_{-8}$\\
Chern-Simons theory
};

\node[box] (K0442)  at (14,6) {
$\ZZ_4$ twisted quantum double\\[-0.1em]
{\normalsize $K=\begin{pmatrix}
0 & 4\\
4 & -2
\end{pmatrix}$}
};

\node[box] (Z2enTC)  at (21,4) {$\mathbb{Z}_4$-enriched \\ $\mathbb{Z}_2$ toric code};

\node[smallbox] (Z16SPT)   at (0,0) {trivial $\mathbb{Z}_{16}$ SPT};
\node[smallbox] (Z8SPT)    at (7.0,0) {$\mathbb{Z}_8$ SPT \\ ($p=4$)};
\node[smallbox] (Z4SPTp1)  at (14,0) {$\ZZ_4$ SPT \\ ($p=1$)};
\node[smallbox] (Z8FSPT)   at (21,0) {$ \mathbb{Z}_8^F$ SPT \\ (Sec.~\ref{sec:Z8F_SPT})};

\draw[->]
([xshift=-0.28cm,yshift=0.18cm]Z16SPT.north) --
node[left=5pt,lab] {gauge $\mathbb{Z}_{16}$}
([xshift=-0.28cm,yshift=-0.18cm]Z16TC.south);

\draw[<-]
([xshift=0.28cm,yshift=0.18cm]Z16SPT.north) --
node[right=5pt,lab,align=center] {{ungauge} \\ $\mathbb{Z}_{16}$ charge}
([xshift=0.28cm,yshift=-0.18cm]Z16TC.south);

\draw[->]
([xshift=-0.28cm,yshift=0.18cm]Z8SPT.north) --
node[left=5pt,lab,align=center] {gauge $\mathbb{Z}_8$}
([xshift=-0.28cm,yshift=-0.18cm]Kmat.south);

\draw[<-]
([xshift=0.28cm,yshift=0.18cm]Z8SPT.north) --
node[right=5pt,lab,align=center]
{{ungauge} \\ $\mathbb{Z}_8$ boson {\small $\begin{pmatrix} 1 \\ 1 \end{pmatrix}$}}
([xshift=0.28cm,yshift=-0.18cm]Kmat.south);

\draw[->]
([xshift=-0.28cm,yshift=0.18cm]Z4SPTp1.north) --
node[left=5pt,lab,align=center, pos=0.65] {gauge $\ZZ_4$}
([xshift=-0.28cm,yshift=-0.18cm]K0442.south);

\draw[<-]
([xshift=0.28cm,yshift=0.18cm]Z4SPTp1.north) --
node[right=5pt,lab,align=center,pos=0.65]
{{ungauge} \\ $\ZZ_4$ boson {\small $\begin{pmatrix} 0 \\ 1 \end{pmatrix}$}}
([xshift=0.28cm,yshift=-0.18cm]K0442.south);

\draw[->]
([xshift=0.24cm,yshift=0.18cm]Z8FSPT.north) --
node[right=6pt,lab,align=center]
{gauge $\mathbb{Z}_2^F$ \\
{(Sec.~\ref{sec:Z4enrichedTC_from_gauging_Z2F_in_ZF8SPT})}}
([xshift=0.24cm,yshift=-0.18cm]Z2enTC.south);

\draw[<-]
([xshift=-0.28cm,yshift=0.18cm]Z8FSPT.north) --
node[left=5pt,lab,align=center] {ungauge $f\psi$}
([xshift=-0.28cm,yshift=-0.18cm]Z2enTC.south);

\draw[->]
([yshift=0.25cm,xshift=0.18cm]Z16SPT.east) --
node[above=2pt,lab] {projection}
([yshift=0.25cm,xshift=-0.18cm]Z8SPT.west);

\draw[->]
([yshift=-0.25cm,xshift=-0.18cm]Z8SPT.west) --
node[below=2pt,lab,align=center] {symmetry\\extension}
([yshift=-0.25cm,xshift=0.18cm]Z16SPT.east);

\draw[->]
([yshift=0.25cm,xshift=0.18cm]Z8SPT.east) --
node[above=2pt,lab] {projection}
([yshift=0.25cm,xshift=-0.18cm]Z4SPTp1.west);

\draw[->]
([yshift=-0.25cm,xshift=-0.18cm]Z4SPTp1.west) --
node[below=2pt,lab,align=center] {symmetry\\extension}
([yshift=-0.25cm,xshift=0.18cm]Z8SPT.east);

\draw[->]
([yshift=-0.25cm,xshift=0.18cm]Z4SPTp1.east) --
node[below=2pt,lab,align=center] {symmetry\\extension}
([yshift=-0.25cm,xshift=-0.18cm]Z8FSPT.west);

\draw[->]
([yshift=+0.25cm,xshift=-0.18cm]Z8FSPT.west) --
node[above=2pt,lab] {projection}
([yshift=+0.25cm,xshift=0.18cm]Z4SPTp1.east);

\draw[->]
([xshift=0.18cm,yshift=-0.10cm]Z16TC.east) --
node[above=4pt,sloped,lab] {condense $e^8m^8~~$}
([xshift=-0.18cm,yshift=0.10cm]Kmat.west);

\draw[->]
([xshift=0.15cm,yshift=-0.10cm]Kmat.east) --
node[pos=0.53,above=5pt,sloped,lab,align=center]
{condense $\ZZ_2$ \\[-0.1em] boson
{\small ${\begin{pmatrix} 0 \\ 4 \end{pmatrix}}$}}
([xshift=-0.15cm,yshift=0.10cm]K0442.west);

{
\draw[->,rounded corners=5pt]
([xshift=0.15cm,yshift=0.12cm]Kmat.north east)
-- (10.0,8.95)
-- node[pos=0.5,below=5pt,lab,align=center]
{ungauge $\ZZ_4$ boson\\[-0.1em]
{\small $\begin{pmatrix}2\\-2\end{pmatrix}$}}
(21,8.95)
-- ([yshift=0.18cm]Z2enTC.north);
}

{
\draw[->]
([xshift=-0.10cm,yshift=-0.10cm]Z2enTC.south west) --
node[pos=0.53,above=3pt,sloped,lab]
{condense $m$}
([xshift=0.10cm,yshift=0.18cm]Z4SPTp1.north east);
}

\coordinate (outerTopLeft)  at ([xshift=-0cm,yshift=0.1cm]Kmat.north);
\coordinate (outerTopBend)  at ([xshift=-0cm,yshift=1.0cm]Kmat.north);
\coordinate (outerRightTop) at ([xshift=17cm,yshift=1.0cm]Kmat.north);
\coordinate (outerRightLow) at (24,0);
\coordinate (outerEntry)    at (Z8FSPT.east);

\draw[->, rounded corners=5pt]
(outerTopLeft)
-- (outerTopBend)
-- node[pos=0.50,above=5pt,lab,align=center]
{ungauge composite $\ZZ_8$ boson\\[-0.1em]
{\footnotesize $\begin{pmatrix} 3 \\ -1 \end{pmatrix}$}$\times\psi$}
(outerRightTop)
-- (outerRightLow)
-- (outerEntry);
\end{tikzpicture}
\caption{
Duality web of the two-dimensional stabilizer models with $\ZZ_{16}$ Pauli operators and generalized Majorana operators constructed in this work.
The generalized Majorana operators are the fermionic clock and shift operators introduced in Sec.~\ref{sec:fermionicclock}.
Here, $\psi$ denotes the transparent physical fermion. {The labels ``ungauge'' and ``condense'' follow the convention in Fig.~\ref{fig:Z_8_commutative_diagram}: ungauging retains the resulting 0-form symmetry, including the full fermionic symmetry in the fSPT constructions. For bosonic charge sectors, it gauges the indicated 1-form symmetry while retaining its dual 0-form symmetry. Condensation does not retain an additional dual symmetry, although it may preserve an existing symmetry.}
Starting from the $\ZZ_{16}$ toric code, condensing $e^8m^8$ gives the bosonic $U(1)_8\times U(1)_{-8}$ Chern--Simons theory, equivalently the gauged $\ZZ_8$ SPT phase with $p=4$ in the $\ZZ_8$ classification.
Condensing {$e^4m^{-4}$} (which is equivalent to first condensing $e^8m^8$ followed by condensing the boson {$(0,4)_{\rm diag}$}) instead gives the $\ZZ_4$ twisted quantum double, equivalently the gauged $\ZZ_4$ SPT phase with $p=1$.
All column-vector anyon labels adjacent to the $U(1)_8\times U(1)_{-8}$ node are written in the diagonal basis $K=\mathrm{diag}(8,-8)$.
The $\ZZ_4$ SPT with $p=1$ admits both bosonic and fermionic symmetry extensions. In the bosonic extension, its cocycle is pulled back along the quotient map $\ZZ_8\to\ZZ_4$. Since the $(2+1)$D bosonic SPT response is quadratic in the background gauge field, the SPT label transforms as $p\mapsto 4p$; hence the $\ZZ_4$ SPT with $p=1$ pulls back to the $\ZZ_8$ SPT with $p=4$. Applying again, the $\ZZ_8$ SPT with $p=4$ pulls back to the $\ZZ_{16}$ SPT with $p=16$, which is trivial.
In the fermionic extension, the same $\ZZ_4$ cocycle is pulled back along $\ZZ_8^F\to \ZZ_8^F/\ZZ_2^F\simeq \ZZ_4$. Equivalently, it is viewed as the $n_2=0$ sector of the $\ZZ_8^F$ supercohomology data~\cite{GW14}. The supercohomology constraint then reduces to $\delta\nu_3=0$; its nontrivial data are encoded in the bosonic cocycle $\nu_3$.
Ungauging the fermionic charge sector generated by $(3,-1)_{\rm diag}$ in the $U(1)_8\times U(1)_{-8}$ theory, with its odd powers bound to local fermionic charges, produces the $\ZZ_8^F$ fermionic SPT phase constructed in Sec.~\ref{sec:Z8F_SPT}.
Gauging its fermion parity then gives the $\ZZ_4$-enriched $\ZZ_2$ toric code discussed in Sec.~\ref{sec:Z4enrichedTC_from_gauging_Z2F_in_ZF8SPT}.
The same $\ZZ_4$-enriched $\ZZ_2$ toric code is also obtained directly from $U(1)_8\times U(1)_{-8}$ by {ungauging} the $\ZZ_4$ boson {$(2,-2)_{\rm diag}$}, as discussed in Sec.~\ref{sec:Z4entrichedTC_from_U88}.
{Further condensing the toric-code flux $m$, with the condensate inherited from $(0,4)_{\rm diag}$, preserves the existing $\ZZ_4$ symmetry and gives the $\ZZ_4$ SPT with $p=1$.}
Thus, this enlarged web makes explicit that the $\ZZ_4$ SPT, the $\ZZ_8$ SPT, the $\ZZ_8^F$ fermionic SPT, and the $\ZZ_4$-enriched $\ZZ_2$ toric code can all be obtained as descendants of the $\ZZ_{16}$ toric code through condensation {and ungauging}.
}
\label{fig:Z_16_commutative_diagram}
\end{figure*}

\section{Majorana-Pauli stabilizer realization of the fermionic toric code}
\label{sec:fermionic_toric_code}

{In this section, we present our main example: a Majorana-Pauli stabilizer model for the fermionic toric code\footnote{Note that this is not to be confused with the fermionic toric code in three spatial dimensions, which does not require physical fermions.}.} {We first describe its anyon content, then give the stabilizer and string operators graphically and explain their origin through anyon condensation.} 

The fermionic toric code is a fermionic analogue of the $\ZZ_2$ toric code. Similar to the double semion topological order~\cite{Levin2005stringnet,LevinGu2012Z2SPT}, it can be viewed as a twisted version of the toric code. The essential difference is that the fermionic toric code can only be realized in the presence of physical fermions: this makes it an intrinsically fermionic topological order rather than an ordinary bosonic topological order.

These phases can be described by Abelian Chern--Simons theory in the $K$-matrix formalism \cite{WenZee1992Kmatrix}. The toric code and double semion order are represented by
\begin{equation}
    K_{\mathrm{TC}}=
    \begin{pmatrix}
        0 & 2 \\
        2 & 0
    \end{pmatrix},
    \qquad
    K_{\mathrm{DS}}=
    \begin{pmatrix}
        0 & 2 \\
        2 & 2
    \end{pmatrix}.
\end{equation}
There are two fermionic toric codes (related by time-reversal), which can be represented by $K$-matrices with odd-diagonal entries~\cite{GuWangWen14}
\begin{equation}
    K_{\mathrm{fTC}}=
    \begin{pmatrix}
        0 & 2 \\
        2 & 1
    \end{pmatrix},
    \qquad
    K_{\overline{\mathrm{fTC}}}=
    \begin{pmatrix}
        0 & 2 \\
        2 & -1
    \end{pmatrix}.
\end{equation}
Here $\overline{\mathrm{fTC}}$ denotes the time-reversal conjugate of $\mathrm{fTC}$. Our lattice model below realizes $\overline{\mathrm{fTC}}$. Equivalently, $K_{\overline{\mathrm{fTC}}}$ is related by a $GL(2,\ZZ)$ transformation to
\begin{equation}
    \begin{pmatrix}
        4 & 0 \\
        0 & -1
    \end{pmatrix},
\end{equation}
and hence may be viewed as the (bosonic) chiral anyon theory $U(1)_4$ stacked with a fermionic invertible anti-chiral $U(1)_{-1}$, i.e. the (time-reversed) integer quantum Hall theory.

We now state the anyon data directly. The theory contains a physical transparent fermion $\psi$, which is a local excitation carrying odd fermion parity. {The equivalent diagonal $K$ presentation includes an invertible fermionic contribution $U(1)_{-1}$ that cancels the chiral central charge of the $U(1)_4$ sector so that the total system is nonchiral.} In addition, it contains a boson $b$ and a spin-$1/8$ excitation $d$, with topological spin
\begin{equation}
    \theta_d=\exp\left(\frac{2\pi i}{8}\right).
\end{equation}
Their fusion rules are
\begin{equation}
    b^2=1=\psi^2,\qquad d^2=b\psi .
\end{equation}
Thus $d$ is an order-four excitation whose square is the composite $b\psi$. The braiding between $b$ and $d$ gives a $-1$ phase, whereas $\psi$ braids trivially with all anyons; this is why $\psi$ is called transparent.

It is also useful to organize the eight simple excitations, including the local fermion $\psi$, by fermion parity. With our choice of convention, the fermion-even sector is generated by $d$:
\begin{equation}
    \{1,d,d^2,d^3\}
    =
    \{1,d,b\psi,bd\psi\}
    \simeq U(1)_4 ,
\end{equation}
while the fermion-odd sector is obtained by fusing these excitations with the physical fermion $\psi$:
\begin{equation}
    \{\psi,d\psi,d^2\psi,d^3\psi\}
    =
    \{\psi,d\psi,b,bd\}.
\end{equation}
Note that in this grading, the boson $b$ lies in the fermion-odd sector. This does not contradict the fact that $b$ is a boson, which refers to its exchange statistics, while ``fermion-odd'' refers to whether the anyon contains a physical fermion or not.

Having fixed the anyon convention, we now turn to the lattice realization. A commuting-projector lattice model for this phase was constructed in Ref.~\cite{GuWangWen14}. Here, we give a simpler stabilizer realization built from $\ZZ_8$ generalized Pauli operators and Majorana operators. We first introduce the notation for these two types of local degrees of freedom.
\subsection{Definition of the stabilizer model}

Our code is defined on the square lattice with one $\ZZ_8$ qudit placed on each edge and one complex fermion placed on each plaquette. The $\ZZ_8$ qudits have generalized Pauli operators $Z$ and $X$, also called clock and shift operators, defined by
\begin{equation}
    X_e = \sum_{j \in \ZZ_8} |j+1\rangle \langle j|,\qquad
    Z_e = \sum_{j \in \ZZ_8} \omega^j |j\rangle \langle j|,
\end{equation}
where addition is defined modulo $8$ and
\begin{equation}
    \omega = \exp\left(\frac{2\pi i}{8}\right).
\end{equation}
These $\ZZ_8$ generalized Pauli operators satisfy
\begin{equation}
    X_e^8 =1,\qquad Z_e^8 =1,
\end{equation}
and, for a pair of edges $e$ and $e'$, obey
\begin{equation}
Z_eX_{e'} =
\begin{cases}
\omega X_{e'}Z_e, & e=e', \\
X_{e'}Z_e, & e \neq e'.
\end{cases}
\label{eq:commutation_X_and_Z}
\end{equation}

We write the complex fermion operator $c_p$ on plaquette $p$ in terms of two Majorana operators $\gamma_p,\gamma'_p$
\begin{equation}
    \gamma_p  = c_p+c_p^\dagger,\qquad
    \gamma_p' = \frac{c_p-c_p^\dagger}{i}.
\end{equation}
They satisfy
\begin{equation}
    \{\gamma_p,\gamma_{p'}\}=2\delta_{p,p'},\quad
    \{\gamma'_p,\gamma'_{p'}\}=2\delta_{p,p'},\quad
    \{\gamma_p,\gamma'_{p'}\}=0 .
\end{equation}
The local fermion parity operator is
\begin{equation}
    P_p=1-2c_p^\dagger c_p=-i\gamma_p\gamma'_p .
\end{equation}
It anticommutes with the two Majorana operators on the same plaquette and commutes with Majorana operators on different plaquettes
\begin{align}
    P_p \gamma_{p'} &= (-1)^{\delta_{p,p'}} \gamma_{p'} P_p, \\
    P_p \gamma'_{p'} &= (-1)^{\delta_{p,p'}} \gamma'_{p'} P_p .
\end{align}

We now define the stabilizer model. The stabilizer generators are
\begin{align}
A_v &=
\begin{tikzpicture}[baseline=-0.6ex, line width=0.5pt]
  \def\L{1.2}
  \draw (-\L,0) -- (\L,0);
  \draw (0,0) -- (0,-\L);
  \draw (0,0) -- (0,\L);
  \draw (0,\L) -- (\L,\L);
  \draw (\L,\L) -- (\L,0);
  \node[op] at (0,0) {$v$};
  \node[op] at (-0.5*\L,0) {$\textcolor{blue}{X^\dagger}$};
  \node[op] at (0,-0.5*\L) {$\textcolor{blue}{X^\dagger}$};
  \node[op] at (0,0.5*\L) {$\textcolor{blue}{XZ}$};
  \node[op] at (0.5*\L,0) {$\textcolor{blue}{XZ^\dagger}$};
  \node[op] at (0.5*\L,\L) {$\textcolor{blue}{Z}$};
  \node[op] at (\L,0.5*\L) {$\textcolor{blue}{Z^\dagger}$};
\end{tikzpicture}
,\quad
B_p =
\raisebox{-0.6cm}{\begin{tikzpicture}[baseline=-0.6ex, line width=0.5pt]
  \def\L{1.2}
  \draw (0,0) rectangle (\L,\L);
  \node[op] at (0.5*\L,\L) {$\textcolor{blue}{Z^6}$};
  \node[op] at (0,0.5*\L) {$\textcolor{blue}{Z^6}$};
  \node[op] at (\L,0.5*\L) {$\textcolor{blue}{Z^2}$};
  \node[op] at (0.5*\L,0) {$\textcolor{blue}{Z^2}$};
  \node[op] at (0.5*\L,0.5*\L) {$\textcolor{red}{P_p}$};
\end{tikzpicture}}
\label{eq:ftc_stabilizers_A_B}
, \\
(-i)\cdot C_e &= ~~
\raisebox{-0.6cm}{\begin{tikzpicture}[baseline=-0.6ex, line width=0.5pt]
  \def\L{1.2}
  \draw (-\L,0) -- (0,0);
  \draw (0,0) -- (0,\L);
  \node[op] at (-0.5*\L,0) {$\textcolor{blue}{Z^2}$};
  \node[op] at (-0.5*\L,0.5*\L) {$\textcolor{red}{\gamma}$};
  \node[op] at (0.0*\L,0.5*\L) {$\textcolor{blue}{X_e^6}$};
  \node[op] at (0.5*\L, 0.53*\L) {$\textcolor{red}{\gamma'}$};
\end{tikzpicture}}
,\qquad 
\raisebox{0.6cm}{\begin{tikzpicture}[baseline=-0.6ex, line width=0.5pt]
  \def\L{1.2}
  \draw (0,0) -- (\L,0);
  \draw (0,0) -- (0,-\L);
  \node[op] at (0.5*\L,0) {$\textcolor{blue}{X_e^2}$};
  \node[op] at (0.5*\L,0.5*\L) {$\textcolor{red}{\gamma}$};
  \node[op] at (0,-0.5*\L) {$\textcolor{blue}{Z^2}$};
  \node[op] at (0.53*\L,-0.5*\L) {$\textcolor{red}{\gamma'}$};
\end{tikzpicture}}
\quad,
\label{eq:ftc_stabilizers_C}
\end{align}
where the blue labels denote Pauli operators on the $\mathbb{Z}_8$ qudits, while the red labels denote plaquette Majorana and fermion parity operators. The two forms of $C_e$ correspond to the two orientations of the edge $e$. Throughout this paper, when translating diagrams into operators, we order the Majorana operators with $\gamma$ to the left of $\gamma'$, namely as $i\gamma_{p_1}\gamma'_{p_2}$. The stabilizer group is
\begin{equation}
    \mathcal{S}_{\overline{\mathrm{fTC}}}
    =
    \left\langle \{A_v\},\{B_p\},\{C_e\}\right\rangle ,
    \label{eq:ftc_stabilizer_group}
\end{equation}
where $A_v$ is order eight, while $B_p$ and $C_e$ are order four. The ground-state subspace is the simultaneous $+1$ eigenspace of all stabilizers
\begin{equation}
    \mathcal{H}_L
    =
    \left\{
    |\Psi\rangle:
    S|\Psi\rangle=|\Psi\rangle,\ 
    \forall S\in \mathcal{S}_{\overline{\mathrm{fTC}}}
    \right\}.
\end{equation}
The commutativity of the stabilizers is straightforward to check. Away from overlapping supports, the generators trivially commute. On overlapping supports, the phases from the $\ZZ_8$ Pauli algebra are cancelled by the fermionic signs from the Majorana algebra. For example, when a $C_e$ operator is moved through a neighboring $B_p$, the Pauli part contributes a factor of $-1$, while the Majorana operator in $C_e$ anticommutes with the local fermion parity $P_p$ in $B_p$, giving a second factor of $-1$. Thus the total commutation phase is trivial. Similar local cancellations show that all generators in Eq.~\eqref{eq:ftc_stabilizer_group} commute.

Since the generators are finite-order unitary operators rather than order-two Hermitian operators, the Hamiltonian is most naturally written using projectors
\begin{equation}
    \Pi_S=\frac{1}{r_S}\sum_{n=0}^{r_S-1}S^n ,
\end{equation}
where $r_S$ is the order of the stabilizer $S$. The Hamiltonian is
\begin{equation}
    H_{\mathrm{fTC}}
    =
    -\sum_v \Pi_{A_v}
    -\sum_p \Pi_{B_p}
    -\sum_e \Pi_{C_e}.
    \label{eq:projector_Hamiltonian_fTC}
\end{equation}

{Let us compute the ground-state degeneracy on the torus. Suppose there are $N_v$ vertices, then $N_p=N_v$ and $N_e=2N_v$, so the total Hilbert-space dimension is $8^{N_e}2^{N_p}=2^{7N_v}$. Each $C_e$ has an $X_e^{\pm2}$ factor on its own edge, and no other $C$ generator acts with $X$ on that edge. The $N_e$ order-four $C_e$ generators are therefore independent, and their subgroup has order $4^{2N_v}$.

The $B_p$ terms contain no $X$ factors, so their subgroup has trivial intersection with the $C_e$ subgroup. A relation among the $B_p$ first requires even exponents to cancel the plaquette fermion parities. Cancellation of the remaining $Z^4$ factors then forces the exponents to agree on neighboring plaquettes. Connectedness leaves only the global relation
\begin{equation}
\prod_p B_p^2=1,
\label{eq:ftc_B_global_relation}
\end{equation}
so the $B_p$ terms contribute $4^{N_v}/2$.

The local relation needed to count the $A_v$ terms can be written graphically using Eqs.~\eqref{eq:ftc_stabilizers_A_B} and~\eqref{eq:ftc_stabilizers_C}:
\begin{equation}
\begin{tikzpicture}[baseline=-.6ex,x=1.1cm,y=1.1cm,
    localstab/.style={inner sep=1.5pt,fill=white}]
  \draw[black!45,line width=.4pt] (-1,-1) rectangle (1,1);
  \draw[black!45,line width=.4pt] (-1,0)--(1,0) (0,-1)--(0,1);
  \node[localstab,text=black] at (0,0) {$v$};
  \node[localstab] at (.5,0) {$C$};
  \node[localstab] at (0,.5) {$C^{-1}$};
  \node[localstab] at (-.5,0) {$C^{-1}$};
  \node[localstab] at (0,-.5) {$C$};
\end{tikzpicture}
\quad=\quad
\begin{tikzpicture}[baseline=-.6ex,x=1.1cm,y=1.1cm,
    localstab/.style={inner sep=1.5pt,fill=white}]
  \draw[black!45,line width=.4pt] (-1,-1) rectangle (1,1);
  \draw[black!45,line width=.4pt] (-1,0)--(1,0) (0,-1)--(0,1);
  \node[localstab] at (0,0) {$A_v^2$};
  \node[localstab] at (.5,.5) {$B$};
  \node[localstab] at (-.5,-.5) {$B$};
\end{tikzpicture}.
\label{eq:ftc_A_square_identity}
\end{equation}
Each diagram denotes the product of the stabilizers at the indicated edges, vertex, or plaquettes. The Pauli factors on both sides agree, while the four Majorana bilinears on the left reduce to the two plaquette parities, including their scalar phases. Thus $A_v^2$ is already fixed by the $B_p,C_e$ constraints, and each $A_v$ supplies at most one further order-two constraint. A product of $A_v$ terms has an even $X$ exponent on every edge only if its vertex exponents have the same parity throughout the connected lattice. The sole remaining relation is
\begin{equation}
\prod_v A_v=1.
\label{eq:ftc_A_global_relation}
\end{equation}
The $A_v$ terms therefore contribute $2^{N_v-1}$. These exact relations also exclude a nontrivial scalar stabilizer. Consequently,
\begin{align}
|\mathcal S_{\overline{\mathrm{fTC}}}|
&=4^{2N_v}\frac{4^{N_v}}{2}\,2^{N_v-1}=2^{7N_v-2},\\
\dim\mathcal H_L&=\frac{2^{7N_v}}{2^{7N_v-2}}=4.
\label{eq:ftc_torus_GSD}
\end{align}
The relation $\prod_pB_p=\prod_pP_p$ fixes total fermion parity to $+1$ in the ground space. The result agrees with the four topological sectors of the fermionic toric code, obtained by identifying excitations that differ by the local transparent fermion $\psi$.}

Next, we describe the representative string operators. Let $c$ be an oriented path on the dual lattice. The elementary string segments for the spin-$1/8$ excitation $d$ and the boson $b$ are
{
\begin{align}
W^d_e&\equiv \begin{tikzpicture}[baseline=-.6ex,>=Stealth,scale=.85]\draw[red,dashed,->] (-.6,0)--(.6,0);\draw[black] (0,-.6)--(0,.6)--(1.2,.6);\node[op,text=blue] at (0,0) {$X_e$};\node[op,text=blue] at (.6,.6) {$Z$};\end{tikzpicture},\qquad \begin{tikzpicture}[baseline=-.6ex,>=Stealth,scale=.85]\draw[red,dashed,->] (0,-.6)--(0,.6);\draw[black] (-.6,0)--(.6,0)--(.6,1.2);\node[op,text=blue] at (0,0) {$X_e^\dagger$};\node[op,text=blue] at (.6,.6) {$Z$};\end{tikzpicture},
\label{eq:d_string_segment}\\
W^b_e&\equiv
\begin{tikzpicture}[baseline=-.6ex,>=Stealth,scale=.85]
  \draw[red,dashed,->] (-.6,0)--(.6,0);
  \draw[black] (0,-.6)--(0,.6);
  \node[op,text=blue] at (0,0) {$X_e^4$};
\end{tikzpicture},\qquad
\begin{tikzpicture}[baseline=-.6ex,>=Stealth,scale=.85]
  \draw[red,dashed,->] (0,-.6)--(0,.6);
  \draw[black] (-.6,0)--(.6,0);
  \node[op,text=blue] at (0,0) {$X_e^4$};
\end{tikzpicture}.
\label{eq:b_string_segment}
\end{align}
The dashed arrows specify the direction of the hopping for the anyon; a reversed segment is represented by its dagger.} A longer string operator $W^a_c$ is obtained by multiplying the corresponding elementary segments along the path $c$. A closed string commutes with all stabilizers, while an open string creates excitations only at its endpoints.

{The fusion rule $d^2=b\psi$ can be seen by squaring a string of the hopping operator of $d$:}
\begin{equation}{
(W^d_c)^2=\resizebox{.77\columnwidth}{!}{\begin{tikzpicture}[scale=.95,baseline=(current bounding box.center),op/.style={inner sep=1.5pt,fill=white,text=blue},>=Stealth]
\draw[red,dashed,->] (-.65,.5)--(4.8,.5);
\foreach \x in {0,1,2,3,4}{\draw[black] (\x,0)--(\x,1);}
\draw[black] (0,1)--(5,1);
\foreach \x in {0,1,2,3,4}{
\node[op] at (\x,.5) {$X^2$};
\node[op] at (\x+.5,1) {$Z^2$};}
\end{tikzpicture}}.
}\end{equation}
{For this horizontal path, let $E(c)$ be the crossed vertical edges and $P(c)$ the plaquettes between consecutive crossed edges. We may simplify the expression using the following stabilizer product 
\begin{equation}
\mathcal R_c=\prod_{e\in E(c)}C_e^{-1}\prod_{p\in P(c)}B_p.
\label{eq:ftc_fusion_stabilizers}
\end{equation}
The result is}
\begin{equation}{
\begin{aligned}&(W^d_c)^2\mathcal R_c\\ &\quad=i\,\resizebox{.85\columnwidth}{!}{\begin{tikzpicture}[
  scale=.95,baseline=(current bounding box.center),>=Stealth,
  op/.style={inner sep=1.5pt,fill=white,text=blue},
  endpoint/.style={fill=white,inner sep=1pt,text=red}
]
\draw[red,dashed,postaction={decorate},
  decoration={markings,mark=at position .58 with {\arrow{Stealth[length=1.7mm,width=1.4mm]}}}]
  (-.65,.5)--(4.8,.5);
\foreach \x in {0,1,2,3,4}{
  \draw[black] (\x,0)--(\x,1);
}
\draw[black] (-1,0)--(0,0);
\draw[black] (4,1)--(5,1);
\node[op] at (-.5,0) {$Z^{-2}$};
\node[op] at (4.5,1) {$Z^2$};
\node[op,xshift=3pt] at (0,.5) {$X^4Z^{-2}$};
\foreach \x in {1,2,3}{
  \node[op] at (\x,.5) {$X^4$};
}
\node[op] at (4,.5) {$X^4Z^2$};
\node[endpoint] at (-.70,.5) {$\gamma_L$};
\node[endpoint] at (4.70,.5) {$\gamma'_R$};
\end{tikzpicture}}.\end{aligned}\label{eq:ftc_explicit_fusion_identity}
}\end{equation}
{The $X^4$ string is precisely $W^b_c$, while the red Majorana operators supply a physical fermion at each endpoint.} {The remaining $Z^{\pm2}$ factors are local endpoint dressings and do not change the anyon type.} {Thus the diagram gives}
\begin{equation}
d\times d=b\psi.
\end{equation}

The exchange and braiding statistics can also be read directly from the string algebra. The $b$ strings have trivial self-statistics. A $b$ string crossing a $d$ string produces the phase
\begin{equation}
    Z X^4=\omega^4 X^4 Z=-X^4 Z,
\end{equation}
so the mutual braiding between $b$ and $d$ is $-1$. {Finally, the T-junction calculation for exchanging two $d$ endpoints gives a single factor of $\omega$ from the local Pauli algebra.} {We present the details of the T-junction calculation in Fig.~\ref{fig:ftc_d_tjunction}.}

\begin{figure*}[t]{
\centering
\begin{tikzpicture}[
  scale=1.20,
  transform shape,
  op/.style={inner sep=1pt, fill=white, text=blue},
  wire/.style={black, line width=.5pt},
  darr/.style={red, dashed, -{Stealth[length=1.7mm]}, line width=.6pt},
  midarr/.style={red, dashed, line width=.6pt, postaction={decorate},
    decoration={markings,mark=at position #1 with {\arrow{Stealth[length=1.7mm,width=1.4mm]}}}}
]
\def\L{0.90}
\def\hL{0.45}

\coordinate (p) at (-5.85,1);
\node[font=\large] at (p) {$p$};
\draw[darr] (-7.45,1) -- (-6,1);
\draw[darr] (-4.2,1) -- (-5.65,1);
\draw[darr] (-5.85,-1) -- (-5.85,0.7);
\node[red, below] at (-6.53,.88) {$\bar\gamma_1$};
\node[red, below] at (-4.47,.88) {$\bar\gamma_3$};
\node[red, left]  at (-5.8,-.18) {$\bar\gamma_2$};

\begin{scope}[shift={(-.10,1.00)}]
  \node[font=\normalsize,inner sep=1pt] (pLeft) at (\hL,0) {$p$};
  \draw[midarr=.36] (-2.30,0) -- (pLeft.west);
  \node at (-2.48,0) {$\cdots$};

  \foreach \x in {-1.80,-.90,0}{
    \draw[wire] (\x,-\hL) -- (\x,\hL)
                -- (\x+\L,\hL);
    \node[op] at (\x,0) {$X$};
    \node[op] at (\x+\hL,\hL) {$Z$};
  }

  \node[below=12pt] at (-.90,0) {$W^d_{\bar\gamma_1}$};
\end{scope}

\begin{scope}[shift={(2.25,1.00)}]
  \node[font=\normalsize,inner sep=1pt] (pRight) at (-\hL,0) {$p$};
  \draw[midarr=.36] (2.30,0) -- (pRight.east);
  \node at (2.48,0) {$\cdots$};

  \foreach \x in {0,.90,1.80}{
    \draw[wire] (\x,-\hL) -- (\x,\hL)
                -- (\x+\L,\hL);
    \node[op] at (\x,0) {$X^{\dagger}$};
    \node[op] at (\x+\hL,\hL) {$Z^{\dagger}$};
  }

  \node[below=12pt] at (.90,0) {$W^d_{\bar\gamma_3}$};
\end{scope}

\begin{scope}[shift={(1.075,-.80)}]
  \node[font=\normalsize,inner sep=1pt] (pBottom) at (0,{.25+\hL}) {$p$};
  \draw[midarr=.62,shorten <=2pt] (pBottom.south) -- (0,-2.08);
  \node at (0,-2.34) {$\vdots$};

  \foreach \y in {.25,-.65,-1.55}{
    \draw[wire] (-\hL,\y) -- (\hL,\y)
                -- (\hL,\y+\L);
    \node[op] at (0,\y) {$X$};
    \node[op] at (\hL,\y+\hL) {$Z^{\dagger}$};
  }

  \node[right=9pt] at (.58,-.65)
    {$\bigl(W^d_{\bar\gamma_2}\bigr)^{\dagger}$};
\end{scope}

\end{tikzpicture}
\caption{Graphical T-junction calculation of the exchange statistics of $d$, following the string-operator method of Ref.~\cite{Ellison22}. The left sketch fixes the three incoming paths at $p$. The other panels show the operators $W^d_{\bar\gamma_1}$, $W^d_{\bar\gamma_3}$, and $(W^d_{\bar\gamma_2})^\dagger$; the downward arrow in the last panel indicates the adjoint of the incoming bottom path. In the comparison
\(W^d_{\bar\gamma_1}(W^d_{\bar\gamma_2})^\dagger W^d_{\bar\gamma_3}=\theta_d\,W^d_{\bar\gamma_3}(W^d_{\bar\gamma_2})^\dagger W^d_{\bar\gamma_1}\), the only noncommuting overlap is the right-hand edge of the junction plaquette: the factors obey $Z^\dagger X^\dagger=\omega X^\dagger Z^\dagger$. Hence $\theta_d=\omega=\exp(2\pi i/8)$ for clockwise-positive exchange. All other factors commute, and these representative $d$ strings contain no Majorana operators, so there is no additional fermionic reordering sign.}
\label{fig:ftc_d_tjunction}
}\end{figure*}

Therefore
\begin{equation}
    \theta_d=\omega=\exp\left(\frac{2\pi i}{8}\right),
    \quad
    \theta_b=1,
    \quad
    \theta_\psi=-1.
\end{equation}
Together with the fusion rule above, this reproduces the anyon data of $\overline{\mathrm{fTC}}$.

\subsection{Construction by anyon condensation}

We now explain how the above stabilizer model arises from anyon condensation. This derivation also clarifies why $\ZZ_8$ qudits and Majorana operators naturally appear in the construction.

\subsubsection{Anyon-level construction}

Start from the $\ZZ_8$ toric code, whose anyons form a $\ZZ_8\times \ZZ_8$ fusion group generated by $e$ and $m$. We use the convention in which the topological spin of $e^p m^q$, with $p,q\in\ZZ_8$, is
\begin{equation}
    \theta_{e^p m^q} = \omega^{pq},
    \qquad
    \omega=\exp\left(\frac{2\pi i}{8}\right).
\end{equation}
Equivalently, the mutual braiding between $e^p m^q$ and $e^{p'}m^{q'}$ is
\begin{equation}
    M_{e^p m^q,e^{p'}m^{q'}} = \omega^{p q'+q p'}.
\end{equation}
We also include a physical transparent fermion $\psi$ with $\theta_\psi=-1$.

The desired fermionic toric code is obtained by condensing the boson
\begin{equation}
    \chi=e^2m^6\psi .
\end{equation}
Indeed,
\begin{equation}
    \theta_\chi = \theta_{e^2m^6}\theta_{\psi} = \omega^{12}\cdot(-1) = 1,
\end{equation}
so $\chi$ is a boson. It has order four, since
\begin{equation}
    \chi^2=e^4m^4, \qquad \chi^4=1 .
\end{equation}
Thus, condensing $\chi$ also condenses its square $e^4m^4$.

An anyon $e^p m^q\psi^s$ remains deconfined after condensation precisely when it braids trivially with $\chi$. Since $\psi$ is transparent, this condition is
\begin{equation}
    M_{e^p m^q,\chi} = M_{e^p m^q,e^2m^6} = \omega^{6p+2q} = 1,
\end{equation}
or equivalently
\begin{equation}
    3p+q=0 \mod 4.
\end{equation}
The deconfined anyons, modulo the identification by fusion with the condensed boson $\chi$, are generated by
\begin{equation}
    d=[em],
    \qquad
    b=[m^4],
    \qquad
    \psi=[\psi].
\end{equation}
Here, square brackets $[\cdot]$ denote the equivalence class after condensation. Their spins are
\begin{equation}
    \theta(d)=\omega=\exp\left(\frac{2\pi i}{8}\right),
    \qquad
    \theta(b)=1,
    \qquad
    \theta(\psi)=-1.
\end{equation}
Moreover,
\begin{equation}
    b^2=[m^8]=1,
    \qquad
    \psi^2=1.
\end{equation}
Since we condensed the boson $\chi$, it is identified with the topological vacuum, i.e. $\chi=1$. Then, we have
\begin{equation}
    e^2m^6=\psi.
\end{equation}
Therefore
\begin{equation}
    d^2=[e^2m^2]
    =
    [m^4\psi]
    =
    b\psi .
\end{equation}
Thus, the condensation produces exactly the anyon theory described at the beginning of this section.

\subsubsection{Lattice-level construction}
\label{sec:fermionic_condensation}

The stabilizer model above is the lattice realization of the same condensation procedure. We start with the $\ZZ_8$ toric code on the square lattice and put one fermion on each plaquette in the unoccupied state, so that $P_p=1$. Before condensation, a convenient generating set is
\begin{equation}
    \raisebox{-0.5\height}{\includegraphics[]{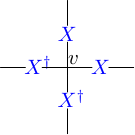}},
    \qquad
    \raisebox{-0.5\height}{\includegraphics[]{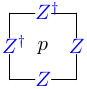}},
    \qquad
    {\color{red}P_p} .
    \label{eq:parent_Z8_TC_stabilizers}
\end{equation}
{We define the $e$ anyon to be the excitation carrying the eigenvalue $\omega$ under the vertex term, and the $m$ anyon to be the excitation carrying the eigenvalue $\omega^{-1}$ under the plaquette term.} {This convention is consistent with the clockwise-positive orientation used above.} The physical fermion is denoted by $\psi$.

To condense
\begin{equation}
    \chi=e^2m^6\psi,
\end{equation}
we add short hopping operators for this boson. On the lattice, these are precisely the $C_e$ operators
\begin{equation}
    (-i) \cdot C_e = ~~
    \raisebox{-0.6cm}{\begin{tikzpicture}[baseline=-0.6ex, line width=0.5pt]
      \def\L{1.2}
      \draw (-\L,0) -- (0,0);
      \draw (0,0) -- (0,\L);
      \node[op] at (-0.5*\L,0) {$\textcolor{blue}{Z^2}$};
      \node[op] at (-0.5*\L,0.5*\L) {$\textcolor{red}{\gamma}$};
      \node[op] at (0.0*\L,0.5*\L) {$\textcolor{blue}{X_e^6}$};
      \node[op] at (0.5*\L, 0.53*\L) {$\textcolor{red}{\gamma'}$};
    \end{tikzpicture}}
    ,\qquad 
    \raisebox{0.6cm}{\begin{tikzpicture}[baseline=-0.6ex, line width=0.5pt]
      \def\L{1.2}
      \draw (0,0) -- (\L,0);
      \draw (0,0) -- (0,-\L);
      \node[op] at (0.5*\L,0) {$\textcolor{blue}{X_e^2}$};
      \node[op] at (0.5*\L,0.5*\L) {$\textcolor{red}{\gamma}$};
      \node[op] at (0,-0.5*\L) {$\textcolor{blue}{Z^2}$};
      \node[op] at (0.53*\L,-0.5*\L) {$\textcolor{red}{\gamma'}$};
    \end{tikzpicture}}~.
   \label{eq:condensate_hopping_ftc}
\end{equation}
Each $C_e$ moves the condensed boson $\chi$ across the edge $e$ (i.e. along the dual lattice). Imposing $C_e=1$ proliferates $\chi$ strings and identifies anyons that differ by fusion with $\chi$.

{Condensation retains the products of parent stabilizers that commute with every $C_e$, namely the centralizer within the parent stabilizer group. The displayed $A_v$ and $B_p$ commute with all $C_e$ and generate this centralizer.
Thus, the condensed stabilizer group is
\begin{equation}
\mathcal S_{\overline{\mathrm{fTC}}}
=\left\langle\{A_v\},\{B_p\},\{C_e\}\right\rangle.
\end{equation}
} In this way, the stabilizer model realizes the anyon-level condensation
\begin{equation}
    \ZZ_8\ \text{toric code} \times \{1,\psi\}
    \xrightarrow{\ \text{condense } e^2m^6\psi\ }
    \overline{\mathrm{fTC}} .
\end{equation}

\section{The duality web for the fermionic toric code}
\label{sec:Z4F_duality_web}

\noindent{
The main result of this section is an explicit implementation of condensation, ungauging, and gauging that preserves the Majorana-Pauli stabilizer description. We construct a $\ZZ_2\times\ZZ_2^F$ fermionic SPT and its fermion-parity-gauged $\ZZ_2$-enriched toric code, and connect them to the fermionic toric code in Fig.~\ref{fig:Z_8_commutative_diagram}. The phase-level gauging, ungauging, and condensation relations provide the organizing framework; our contribution is their realization through local commuting checks and explicit symmetry operators. In particular, the fermionic SPT phase itself was previously realized in Refs.~\cite{TV18,EF19}.

We first describe the phases and arrows in the diagram, then give the stabilizer constructions in the following subsections. The longer checks of the remaining bosonic ungauging routes are collected in Appendix~\ref{app:Z8_duality_condensation_checks}. Appendix~\ref{app:stable_equivalence_Z2_SET} gives the stable-equivalence construction relating the two microscopic realizations of the enriched toric code.}

The upper part of the diagram starts from the $\ZZ_8$ toric code. As shown in Sec.~\ref{sec:fermionic_toric_code}, condensing the order-four boson $e^2m^6\psi$ produces the fermionic toric code. Starting from the fermionic toric code, one can further ungauge the charge subgroup generated by the boson
\begin{equation}
    b=d^2\psi,
\end{equation}
(equivalently, gauge the 1-form symmetry generated by $b$ while retaining the dual 0-form symmetry) to obtain a fermionic SPT phase protected by $\ZZ_2\times\ZZ_2^F$. Conversely, gauging the resulting dual bosonic $\ZZ_2$ symmetry of this fermionic SPT recovers the fermionic toric code.

Gauging the fermion parity $\ZZ_2^F$ of the $\ZZ_2\times\ZZ_2^F$ fermionic SPT gives its bosonic shadow. In the present case, this bosonic shadow is a $\ZZ_2$-enriched toric code, whose anyons $e$ and $m$ are fractionalized by the remaining $\ZZ_2$ symmetry. Thus, the fermionic toric code, the $\ZZ_2\times\ZZ_2^F$ fSPT, and the $\ZZ_2$-enriched toric code are related by the two standard operations: gauging a bosonic symmetry and gauging fermion parity.

There is also a second fermion-parity gauging operation in the diagram. One may gauge $\ZZ_2^F$ directly in the fermionic toric code. This gives a bosonic topological order described by the $K$-matrix
\begin{equation}
    K=
    \begin{pmatrix}
        0 & 4 \\
        4 & -4
    \end{pmatrix}~,
\end{equation}
which is a $\ZZ_4$ twisted quantum double. Alternatively, by a $GL(2,\ZZ)$ change of basis, we may also represent this anyon theory as 
\begin{equation}
    K'=
    \begin{pmatrix}
        4 & 0 \\
        0 & -4
    \end{pmatrix}
    =
    \begin{pmatrix}
        1 & 0 \\
        1 & 1
    \end{pmatrix}^T
    \begin{pmatrix}
        0 & 4 \\
        4 & -4
    \end{pmatrix}
    \begin{pmatrix}
        1 & 0 \\
        1 & 1
    \end{pmatrix}.
\end{equation}
Thus, this bosonic topological order can equivalently be written as $U(1)_4\times U(1)_{-4}$.

The same bosonic topological order can also be obtained directly from the $\ZZ_8$ toric code by condensing the order-two boson $e^4m^4$. Equivalently, it can be obtained by gauging a bosonic $\ZZ_4$ SPT. Since
\begin{equation}
    H^3(\ZZ_4,\RR/\ZZ)\cong \ZZ_4,
\end{equation}
the gauged $\ZZ_4$ SPTs are the $\ZZ_4$ twisted quantum doubles. In the standard $K$-matrix convention,
\begin{equation}
    K_p=
    \begin{pmatrix}
        0 & 4 \\
        4 & -2p
    \end{pmatrix},
    \qquad
    p\in \ZZ_4 .
\end{equation}
The phase appearing in Fig.~\ref{fig:Z_8_commutative_diagram} corresponds to the element $p=2$, giving
\begin{equation}
    K_{p=2}=
    \begin{pmatrix}
        0 & 4 \\
        4 & -4
    \end{pmatrix}
    \sim
    \begin{pmatrix}
        4 & 0 \\
        0 & -4
    \end{pmatrix}
    =
    U(1)_4\times U(1)_{-4}.
\end{equation}
In this sense, the duality web relates the same set of phases through several complementary operations. These are condensation in the $\ZZ_8$ toric code, ungauging charge subgroups while retaining the dual symmetries, gauging bosonic symmetries of SPT phases, and gauging fermion parity of fermionic phases. In the following subsections, we implement these relations explicitly at the level of stabilizer models.

\subsection{\texorpdfstring{$\ZZ_2\times\ZZ_2^F$}{Z2 x Z2F} SPT construction}
\label{sec:Z2Z2F_SPT}

{We first give the stabilizer realization of the $\ZZ_2\times\ZZ_2^F$ fermionic SPT introduced above. The construction starts from the fermionic toric code of Sec.~\ref{sec:fermionic_toric_code}; the boundary symmetry calculation below identifies the resulting SPT data.}

We can arrive at the fermionic SPT by starting from the fermionic toric code and ungauging the charge subgroup generated by the boson
\begin{equation}
    b=d^2\psi .
\end{equation}
Note that $b$ is a boson in the sense of exchange statistics, even though it lies in the fermion-parity-odd sector as defined in Sec.~\ref{sec:fermionic_toric_code}. Since $b$ has mutual braiding $-1$ with the anyon $d$, condensing $b$ confines $d$. Thus a bare condensation of $b$ removes the intrinsic topological order of the fermionic toric code, leaving only the physical transparent fermion. After forgetting any additional symmetry data, the resulting phase is an invertible fermionic phase with only $\ZZ_2^F$ manifest.

To retain the dual 0-form symmetry, we ungauge the charge subgroup generated by $b$; equivalently, we gauge its $\ZZ_2$ 1-form symmetry. To facilitate the gauging, we enlarge the Hilbert space by placing an auxiliary $\ZZ_2$ qubit on each plaquette, initialized in the state $X=1$. This acts as a dynamical 2-form gauge field for gauging the 1-form symmetry. We then impose the Gauss law for gauging this 1-form symmetry as
\begin{equation}
    D_e
    =
    {\color{orange}Z_{L(e)}}\,W^b_e\,{\color{orange}Z_{R(e)}}~,
\end{equation}
where $L(e)$ and $R(e)$ denote the two plaquettes adjacent to $e$. Pictorially,
\begin{equation}
D_e :=
\left\{~~
\tikz[baseline=-0.6ex]{
  \def\L{\Ledge}
  \draw (0,-0.5*\L) -- (0,0.5*\L);
  \node[op] at (-0.5*\L,0) {$\color{orange}Z$};
  \node[op] at (0,0) {$\color{blue}X_e^4$};
  \node[op] at (0.5*\L,0) {$\color{orange}Z$};
}
\;,\quad
\tikz[baseline=-0.6ex]{
  \def\L{\Ledge}
  \draw (-0.5*\L,0) -- (0.5*\L,0);
  \node[op] at (0,0.5*\L) {$\color{orange}Z$};
  \node[op] at (0,0) {$\color{blue}X_e^4$};
  \node[op] at (0,-0.5*\L) {$\color{orange}Z$};
}
~~\right\}.
\end{equation}
Alternatively, one can think of this auxiliary system as a $\ZZ_2$ paramagnet equipped with the (soon to be) dual $\ZZ_2$ $0$-form symmetry $\prod_p {\color{orange}X_p}$. Imposing the Gauss law implements ungauging by identifying $b$ with a charge of this dual symmetry, which is created by $Z$ on each plaquette. After ungauging, the boson $b$ is thus a local $\ZZ_2$ charge of the dual symmetry (see Ref.~\cite{Ellison22} for an extended discussion of this point).

We impose the ungauging constraints energetically by adding the commuting stabilizer term
\begin{equation}
    H_D=-\sum_e D_e .
\end{equation}
Equivalently, in the ground-state subspace,
\begin{equation}
    D_e|\Psi\rangle=|\Psi\rangle ,
    \qquad \forall e .
\end{equation}
Thus, when we write $D_e=1$ below, we mean equality within the low-energy stabilizer subspace. We do not impose $D_e=1$ as a kinematic hard constraint on the microscopic Hilbert space; states with $D_e=-1$ remain in the Hilbert space but are gapped by the Hamiltonian. In the infinite-coupling limit for $H_D$, this energetic constraint is equivalent to restricting to the $D_e=+1$ subspace. In this sense, proliferating the $D_e$ operators implements ungauging of the charge subgroup generated by $b$.

We now determine the stabilizers that remain after the $D_e$ ungauging constraints are imposed. We start from the fermionic toric-code stabilizers $A_v,B_p,C_e$ in Eqs.~\eqref{eq:ftc_stabilizers_A_B} and \eqref{eq:ftc_stabilizers_C}, together with the product-state stabilizers $\color{orange}X_p$ for the ancillary plaquette spins. The stabilizer group after ungauging is generated by combinations of these operators that commute with every $D_e$. {The plaquette and edge stabilizers $B_p$ and $C_e$ already commute with the $D_e$'s. The vertex stabilizer is dressed by the ancillary $X$ on its northeast plaquette, as displayed below; its two anticommutations with the adjacent $D_e$ terms cancel those of the vertex operator.} The resulting commuting stabilizers, in addition to the $D_e$ constraints, are

\begin{align}
A_v' &=
\begin{tikzpicture}[baseline=-0.6ex, line width=0.5pt]
  \def\L{1.2}
  \draw (-\L,0) -- (\L,0);
  \draw (0,0) -- (0,-\L);
  \draw (0,0) -- (0,\L);
  \draw (0,\L) -- (\L,\L);
  \draw (\L,\L) -- (\L,0);
  \node[op] at (0,0) {$v$};
  \node[op] at (-0.5*\L,0) {$\textcolor{blue}{X^\dagger}$};
  \node[op] at (0,-0.5*\L) {$\textcolor{blue}{X^\dagger}$};
  \node[op] at (0,0.5*\L) {$\textcolor{blue}{XZ}$};
  \node[op] at (0.5*\L,0) {$\textcolor{blue}{XZ^\dagger}$};
  \node[op] at (0.5*\L,\L) {$\textcolor{blue}{Z}$};
  \node[op] at (\L,0.5*\L) {$\textcolor{blue}{Z^\dagger}$};
  \node[op] at (0.5*\L,0.5*\L) {$\textcolor{orange}{X}$};
\end{tikzpicture}
,\quad
B_p =
\raisebox{-0.6cm}{\begin{tikzpicture}[baseline=-0.6ex, line width=0.5pt]
  \def\L{1.2}
  \draw (0,0) rectangle (\L,\L);
  \node[op] at (0.5*\L,\L) {$\textcolor{blue}{Z^6}$};
  \node[op] at (0,0.5*\L) {$\textcolor{blue}{Z^6}$};
  \node[op] at (\L,0.5*\L) {$\textcolor{blue}{Z^2}$};
  \node[op] at (0.5*\L,0) {$\textcolor{blue}{Z^2}$};
  \node[op] at (0.5*\L,0.5*\L) {$\textcolor{red}{P_p}$};
\end{tikzpicture}}
, \\
(-i)\cdot C_e &= ~~
\raisebox{-0.6cm}{\begin{tikzpicture}[baseline=-0.6ex, line width=0.5pt]
  \def\L{1.2}
  \draw (-\L,0) -- (0,0);
  \draw (0,0) -- (0,\L);
  \node[op] at (-0.5*\L,0) {$\textcolor{blue}{Z^2}$};
  \node[op] at (-0.5*\L,0.5*\L) {$\textcolor{red}{\gamma}$};
  \node[op] at (0.0*\L,0.5*\L) {$\textcolor{blue}{X_e^6}$};
  \node[op] at (0.5*\L, 0.53*\L) {$\textcolor{red}{\gamma'}$};
\end{tikzpicture}}
,\qquad 
\raisebox{0.6cm}{\begin{tikzpicture}[baseline=-0.6ex, line width=0.5pt]
  \def\L{1.2}
  \draw (0,0) -- (\L,0);
  \draw (0,0) -- (0,-\L);
  \node[op] at (0.5*\L,0) {$\textcolor{blue}{X_e^2}$};
  \node[op] at (0.5*\L,0.5*\L) {$\textcolor{red}{\gamma}$};
  \node[op] at (0,-0.5*\L) {$\textcolor{blue}{Z^2}$};
  \node[op] at (0.53*\L,-0.5*\L) {$\textcolor{red}{\gamma'}$};
\end{tikzpicture}}
\quad,
\end{align}
where the blue labels denote Pauli operators on the $\mathbb{Z}_8$ qudits, the orange label denotes the ancillary $\mathbb{Z}_2$ face qubit, and the red labels denote plaquette fermionic operators.
The model has an onsite $\ZZ_2\times\ZZ_2^F$ symmetry. The bosonic
$\ZZ_2$ symmetry is generated by
\begin{equation}
    U(g)=
    \left(\prod_p {\color{orange} X_p}\right)^g,
    \qquad g\in\ZZ_2=\{0,1\},
\end{equation}
where $p$ refers to each plaquette, and the fermion-parity symmetry is generated by
\begin{equation}
    P_F=\prod_p {\color{red} P_p}.
\end{equation}
The operator $U(g)$ is a genuine global symmetry, commuting with all stabilizers. Similarly, $P_F$ commutes with all stabilizers because every stabilizer is fermion-parity even. Hence, the ungauged model realizes an invertible fermionic phase protected by $\ZZ_2\times\ZZ_2^F$.

We next identify which $\ZZ_2\times\ZZ_2^F$ fSPT this model realizes by extracting its supercohomology data from the anomalous boundary symmetry action. Recall that a supercohomology fermionic SPT protected by $G\times \ZZ_2^F$ is described by cochains~\cite{GW14}
\begin{equation}
    \nu_3\in C^3(G,\RR/\ZZ), \quad n_2\in Z^2(G,\ZZ_2),
\end{equation}
satisfying
\begin{equation}
    \delta\nu_3=\frac{1}{2}n_2\cup n_2 .
    \label{eq:cohomology_equation}
\end{equation}
This data do not include possible Kitaev-chain decorations \cite{GaiottoJohnsonFreyd19,WangGu2020Construction,Barkeshli2022Classification}, as they are absent in the stabilizer models considered here.

We can actually extract this topological data of $\nu_3$ and $n_2$ by looking at its anomalous symmetry action on the boundary, following the description of Else-Nayak~\cite{EN15}. {We act with the $\ZZ_2$ symmetry on the lower half-plane $R$:}
\begin{equation}
    U_R(1):=\prod_{p\in R} {\color{orange} X_p}.
\end{equation}
{
Multiplying by the bulk vertex stabilizers cancels the ancillary $X_p$ factors and the Pauli operators on interior edges, leaving the boundary string shown below:
\begin{align}
&\vcenter{\hbox{
{\color{black}\begin{tikzpicture}[scale=1.2,baseline=5pt]
\foreach \x in {0,1,2,3,4}{
    \draw[dashed] (\x,0) -- (\x,4);
}
\foreach \y in {0,1,2,3}{
    \draw[dashed] (-0.3,\y) -- (4.3,\y);
}
\draw[line width=0.8pt] (-0.3,4) -- (4.3,4);
\foreach \x in {0.5,1.5,2.5,3.5}{
    \foreach \y in {0.5,1.5,2.5,3.5}{
        \node at (\x,\y) {$\textcolor{orange}{X}$};
    }
}
\end{tikzpicture}}
}} \\[1em]
\cong\quad
&\vcenter{\hbox{
{\color{black}\begin{tikzpicture}[scale=1.2,baseline=5pt]
\foreach \x in {0,1,2,3,4}{
    \draw[dashed] (\x,0) -- (\x,-4);
}
\foreach \y in {-1,-2,-3,-4}{
    \draw[dashed] (-0.3,\y) -- (4.3,\y);
}
\draw[line width=0.8pt] (-0.3,0) -- (4.3,0);
\foreach \x in {0,1,2,3,4}{
    \node[op] at (\x,-0.5) {$\textcolor{blue}{X}$};
}
\foreach \x in {0.5,1.5,2.5,3.5}{
    \node[op] at (\x,0) {$\textcolor{blue}{Z}$};
}
\end{tikzpicture}}
}} \,
:= \widetilde{U}(1)~.
\end{align}
The $\ZZ_2$ group law implies that the full boundary operator satisfies
\begin{equation}
    \widetilde{U}(1)^2\sim \mathbf{1},
\end{equation}
where $\sim$ denotes equality up to multiplication by bulk stabilizers.

We now restrict the boundary symmetry to a finite interval $l=[a,b]$ and denote the truncated operator by $\widetilde{U}_l(1)$. We choose the truncation shown below:
\begin{eqs}
   \widetilde{U}_l (1)=\;&~~
\raisebox{0.8em}{{\color{black}\begin{tikzpicture}[
  baseline=(current bounding box.center),
  scale=0.85,
  op/.style={inner sep=1pt, fill=white, text=blue},
  redop/.style={inner sep=1pt, fill=white, text=red},
  orangeop/.style={inner sep=1pt, fill=white, text=orange},
  wire/.style={black, line width=0.8pt},
  dashline/.style={gray, dashed, line width=0.55pt}
]
\def\L{1.2}

\begin{scope}[shift={(-2.3,0)}]
  \draw[wire] (-0.7*\L,\L) -- (1.9*\L,\L);
  \draw[dashline] (0,\L) -- (0,-0.15*\L);
  \draw[dashline] (\L,\L) -- (\L,-0.15*\L);
  \draw[dashline] (-0.7*\L,0) -- (1.9*\L,0);
  \node at (0,1.3*\L) {$a$};
  \fill (0,\L) circle (2pt);
  \node[op] at (0.5*\L,\L) {$Z$};
  \node[op] at (1.5*\L,\L) {$Z$};
  \node[op] at (0,0.5*\L) {$X$};
  \node[op] at (\L,0.5*\L) {$X$};
\end{scope}

\node at (0.6,0.5*\L) {$\cdots$};

\begin{scope}[shift={(2.0,0)}]
  \draw[wire] (-0.7*\L,\L) -- (2.0*\L,\L);
  \draw[dashline] (0,\L) -- (0,-0.15*\L);
  \draw[dashline] (\L,\L) -- (\L,-0.15*\L);
  \draw[dashline] (-0.7*\L,0) -- (1.9*\L,0);
  \node at (2*\L,1.3*\L) {$b$};
  \fill (2*\L,\L) circle (2pt);
  \node[op] at (0.5*\L,\L) {$Z$};
  \node[op] at (1.5*\L,\L) {$Z$};
  \node[op] at (0,0.5*\L) {$X$};
  \node[op] at (\L,0.5*\L) {$X$};
\end{scope}
\end{tikzpicture}}}
\label{eq:tildeU(1)_l}
\end{eqs}
Since $\widetilde{U}_l(1)$ now has two endpoints, the relation $\widetilde{U}(1)^2 \sim \mathbf{1}$ no longer holds exactly after truncation. Instead, the truncated operators satisfy
\begin{equation}
\widetilde{U}_l(g)\widetilde{U}_l(h)
\sim
\Omega(g,h)\widetilde{U}_l([g+h]_2), \quad g, h \in \mathbb{Z}_2,
\label{eq:EN_composition_Z2}
\end{equation}
Here $\Omega(g,h)$ is supported near the two endpoints. For $g=h=1$, it factorizes as 
\begin{equation}
    \Omega(1,1)=\Omega_a(1,1)\,\Omega_b(1,1),
\end{equation}
where $\Omega_a$ and $\Omega_b$ are localized near the left and right endpoints respectively. 

In the Else--Nayak procedure, we extract the cocycle from the action of $\widetilde U_l$ on the left endpoint operator $\Omega_a$. The stabilizer reduction near this endpoint must preserve that action. Local choices of truncation and dressing near the distant right endpoint do not affect this calculation; the diagrams below use one such choice.

For $g=h=1$, squaring the truncated string and removing its interior with bulk stabilizers leaves
\begin{equation}
\begin{aligned}
\Omega(1,1)=\;&~~
\raisebox{0.8em}{{\color{black}\begin{tikzpicture}[
  baseline=(current bounding box.center),
  scale=0.85,
  op/.style={inner sep=1pt, fill=white, text=blue},
  redop/.style={inner sep=1pt, fill=white, text=red},
  orangeop/.style={inner sep=1pt, fill=white, text=orange},
  wire/.style={black, line width=0.8pt},
  dashline/.style={gray, dashed, line width=0.55pt}
]
\def\L{1.2}

\begin{scope}[shift={(-2.3,0)}]
  \draw[wire] (-0.7*\L,\L) -- (1.9*\L,\L);
  \draw[dashline] (0,\L) -- (0,-0.15*\L);
  \draw[dashline] (\L,\L) -- (\L,-0.15*\L);
  \draw[dashline] (-0.7*\L,0) -- (1.9*\L,0);
  \node at (0,1.3*\L) {$a$};
  \fill (0,\L) circle (2pt);
  \node[op] at (0.5*\L,\L) {$Z^2$};
  \node[op] at (1.5*\L,\L) {$Z^2$};
  \node[op] at (0,0.5*\L) {$X^2$};
  \node[op] at (\L,0.5*\L) {$X^2$};
\end{scope}

\node at (0.6,0.5*\L) {$\cdots$};

\begin{scope}[shift={(2.0,0)}]
  \draw[wire] (-0.7*\L,\L) -- (2.0*\L,\L);
  \draw[dashline] (0,\L) -- (0,-0.15*\L);
  \draw[dashline] (\L,\L) -- (\L,-0.15*\L);
  \draw[dashline] (-0.7*\L,0) -- (1.9*\L,0);
  \node at (2*\L,1.3*\L) {$b$};
  \fill (2*\L,\L) circle (2pt);
  \node[op] at (0.5*\L,\L) {$Z^2$};
  \node[op] at (1.5*\L,\L) {$Z^2$};
  \node[op] at (0,0.5*\L) {$X^2$};
  \node[op] at (\L,0.5*\L) {$X^2$};
\end{scope}
\end{tikzpicture}}}
\\[0.8em]
\sim\;&
\raisebox{0.8em}{{\color{black}\begin{tikzpicture}[
  baseline=(current bounding box.center),
  scale=0.85,
  op/.style={inner sep=1pt, fill=white, text=blue},
  redop/.style={inner sep=1pt, fill=white, text=red},
  orangeop/.style={inner sep=1pt, fill=white, text=orange},
  wire/.style={black, line width=0.8pt},
  dashline/.style={gray, dashed, line width=0.55pt}
]
\def\L{1.2}

\begin{scope}[shift={(-2.3,0)}]
  \draw[wire] (-0.7*\L,\L) -- (1.9*\L,\L);
  \draw[dashline] (0,\L) -- (0,-0.2*\L);
  \draw[dashline] (\L,\L) -- (\L,-0.2*\L);
  \draw[dashline] (-0.7*\L,0) -- (1.9*\L,0);
  \node at (0,1.3*\L) {$a$};
  \fill (0,\L) circle (2pt);
  \node[op] at (-0.5*\L,0.5*\L) {${\color{red}\gamma}{\color{orange}Z}$};
  \node[op] at (0.15*\L,0.55*\L) {$Z^{-2}$};
  \node[op] at (-0.55*\L,0) {$Z^{-2}$};
\end{scope}

\node at (0.6,0.5*\L) {$\cdots$};

\begin{scope}[shift={(2.0,0)}]
  \draw[wire] (-0.7*\L,\L) -- (2.0*\L,\L);
  \draw[dashline] (0,\L) -- (0,-0.2*\L);
  \draw[dashline] (\L,\L) -- (\L,-0.2*\L);
  \draw[dashline] (2*\L,\L) -- (2*\L,-0.2*\L);
  \draw[dashline] (-0.7*\L,0) -- (1.9*\L,0);
  \node at (2*\L,1.3*\L) {$b$};
  \fill (2*\L,\L) circle (2pt);
  \node[op] at (1.55*\L,0*\L) {$Z^2$};
  \node[op] at (2.1*\L,0.55*\L) {$Z^2$};
  \node[op] at (1.5*\L,0.5*\L) {${\color{red}\gamma}{\color{orange}Z}$};
\end{scope}
\end{tikzpicture}}}
\; .
\end{aligned}
\end{equation}
We choose the left endpoint representative
\begin{equation}
\Omega_a(1,1)=
\raisebox{0.8em}{{\color{black}\begin{tikzpicture}[
  baseline=(current bounding box.center),
  scale=0.85,
  op/.style={inner sep=1pt, fill=white, text=blue},
  redop/.style={inner sep=1pt, fill=white, text=red},
  orangeop/.style={inner sep=1pt, fill=white, text=orange},
  dashline/.style={gray, dashed, line width=0.55pt}
]
\def\L{1.6}

\coordinate (c) at (1.0*\L,0);

\draw[dashline] (c) -- ++(0,\L);

\draw[dashline] (c) -- ++(-\L,0);

\node at (1.0*\L,1.2*\L) {$a$};
\fill (\L,\L) circle (2pt);

\node[op] at (0.5*\L,0.5*\L) {${\color{red}\gamma}{\color{orange}Z}$};
\node[op] at (1.14*\L,0.55*\L) {$Z^{-2}$};

\node[op] at (0.6*\L,0.05*\L) {$Z^{-2}$};

\end{tikzpicture}}}
\end{equation}
The operator $\Omega_a(1,1)$ has odd fermion parity. This means that two copies of the boundary $\ZZ_2$ symmetry fuse to a physical fermion at the endpoint. Therefore
\begin{equation}
    n_2(1,1)=1 .
\end{equation}

For the normalized endpoint representatives used here, the remaining supercohomology data is obtained from the action of the boundary symmetry on $\Omega_a$:
\begin{equation}
    \widetilde{U}_l(g)\Omega_a(h,k)\widetilde{U}_l^{\dagger}(g)
    =
    \exp\bigl(2\pi i\,\nu_3(g,h,k)\bigr)\,
    \Omega_a(h,k).
    \label{eq:Else_Nayak_nu3}
\end{equation}
Only the boundary $X$ overlapping the vertical $Z^{-2}$ in $\Omega_a(1,1)$ contributes a nontrivial conjugation phase, namely $i$. Thus
\begin{equation}
    \exp\bigl(2\pi i\,\nu_3(g,h,k)\bigr)
    =
    \exp\!\left[
        \frac{2\pi i}{8}\,
        g\bigl(h+k-[h+k]_2\bigr)
    \right].
    \label{eq:Z2Z2F_nu3_phase}
\end{equation}
Equivalently, if $a\in H^1(\ZZ_2,\ZZ_2)$ is the canonical generator with $a(g)=g$, and if we use the same symbol for its integer lift, then the supercohomology data can be written as
\begin{equation}
    \nu_3=\frac{1}{8}a\cup \delta a, \quad n_2=a\cup a~.
    \label{eq:Z2xZ2f_v3_n2}
\end{equation}
They satisfy the supercohomology {constraint in Eq.}~\eqref{eq:cohomology_equation}. The solution {in Eq.}~\eqref{eq:Z2xZ2f_v3_n2} is the $\nu=2$ element in the $\ZZ_8$ classification of $\ZZ_2\times\ZZ_2^F$ fermionic SPT phases.
}

\subsection{\texorpdfstring{$\ZZ_2$}{Z2}-enriched \texorpdfstring{$\ZZ_2$}{Z2} toric code}
\label{sec:Z2_enriched_TC}

Next, we gauge fermion parity in the
$\ZZ_2\times\ZZ_2^F$ fermionic SPT constructed above. The result is a bosonic theory, often called the bosonic shadow of the fermionic SPT phase~\cite{BGK17}. In the absence of additional global symmetry, this operation is described by Kitaev's 16-fold way~\cite{Kitaev2006anyon}: gauging $\ZZ_2^F$ in an invertible fermionic phase produces a bosonic topological order whose anyon data are determined by the corresponding fermionic invertible phase, or more precisely, the chiral central charge $c_-$ modulo $8$. In this work, we focus on the non-chiral case $c_-=0$, for which the intrinsic bosonic topological order is the ordinary $\ZZ_2$ toric code.

The present construction has an additional bosonic symmetry $G=\ZZ_2$. Gauging fermion parity does not affect this symmetry, so it remains as a genuine global symmetry of the bosonic shadow. Therefore, the resulting phase is a symmetry-enriched topological (SET) phase: $\ZZ_2$-enriched $\ZZ_2$ toric code~\cite{Barkeshli2022Classification}. Its intrinsic anyon content is still $\ZZ_2$ toric code, while the remaining global $\ZZ_2$ symmetry acts projectively on the anyons. As we show below, both $e$ and $m$ carry nontrivial projective $\ZZ_2$ quantum numbers.

We now briefly review the lattice bosonization map used to gauge fermion parity. We follow the exact bosonization prescription of Ref.~\cite{CKR18}. On the fermionic side, there is one complex fermion on each plaquette. On the bosonic side, there is one $\ZZ_2$ gauge qubit on each edge, with Pauli operators $\widetilde X_e$ and $\widetilde Z_e$. The map sends fermion-parity-even operators to gauge-invariant bosonic operators. For the local operators used below, the map is
\begin{eqs}
i\times
\tikz[baseline=-0.6ex]{
  \def\L{\Ledge}
  \draw (0,0) -- (\L,0);
  \node[op, above] at (0.5*\L,0.25*\L) {$\color{red}\gamma_{L(e)}$};
  \node[op, below] at (0.5*\L,-0.15*\L) {$\color{red}\gamma'_{R(e)}$};
  \node[op] at (0.5*\L,0.15) {$e$};
}\quad
&
\tikz[baseline=-0.6ex]{
  \draw[<->] (0,0) -- (0.8,0);
}
\quad 
\raisebox{1.5em}{\tikz[baseline=-0.6ex]{
  \def\L{\Ledge}
  \draw (0,0) -- (\L,0);
  \draw (0,0) -- (0,-\L);
  \node[op, above] at (0.5*\L,-0.2) {$\color{magenta}\tilde X_e$};
  \node[op, left] at (0.15,-0.5*\L) {$\color{magenta}\tilde Z$};
}}
\quad, \\[1.4em]
i\times
\tikz[baseline=-0.6ex]{
  \def\L{\Ledge}
  \draw (0,-0.5*\L) -- (0,0.5*\L);
  \node[op, left] at (-0.1*\L,0) {$\color{red}\gamma_{L(e)}$};
  \node[op, right] at (0.2*\L,0) {$\color{red}\gamma'_{R(e)}$};
  \node[op, right] at (0,0.28) {$e$};
}\quad
&
\tikz[baseline=-0.6ex]{
  \draw[<->] (0,0) -- (0.8,0);
}
\quad
\raisebox{-0.6cm}{\tikz[baseline=-0.6ex]{
  \def\L{\Ledge}
  \draw (0,0) -- (\L,0);
  \draw (\L,0) -- (\L,\L);
  \node[op, above] at (0.5*\L,-0.2) {$\color{magenta}\tilde Z$};
  \node[op, right] at (0.82*\L,0.5*\L) {$\color{magenta}\tilde X_e$};
}}
\quad, \\[1.4em]
{\color{red} P_p = -i\gamma_p\gamma_p'}
\quad
&
\tikz[baseline=-0.6ex]{
  \draw[<->] (0,0) -- (0.8,0);
}
\quad
\raisebox{-0.6cm}{\tikz[baseline=-0.6ex]{
  \def\L{\Ledge}
  \draw (0,0) rectangle (\L,\L);
  \node[op, above] at (0.5*\L,0.87*\L) {$\color{magenta}\tilde Z$};
  \node[op, below] at (0.5*\L,0.17*\L) {$\color{magenta}\tilde Z$};
  \node[op, left] at (0.14*\L,0.5*\L) {$\color{magenta}\tilde Z$};
  \node[op, right] at (0.86*\L,0.5*\L) {$\color{magenta}\tilde Z$};
  \node[op] at (0.5*\L,0.5*\L) {$p$};
}}
\quad .
\label{eq:bosonization_review}
\end{eqs}
Here $L(e)$ and $R(e)$ denote the two plaquettes adjacent to the oriented edge $e$. The first two lines are the images of nearest-neighbor Majorana bilinears, while the last line is the image of the local fermion parity
operator.

{The local bosonization map acts on the fermion-parity-even operator algebra. On a closed lattice, the product of the displayed plaquette parity images is one, so the map selects the even total fermion-parity sector,}
\begin{equation}
    P_F=\prod_p P_p
    =
    \prod_p(-i\gamma_p\gamma'_p)
    =
    1 .
\end{equation}
{The bosonic images must also satisfy a $\ZZ_2$ Gauss-law constraint at each vertex}
\begin{equation}
    G_v :=
    \begin{tikzpicture}[baseline=-0.6ex, line width=0.5pt]
        \def\L{\Ledge}
        \draw (-\L,0) -- (\L,0);
        \draw (0,0) -- (0,-\L);
        \draw (0,0) -- (0,\L);
        \draw (0,\L) -- (\L,\L);
        \draw (\L,\L) -- (\L,0);
        \node[op] at (0,0) {$v$};
        \node[op] at (-0.5*\L,0) {$\color{magenta}\tilde{X}$};
        \node[op] at (0,-0.5*\L) {$\color{magenta}\tilde{X}$};
        \node[op] at (0,0.5*\L) {$\color{magenta}\tilde{X}\tilde{Z}$};
        \node[op] at (0.5*\L,0) {$\color{magenta}\tilde{X}\tilde{Z}$};
        \node[op] at (0.5*\L,\L) {$\color{magenta}\tilde{Z}$};
        \node[op] at (\L,0.5*\L) {$\color{magenta}\tilde{Z}$};
    \end{tikzpicture}
    =1 .
    \label{eq:bosonization_vertex_constraint}
\end{equation}
{These local gauge constraints require}
\begin{equation}
    G_v|\Psi\rangle=|\Psi\rangle,
    \qquad \forall v .
\end{equation}
We denote by
\begin{equation}
    \mathcal G=\langle \{G_v\}\rangle
\end{equation}
the group generated by these Gauss-law constraints. Modulo $\mathcal G$, the bosonic operators on the right-hand side of Eq.~\eqref{eq:bosonization_review} generate the local gauge-invariant operator algebra, which reproduces the fermion-parity-even operator algebra on the fermionic side. {On a torus, a Hilbert-space isomorphism additionally requires matching the noncontractible Wilson-loop sectors to the chosen fermionic boundary conditions. Retaining all such sectors instead gives the fermion-parity-gauged theory.}

Applying this bosonization map to the stabilizers of the $\ZZ_2\times\ZZ_2^F$ fermionic SPT gives a purely bosonic stabilizer Hamiltonian. The resulting stabilizer generators are
\begin{eqs}
A'_v &=
\begin{tikzpicture}[baseline=-0.6ex, line width=0.5pt]
  \def\L{1.2}
  \draw (-\L,0) -- (\L,0);
  \draw (0,0) -- (0,-\L);
  \draw (0,0) -- (0,\L);
  \draw (0,\L) -- (\L,\L);
  \draw (\L,\L) -- (\L,0);
  \node[op] at (0,0) {$v$};
  \node[op] at (-0.5*\L,0) {$\textcolor{blue}{X^\dagger}$};
  \node[op] at (0,-0.5*\L) {$\textcolor{blue}{X^\dagger}$};
  \node[op] at (0,0.5*\L) {$\textcolor{blue}{XZ}$};
  \node[op] at (0.5*\L,0) {$\textcolor{blue}{XZ^\dagger}$};
  \node[op] at (0.5*\L,\L) {$\textcolor{blue}{Z}$};
  \node[op] at (\L,0.5*\L) {$\textcolor{blue}{Z^\dagger}$};
  \node[op] at (0.5*\L,0.5*\L) {$\textcolor{orange}{X}$};
\end{tikzpicture}
,\quad
B'_p =
\raisebox{-0.6cm}{\begin{tikzpicture}[baseline=-0.6ex, line width=0.5pt]
  \def\L{1.2}
  \draw (0,0) rectangle (\L,\L);
  \node[op] at (0.58*\L,1.2*\L) {$\textcolor{magenta}{\tilde{Z}}\textcolor{blue}{Z^{-2}}$};
  \node[op] at (-0.1*\L,0.5*\L) {$\textcolor{magenta}{\tilde{Z}}\textcolor{blue}{Z^{-2}}$};
  \node[op] at (1.15*\L,0.5*\L) {$\textcolor{magenta}{\tilde{Z}}\textcolor{blue}{Z^2}$};
  \node[op] at (0.5*\L,-0.17*\L) {$\textcolor{magenta}{\tilde{Z}}\textcolor{blue}{Z^2}$};
  \node[op] at (0.5*\L,0.5*\L) {${p}$};
\end{tikzpicture}}
, \\
C'_e &= ~~
\raisebox{-0.6cm}{\begin{tikzpicture}[baseline=-0.6ex, line width=0.5pt]
  \def\L{1.2}
  \draw (-\L,0) -- (0,0);
  \draw (0,0) -- (0,\L);
  \node[op] at (-0.5*\L,0.05*\L) {$\textcolor{magenta}{\tilde{Z}}\textcolor{blue}{Z^2}$};
  \node[op] at (0.15*\L,0.5*\L) {$\textcolor{magenta}{\tilde{X}}\textcolor{blue}{X_e^{-2}}$};
\end{tikzpicture}}
,\qquad 
\raisebox{0.6cm}{\begin{tikzpicture}[baseline=-0.6ex, line width=0.5pt]
  \def\L{1.2}
  \draw (0,0) -- (\L,0);
  \draw (0,0) -- (0,-\L);
  \node[op] at (0.5*\L,0) {$\textcolor{magenta}{\tilde{X}}\textcolor{blue}{X_e^2}$};
  \node[op] at (0,-0.5*\L) {$\textcolor{magenta}{\tilde{Z}}\textcolor{blue}{Z^2}$};
\end{tikzpicture}}
\quad ,\\
D_e &= ~~
\tikz[baseline=-0.6ex]{
  \def\L{\Ledge}
  \draw (0,-0.5*\L) -- (0,0.5*\L);
  \node[op] at (-0.5*\L,0) {$\color{orange}Z$};
  \node[op] at (0,0) {$\color{blue}X_e^4$};
  \node[op] at (0.5*\L,0) {$\color{orange}Z$};
}
\;,\quad
\tikz[baseline=-0.6ex]{
  \def\L{\Ledge}
  \draw (-0.5*\L,0) -- (0.5*\L,0);
  \node[op] at (0,0.5*\L) {$\color{orange}Z$};
  \node[op] at (0,0) {$\color{blue}X_e^4$};
  \node[op] at (0,-0.5*\L) {$\color{orange}Z$};
}\;,
\label{eq:Z4F_SET_stabilizers}
\end{eqs}
where the \(\textcolor{blue}{X}\) and \(\textcolor{blue}{Z}\) operators denote \(\ZZ_8\) qudit Pauli operators, while \(\textcolor{orange}{X},\textcolor{orange}{Z}\) and \(\textcolor{magenta}{\widetilde X},\textcolor{magenta}{\widetilde Z}\) denote ordinary \(\ZZ_2\) Pauli operators on the auxiliary qubits and the bosonization gauge qubits, respectively.
Let
\begin{equation}
    \mathcal S_{\ZZ_2\text{-SET}} =
    \big\langle
        \{A'_v\},\{B'_p\},\{C'_e\},\{D_e\}
    \big\rangle
\end{equation}
be the stabilizer group generated by the operators above. Note that $G_v \in \mathcal S_{\ZZ_2\text{-SET}}$ is generated by the terms in Eq.~\eqref{eq:Z4F_SET_stabilizers}, so we do not need to include it as an independent stabilizer generator.

The remaining global $\ZZ_2$ symmetry is generated by
\begin{equation}
    U(1)=\prod_p {\color{orange}X_p}.
\end{equation}
Thus gauging fermion parity converts the $\ZZ_2\times\ZZ_2^F$ fermionic SPT into a bosonic SET phase whose intrinsic topological order is the ordinary $\ZZ_2$ toric code and whose remaining global symmetry is $\ZZ_2$.

The intrinsic anyon content is that of the toric code,
\begin{equation}
    \{1,e,m,f\}.
\end{equation}
One convenient choice for the string operators is
{\begin{eqs}
    m&=
\raisebox{-0.6cm}{\tikz[baseline=-0.6ex]{
  \def\L{\Ledge}
  \draw (0,0) -- (\L,0);
  \draw (\L,0) -- (\L,\L);
  \draw[dashed, magenta, -{Latex[length=2mm]}] (0.5*\L,0.7*\L) -- (0.5*\L, -0.4*\L);
  \node[op, above] at (0.5*\L,-0.2) {$\color{magenta}\widetilde X_e$};
  \node[op, right] at (0.82*\L,0.5*\L) {$\color{blue}Z^2$};
}},\quad
\raisebox{-0.6cm}{\tikz[baseline=-0.6ex]{
  \def\L{\Ledge}
  \draw (0,0) -- (0,\L);
  \draw (0,\L) -- (\L,\L);
  \draw[dashed, magenta, -{Latex[length=2mm]}] (-0.4*\L,0.5*\L) -- (0.7*\L, 0.5*\L);
  \node[op, above] at (0,0.3*\L) {$\color{magenta}\widetilde X_e$};
  \node[op, right] at (0.3*\L,\L) {$\color{blue}Z^2$};
}},
\\
   e&=
\raisebox{0cm}{\tikz[baseline=-0.6ex]{
  \def\L{\Ledge}
  \draw (0,0) -- (\L,0);
  \draw[dashed, magenta, -{Latex[length=2mm]}] (0,-0.1*\L) -- (1.3*\L, -0.1*\L);
  \node[op] at (0.5*\L,0) {$\color{blue}{Z^2}\color{magenta}\widetilde Z_e$};
}},\quad
\raisebox{-0.6cm}{\tikz[baseline=-0.6ex]{
  \def\L{\Ledge}
  \draw (0,0) -- (0,\L);
  \draw[dashed, magenta, -{Latex[length=2mm]}] (0.1*\L,1*\L) -- (0.1*\L, -0.2*\L);
  \node[op, above] at (0,0.3*\L) {$\color{blue}{Z^2}\color{magenta}\widetilde Z_e$};
}},\\
f&=\raisebox{-0.6cm}{\tikz[baseline=-0.6ex]{
  \def\L{\Ledge}
  \draw (0,0) -- (\L,0);
  \draw (\L,0) -- (\L,\L);
  \draw[dashed, magenta, -{Latex[length=2mm]}] (0.5*\L,0.7*\L) -- (0.5*\L, -0.4*\L);
  \node[op, above] at (0.5*\L,-0.2) {$\color{magenta}\widetilde X_e$};
  \node[op, right] at (0.82*\L,0.5*\L) {$\color{magenta}\widetilde{Z}$};
}},\quad
\raisebox{-0.6cm}{\tikz[baseline=-0.6ex]{
  \def\L{\Ledge}
  \draw (0,0) -- (0,\L);
  \draw (0,\L) -- (\L,\L);
  \draw[dashed, magenta, -{Latex[length=2mm]}] (-0.4*\L,0.5*\L) -- (0.7*\L, 0.5*\L);
  \node[op, above] at (0,0.3*\L) {$\color{magenta}\widetilde X_e$};
  \node[op, right] at (0.3*\L,\L) {$\color{magenta}\widetilde{Z}$};
}}.
\end{eqs}}\noindent
With this choice of the string-operators, the fusion rule $f=e\times m$ can be verified upon multiplication by stabilizers and local endpoint operators.

Symmetry enrichment can be diagnosed from the fusion of symmetry defects. For a finite region $R$, the restricted symmetry operator $U_R(g)$ creates a $g$-defect line along the boundary $\partial R$. Since the global symmetry does not permute the toric code anyons in the present model, the fusion of two such defect lines can differ from the defect for the product group element by an Abelian anyon string \cite{Barkeshli14, Tarantino2015, Barkeshli2022Classification}
\begin{equation}
    U_R(g)U_R(h) \sim W_{t(g,h)}(\partial R)\, U_R(gh).
    \label{eq:SET_defect_fusion}
\end{equation}
Here $W_{t(g,h)}(\partial R)$ denotes a closed string operator of the anyon $t(g,h)$ along $\partial R$, and
\begin{equation}
    t(g,h)\in \mathcal{A}
\end{equation}
is the symmetry fractionalization cocycle with $\mathcal{A}$ being the Abelian anyon group of the intrinsic topological order.

This defect fusion description is equivalent to the usual projective symmetry action on anyons. If the region $R$ encloses an anyon $x$, then the closed string $W_{t(g,h)}(\partial R)$ acts on the state by the mutual braiding phase between $x$ and $t(g,h)$. Hence, the local symmetry action on $x$ satisfies \cite{Barkeshli14}
\begin{equation}
    U_x(g)U_x(h) = \eta_x(g,h)\,U_x(gh),
    \quad \eta_x(g,h)=M_{x,t(g,h)},
    \label{eq:SET_eta_from_omega}
\end{equation}
where $M_{\alpha, \beta}$ is the braiding phase of anyons $\alpha$ and $\beta$ . Thus, the anyon-valued defect fusion rule \(t(g,h)\) and the projective phase \(\eta_x(g,h)\) describe the same symmetry fractionalization data.

In the present model, the symmetry does not permute the toric code anyons, so the fractionalization class is an element of
$H^2(\ZZ_2,\mathcal A)$ with
\[
    \mathcal A=\{1,e,m,f\}.
\]
{To compute this class, consider the truncated $\ZZ_2$ symmetry operator $\widetilde U_l(1)$ on a boundary interval $l=[a,b]$, with the same truncation as in Eq.~\eqref{eq:tildeU(1)_l}.} Since the group law is $1+1=0$ in $\ZZ_2$, the operator
\[
    \Omega(1,1)
    :=
    \widetilde U_{l}(1)^2
\]
measures the obstruction to the local symmetry action squaring to the identity. Using local stabilizers, we find
\begin{equation}
\begin{aligned}
&\Omega(1,1)
\\
\raisebox{-0.5em}{$=$}\;&\quad~~
\begin{tikzpicture}[
  baseline=(current bounding box.center),
  scale=0.95,
  op/.style={inner sep=1pt, fill=white, text=blue},
  wire/.style={black, line width=0.8pt},
  dashline/.style={gray, dashed, line width=0.55pt}
]
\def\L{1.2}

\begin{scope}[shift={(-2.3,0)}]
  \draw[wire] (-1*\L,\L) -- (1.9*\L,\L);
  \draw[dashline] (-1*\L,\L) -- (-1*\L,-0.15*\L);
  \draw[dashline] (0,\L) -- (0,-0.15*\L);
  \draw[dashline] (\L,\L) -- (\L,-0.15*\L);
  \draw[dashline] (-1*\L,0) -- (1.9*\L,0);
  \node at (0,1.3*\L) {$a$};
  \fill (0,\L) circle (2pt);
  \node[op] at (0.5*\L,\L) {$Z^2$};
  \node[op] at (1.5*\L,\L) {$Z^2$};
  \node[op] at (0,0.5*\L) {$X^2$};
  \node[op] at (\L,0.5*\L) {$X^2$};
\end{scope}

\node at (0.3,0.5*\L) {$\cdots$};

\begin{scope}[shift={(1.4,0)}]
  \draw[wire] (-0.6*\L,\L) -- (2.0*\L,\L);
  \draw[dashline] (0,\L) -- (0,-0.15*\L);
  \draw[dashline] (\L,\L) -- (\L,-0.15*\L);
  \draw[dashline] (-0.6*\L,0) -- (2*\L,0);
  \node at (2*\L,1.3*\L) {$b$};
  \fill (2*\L,\L) circle (2pt);
  \node[op] at (-0.5*\L,\L) {$Z^2$};
  \node[op] at (0.5*\L,\L) {$Z^2$};
  \node[op] at (1.5*\L,\L) {$Z^2$};
  \node[op] at (0,0.5*\L) {$X^2$};
  \node[op] at (\L,0.5*\L) {$X^2$};
\end{scope}
\end{tikzpicture}
\\
\raisebox{-1em}{$\sim$}\;&
\begin{tikzpicture}[
  baseline=(current bounding box.center),
  >=Latex,
  scale=0.95,
  wire/.style={black, line width=0.8pt},
  dashline/.style={gray, dashed, line width=0.55pt},
  blueop/.style={inner sep=1pt, fill=white, text=blue},
  pinkop/.style={inner sep=1pt, fill=white, text=magenta},
  orangeop/.style={inner sep=1pt, fill=white, text=orange},
  redarrow/.style={red, -{Latex[length=2mm]}, line width=0.6pt}
]
\def\L{1.2}

\begin{scope}[shift={(-2.3,0)}]
  \draw[wire] (-1*\L,\L) -- (1.9*\L,\L);
  \draw[dashline] (-\L,\L) -- (-\L,-0.2*\L);
  \draw[dashline] (0,\L) -- (0,-0.2*\L);
  \draw[dashline] (\L,\L) -- (\L,-0.2*\L);
  \draw[dashline] (-1*\L,0) -- (1.9*\L,0);

  \node at (0,1.3*\L) {$a$};
  \fill (0,\L) circle (2pt);
  \coordinate (Aend) at (0.25*\L,1.3*\L);

  \node[pinkop] at (0.5*\L,1*\L) {$\widetilde{Z}$};
  \node[pinkop] at (1.5*\L,1*\L) {$\widetilde{Z}$};

  \node[orangeop] at (-0.5*\L,0.5*\L) {$Z$};
  \node[op] at (0*\L,0.5*\L) {${\color{magenta}\widetilde{X}}$};
  \node[pinkop] at (1*\L,0.55*\L) {$\widetilde{X}$};

  \node[op]   at (-0.5*\L,\L) {${\color{magenta}\widetilde{Z}}{\color{blue}Z^{-2}}$};
  \node[op]   at (-1.1*\L,0.55*\L) {${\color{magenta}\widetilde{Z}}{\color{blue}Z^{-2}}$};
\end{scope}

\node at (0.3,0.5*\L) {$\cdots$};

\begin{scope}[shift={(1.4,0)}]
  \draw[wire] (-0.6*\L,\L) -- (2.0*\L,\L);
  \draw[dashline] (0,\L) -- (0,-0.2*\L);
  \draw[dashline] (\L,\L) -- (\L,-0.2*\L);
  \draw[dashline] (-0.6*\L,0) -- (1.9*\L,0);

  \node at (2*\L,1.3*\L) {$b$};
  \fill (2*\L,\L) circle (2pt);
  \coordinate (Bend) at (1.75*\L,1.3*\L);

  \node[pinkop] at (-0.5*\L,1*\L) {$\widetilde{Z}$};
  \node[pinkop] at (0.5*\L,1*\L) {$\widetilde{Z}$};
  \node[op] at (1.5*\L,1*\L) {$\color{blue}Z^2$};

  \node[pinkop]   at (0,0.55*\L) {$\widetilde{X}$};
  \node[op]   at (\L,0.55*\L) {${\color{magenta}\widetilde{Y}}{\color{blue}Z^2}$};
  \node[orangeop] at (1.5*\L,0.5*\L) {$Z$};
\end{scope}

\draw[redarrow] (Aend) -- (Bend)
  node[midway, above=2pt, fill=white, inner sep=1pt, text=red] {$f$};

\end{tikzpicture}
\; .
\end{aligned}
\end{equation}
Therefore, two $\ZZ_2$ symmetry defects fuse to the worldline of fermion $f$ in the toric code, and the symmetry fractionalization class is
\begin{equation}
    t(1,1)=f .
\end{equation}
Equivalently, the local $\ZZ_2$ symmetry action on an anyon $x$ squares to the mutual braiding phase with $f$
\begin{equation}
    \eta_x(1,1)=M_{x,f},
    \qquad
    x\in\{e,m,f\}.
\end{equation}
Using the toric-code braiding data, we obtain
\begin{equation}
    \eta_e(1,1)=-1,\qquad
    \eta_m(1,1)=-1,\qquad
    \eta_f(1,1)=+1 .
\end{equation}
Thus, both $e$ and $m$ carry nontrivial projective representations of the global $\ZZ_2$ symmetry, while their bound state $f=e\times m$ carries a linear representation.

The calculation above determines the symmetry fractionalization class $t$, which controls the projective action of the global $\mathbb{Z}_2$ symmetry on the deconfined toric code anyons. The bosonic shadow also inherits the $\nu_3$ data of the parent $\mathbb{Z}_2\times\mathbb{Z}_2^F$ fSPT. In the SET language, $\nu_3$ should be interpreted as the associator phase of the $\mathbb{Z}_2$ symmetry defects. $t(g,h)$ then records the anyon-valued defect fusion rule
\begin{equation}
    \sigma_g\times\sigma_h = t(g,h)\,\sigma_{gh},
\end{equation}
whereas $\nu_3$ records the phase relating the two ways of fusing three defects
\begin{equation}
    (\sigma_g\times\sigma_h)\times\sigma_k
    =
    \exp\bigl(2\pi i\,\nu_3(g,h,k)\bigr)\,
    \sigma_g\times(\sigma_h\times\sigma_k).
\end{equation}
For the present model,
\begin{equation}
    \nu_3=\frac{1}{8}a\cup\delta a .
\end{equation}
In particular, for \(g=h=k=1\),
\begin{equation}
    \exp\Bigl(2\pi i\,\nu_3(1,1,1)\Bigr) = \exp\left(\frac{\pi i}{2}\right) = i .
\end{equation}
This phase does not change the toric-code anyon content or the fractionalization class $t(1,1)=f$, but it fixes the symmetry defect associativity data.  Equivalently, it is the SPT-like defect data that becomes visible if one gauges the remaining global $\mathbb{Z}_2$ symmetry.

{We have constructed the same $\ZZ_2$-enriched $\ZZ_2$ toric-code phase by two microscopic routes: gauging fermion parity in the $\ZZ_2\times\ZZ_2^F$ fermionic SPT and ungauging the appropriate bosonic charge subgroup in the $U(1)_4\times U(1)_{-4}$ theory. Although the two stabilizer models use different microscopic Hilbert spaces, we prove in Appendix~\ref{app:stable_equivalence_Z2_SET} that they are stably equivalent: after adjoining product-state ancillas, they are related by a finite-depth local unitary and therefore realize the same symmetry-enriched topological phase.}

\subsection{Condensation and ungauging from the \texorpdfstring{$U(1)_4\times U(1)_{-4}$}{U(1)4 x U(1)-4} Chern-Simons theory}
\label{sec:U1_4_U1_minus4_TQD}

We now describe the bosonic shadow of the fermionic toric code, which is the middle node in Fig.~\ref{fig:Z_8_commutative_diagram}. There are two equivalent ways to obtain this phase.

First, one can gauge fermion parity in the fermionic toric code. Since the fermionic toric code constructed in Sec.~\ref{sec:fermionic_toric_code} may be viewed as
\begin{equation}
    U(1)_4\times U(1)_{-1},
    \text{ and equivalently,}
    ~~K=
    \begin{pmatrix}
        4 & 0 \\
        0 & -1
    \end{pmatrix}~,
\end{equation}
the fermion-parity gauging acts only on the invertible fermionic $U(1)_{-1}$ sector. By Kitaev's 16-fold way~\cite{Kitaev2006anyon}, gauging fermion parity in an invertible fermionic phase with chiral central charge $c_-=1$ produces the bosonic theory $U(1)_4$. Equivalently, $U(1)_1$ is a complex-fermion integer quantum Hall phase, or two copies of the $p+ip$ Majorana phase, so it corresponds to the $\nu=2$ element of the 16-fold way. 
Gauging its fermion parity gives a $\ZZ_2^F$ flux {$d$} with
\begin{equation}
    d^2=f,
    \qquad
    \theta_d=\exp\left(\frac{\pi i}{4}\right),
\end{equation}
which is precisely the generator of $U(1)_4$. Since $U(1)_{-1}$ is the time-reversal conjugate of $U(1)_1$, gauging its fermion-parity gives the time-reversal conjugate bosonic theory $U(1)_{-4}$. In this case, the $\ZZ_2^F$ flux has
\begin{equation}
    {\bar d}^2=f,
    \qquad
    \theta_{\bar d}=\exp\left(-\frac{\pi i}{4}\right),
\end{equation}
which agrees with the generator of the $K=-4$ Chern--Simons theory. Indeed, in $U(1)_{-4}$, the generator $\bar d$ has spin $\exp(-\pi i/4)$, while $\bar d^2$ has spin $-1$ and is identified with gauge charge, which is a fermion $f$. Thus, gauging the physical fermion parity converts
\begin{equation}
    U(1)_4\times U(1)_{-1}
    \xrightarrow{\ \text{gauge }\ZZ^F_2 }
    U(1)_4\times U(1)_{-4}.
\end{equation}
The resulting bosonic shadow therefore has diagonal \(K\)-matrix
\begin{equation}
    K_{\mathrm{diag}} =
    \begin{pmatrix}
        4 & 0 \\
        0 & -4
    \end{pmatrix}.
\end{equation}

The second way to obtain the bosonic shadow phase is directly from the $\ZZ_8$ toric code by condensing the order-two boson~\cite{Ellison22}
\begin{equation}
    \beta=e^4m^4 .
\end{equation}
Indeed,
\begin{equation}
    \theta_\beta=\omega^{16}=1, \qquad \beta^2=1,
\end{equation}
where $\omega = \exp(2\pi i/8)$. An anyon $e^pm^q$ remains deconfined after this condensation precisely when it braids trivially with $\beta$
\begin{equation}
    M_{e^pm^q,\beta} = \omega^{4(p+q)} = (-1)^{p+q} = 1 .
\end{equation}
Thus, the deconfined anyons satisfy
\begin{equation}
    p+q=0 \mod 2,
\end{equation}
and anyons differing by fusion with $\beta=e^4m^4$ are identified.

A convenient set of generators for the condensed theory is
\begin{equation}
    x=[em], \qquad y=[em^{-1}],
\end{equation}
where $[\cdots]$ denotes the equivalence class after condensing $\beta=e^4m^4$. These generators obey
\begin{equation}
    x^4=1,\qquad y^4=1,\qquad B(x,y)=1,
\end{equation}
and their topological spins are
\begin{equation}
    \theta_x = \exp\left(\frac{2\pi i}{8}\right),
    \qquad
    \theta_y = \exp\left(-\frac{2\pi i}{8}\right).
\end{equation}
Thus, an arbitrary anyon can be written as
\begin{equation}
    x^a y^b,\qquad a,b\in\ZZ_4 .
\end{equation}
Throughout this paragraph, unless otherwise stated, we use the diagonal basis
\begin{equation}
    K_{\rm diag} =
    \begin{pmatrix}
        4&0\\
        0&-4
    \end{pmatrix},
\end{equation}
and identify
\begin{equation}
    x^a y^b
    \quad \longleftrightarrow \quad
    \begin{pmatrix}
        a\\ b
    \end{pmatrix}_{\rm diag}.
\end{equation}
In this basis, the spin and mutual braiding are
\begin{equation}
    \theta_{(a,b)_{\rm diag}}
    =
    \exp\left[
        \frac{\pi i}{4}(a^2-b^2)
    \right],
\end{equation}
and
\begin{equation}
    M_{(a,b)_{\rm diag},(a',b')_{\rm diag}} =
    \exp\left[
        \frac{2\pi i}{4}(aa'-bb')
    \right].
\end{equation}
This is precisely the anyon theory of
\begin{equation}
    U(1)_4\times U(1)_{-4}.
\end{equation}

At the lattice level, the condensation of $\beta=e^4m^4$ is implemented by adding short hopping operators for this boson. These are the purely Pauli analogues of the fermionic-toric-code condensate operators. One replaces the $e^2m^6\psi$ hopping operator by the corresponding $e^4m^4$ hopping operator, with no Majorana insertion.
{
Taking the centralizer of these hopping operators within the parent
$\mathbb Z_8$ toric-code stabilizer group and adjoining the hopping
operators themselves gives a Pauli stabilizer model for the
$U(1)_4\times U(1)_{-4}$ twisted quantum double, as in the
Pauli-stabilizer construction of Ref.~\cite{Ellison22}.}

We now explain the condensation and ungauging arrows adjacent to this middle node in Fig.~\ref{fig:Z_8_commutative_diagram}. Different condensable anyons in the same $U(1)_4\times U(1)_{-4}$ theory remove different amounts of topological order and therefore lead to different neighboring phases.

First, let us consider the route from $U(1)_4\times U(1)_{-4}$ to the fermionic toric code.  We adjoin a physical transparent fermion $\psi$ and condense the composite boson
\begin{equation}
    y^2\psi =
    \begin{pmatrix}
        0\\2
    \end{pmatrix}_{\rm diag}
    \times \psi .
\end{equation}
Here $y^2$ is an emergent fermion, because
\begin{equation}
    \theta_{y^2}=-1.
\end{equation}
Therefore, $y^2\psi$ is bosonic and can be condensed. In the original $\ZZ_8$ toric-code notation,
\begin{equation}
    y^2=[e^2m^{-2}]=[e^2m^6],
\end{equation}
so this is precisely the same anyon that appears in the direct condensation in the $\ZZ_8$ toric code 
\begin{equation}
    e^2m^6\psi .
\end{equation}
Physically, condensing $y^2\psi$ identifies the emergent fermion in the anti-chiral $U(1)_{-4}$ sector with the physical transparent fermion. Equivalently, it converts the bosonic $U(1)_{-4}$ sector into the invertible fermionic $U(1)_{-1}$ sector. Hence,
\begin{equation*}
    U(1)_4\times U(1)_{-4}
    \xrightarrow{\ \text{condense }y^2\psi\ }
    U(1)_4\times U(1)_{-1}
    \simeq
    \overline{\mathrm{fTC}} .
\end{equation*}
In $K$-matrix form,
\begin{equation}
    \begin{pmatrix}
        4&0\\
        0&-1
    \end{pmatrix}
    \sim
    \begin{pmatrix}
        0&2\\
        2&-1
    \end{pmatrix},
\end{equation}
which is the fermionic toric code realized in Sec.~\ref{sec:fermionic_toric_code}.

{The remaining two ungauging routes from $U(1)_4\times U(1)_{-4}$ are verified in Appendix~\ref{app:Z8_duality_condensation_checks}. Ungauging the order-two charge subgroup generated by $q^2=x^2y^2$ leaves a $\ZZ_2$ toric code with the fractionalization class $t(1,1)=f_{\rm TC}$, and hence gives the $\ZZ_2$-enriched toric code constructed above. Ungauging instead the full $\ZZ_4$ charge algebra generated by $q=xy$ removes all intrinsic topological order and gives the $p=2$ $\ZZ_4$ SPT. Together with the $y^2\psi$ condensation discussed above, these checks establish the three outgoing condensation and ungauging arrows and the basis conventions used in Fig.~\ref{fig:Z_8_commutative_diagram}.}

\section{Fermionic clock and shift operators}
{\label{sec:fermionicclock}
This section constructs fermionic clock and shift operators for every even order $D=2d$ on a local Hilbert space consisting of one $d$-state qudit and one complex fermion. The central result is a parity-even shift operator $\Pi$ satisfying
\begin{equation}
    \Pi^d=P=(-1)^F,\qquad \Pi^{2d}=1.
\end{equation}
Thus the order-two element of the cyclic symmetry is physical fermion parity. Together with the parity-odd operators $\Gamma$ and $\Gamma'$, this provides the local algebra needed for ungauging higher-order fermionic charge subgroups in Secs.~\ref{sec:Z8F_and_duality_web} and~\ref{sec:general_constructions}.

Related algebraic structures appeared in Ref.~\cite{HFH18}, and explicit order-four realizations were given in Refs.~\cite{TV18,Chew18}. Following the construction principle in Appendix C of Ref.~\cite{Shirley22}, we give a uniform realization for all even $D$ and incorporate it into the stabilizer framework. We then use the same operators to formulate exact bosonization for $\ZZ_D^F$ symmetry. The group-theoretic discussion below specifies how this cyclic building block fits into a general finite Abelian fermionic symmetry.}

Previously, we have considered the fermionic symmetry group $\ZZ_2 \times \ZZ_2^F$. We now take one step further and consider the finite fermionic symmetry groups $G_f$ whose fermion parity $\ZZ_2^F$ is extended by a finite Abelian group $G_b$. In other words, we consider the symmetry groups $G_f$ fitting into the short exact sequence
\begin{equation}
    1 \rightarrow \ZZ_2^F \rightarrow G_f \rightarrow G_b \rightarrow 1
\end{equation}
whose central extension is given by a 2-cocycle class $w_2$, i.e. $w_2 \in H^2(G_b,\ZZ_2)$~\cite{Barkeshli2022Classification}. Explicitly, the product rule in $G_f$ is given by
\begin{equation}
    (g,a)\times(h,b)=(g+h,a+b+w_2(g,h))
    \label{eq:group_law_Gf}
\end{equation}
for $g,h\in G_b$, $a,b\in\ZZ_2^F$, and $w_2\in H^2(G_b,\ZZ_2)$. {When the full group $G_f$ is Abelian, one can choose cyclic generators such that $w_2$ is supported on at most one factor, as proved in Appendix~\ref{sec:sufficiency_non_split_ZF2n}. This statement concerns the symmetry extension, not the independent SPT response data.} This allows us to write a finite Abelian fermionic symmetry group as a product of a fermionic cyclic group and a finite Abelian group
\begin{equation}
    G_f = \ZZ^F_{2d} \times \prod_{i \in \mathcal{I}} \ZZ_{N'_i}
    \label{eq:abelian_fermionic_symmetry_group}
\end{equation}
for $d \in \ZZ_+$ and some index set $\mathcal{I}$. We will also define $D=2d$ moving forward. The extension class $w_2$ is then supported in the $\ZZ^F_{D}$ sector.

{The cyclic factor in Eq.~\eqref{eq:abelian_fermionic_symmetry_group} fits into
\begin{equation}
    1 \rightarrow \ZZ_2^F \rightarrow \ZZ^F_D \rightarrow \ZZ_d \rightarrow 1.
\end{equation}
The relation $\Pi^d=P$ realizes this extension directly on the local Hilbert space. It supplies the higher-order local charged fermions needed for the ungauging constructions below, while retaining explicit control of physical fermion parity.}

More generally, for a finite Abelian fermionic group $G_f$, we can construct the clock and shift operators compatible with the symmetry group as follows. Since a finite Abelian fermionic group has the decomposition of Eq.~\eqref{eq:abelian_fermionic_symmetry_group}, the set of compatible clock and shift operators is simply a tensor product of the $\ZZ^F_D$ fermionic clock and shift operators constructed in the next subsection together with the usual $\ZZ_{N'_i}$ generalized Pauli operators for each cyclic group $\ZZ_{N'_i}$.

Next, we present the construction of fermionic clock and shift operators in Sec.~\ref{sec:fermionic_clock_and_shift_operators_constructions}. Although the main focus of the latter parts of this work is to present stabilizer models for fermionic phases of matter using these clock and shift operators (see Sec.~\ref{sec:Z8F_and_duality_web} and Sec.~\ref{sec:general_constructions}), we discuss how these operators also enable us to define exact bosonization for any $\ZZ^F_D$ group and present the mapping in Sec.~\ref{sec:ZDF_exact_Bosonization}.

\subsection{The construction}
\label{sec:fermionic_clock_and_shift_operators_constructions}

{We now give the local realization of the fermionic clock and shift algebra announced above. The construction extends the ordinary Majorana algebra by a bosonic clock degree of freedom, so its operators can be used as local building blocks for the stabilizer models.}

Let us also remark that the algebra of these fermionic clock and shift operators should not be confused with $\mathbb Z_D$ parafermions. For example, $\mathbb Z_4$ parafermions exhibit semionic statistics, rather than fermionic statistics\footnote{Nevertheless as edge modes of a bulk 1D phase, they are closely related by a non-local transformation. See Ref.\cite{Chew18} for a detailed discussion}.

First, define the local Hilbert space to be a tensor product
\begin{equation}
    \mathcal H_{\rm loc} = \mathcal H_{\mathbb Z_d}\otimes \mathcal H_f ,
\end{equation}
where $\mathcal H_{\mathbb Z_d}$ is a $d$-dimensional bosonic qudit Hilbert space and $\mathcal H_f$ is the Hilbert space of a single complex fermion. Thus the total local Hilbert space has dimension
$2d=D$.

On the bosonic qudit, we define the usual clock and shift operators
$Z$ and $X$ by
\begin{equation}
    X^d=Z^d=1, \qquad ZX=\omega XZ, \qquad \omega=e^{2\pi i/d}.
\end{equation}
Choosing the basis in which $Z$ is diagonal, we write
\begin{align}
    X &= \sum_{n=0}^{d-1} |n+1\rangle\langle n|,  & Z &= \sum_{n=0}^{d-1} \omega^n |n\rangle\langle n|,
\end{align}
where the label $n+1$ is understood modulo $d$. We will also define the operator
\begin{equation}
    \sqrt Z = \sum_{n=0}^{d-1} \zeta^n |n\rangle\langle n|, \qquad \zeta:= e^{2\pi i/D}=e^{\pi i/d}.
\end{equation}
Therefore,
\begin{equation}
    (\sqrt Z)^2=Z, \qquad \zeta^2=\omega, \qquad \zeta^d=-1.
\end{equation}
On the fermionic Hilbert space, let $c$ and $c^\dagger$ be the annihilation and creation operators of a single complex fermion. We define the two Majorana operators
\begin{equation}
    \gamma = c+c^\dagger, \qquad \gamma'=-i(c-c^\dagger),
\end{equation}
and the fermion parity operator
\begin{align}
    P=(-1)^F=1-2c^\dagger c=-i\gamma\gamma'.
\end{align}
These obey
\begin{align}
    \gamma^2&=(\gamma')^2=1, \qquad \{\gamma,\gamma'\}=0, \\
    \quad P\gamma&=-\gamma P, \qquad \qquad P\gamma'=-\gamma'P .
\end{align}
We now combine the bosonic clock degree of freedom with the fermionic degree of freedom to define the fermionic clock operators
\begin{align}
    \Gamma &= \sqrt Z\,\gamma, & \Gamma' &= \sqrt Z^\dagger \,\gamma',
\end{align}
which are parity-odd, and
\begin{equation}
    \Pi = X P^{|d-1\rangle\langle d-1|} \, ,
\end{equation}
where explicitly,
\begin{equation}
P^{|d-1\rangle\langle d-1|} = \sum_{n\ne d-1}|n\rangle\langle n| \otimes \mathbbm 1 +\,|d-1\rangle\langle d-1| \otimes P \,.
\end{equation}
$\Pi$ acts as a $\mathbb Z_d$ shift on the bosonic qudit. It picks up the fermion parity only when acting on the state $|d-1\rangle$. It then follows that
\begin{equation}
    \Pi^d=P, \qquad \Pi^D=\Pi^{2d}=1.
\end{equation}
This is the defining fermionic refinement of the ordinary clock-shift algebra. The order-two central element of the $\mathbb Z_D$ clock variable is now identified with physical fermion parity.

The squares of the operators $\Gamma$ and $\Gamma'$ are  parity-even operators. Indeed, 
\begin{equation}
    \Gamma^2=Z, \qquad \Gamma'^2=Z^\dagger .
\end{equation}
Moreover,
\begin{equation}
    -i\Gamma\Gamma' = -i(\sqrt Z\gamma)(\sqrt Z^\dagger\gamma') = -i\gamma\gamma' = P.
\end{equation}
The commutation relations between $\Gamma,\Gamma'$, and $\Pi$ are
\begin{align}
    \Gamma \Pi &= \zeta \, \Pi \Gamma, \\
    \Gamma' \Pi &= \zeta^{-1} \, \Pi \Gamma',
\end{align}
which can be readily checked using the above definition.

In summary, the $\ZZ^F_D$ fermionic clock and shift operators $\Gamma$ and $\Pi$ satisfy the following relations.
\begin{align}
    \Pi^d &= P, & \Pi^D &= 1, \\
    \Gamma^2 &= Z, & \Gamma'^2 &= Z^\dagger, \\
    \Gamma\Pi &= \zeta \Pi\Gamma, & \Gamma'\Pi &=\zeta^{-1} \,\Pi\Gamma', \\
    -i\Gamma\Gamma' &= P, & \Gamma^D&=\Gamma'^D=1.
\end{align}
$\Gamma$ and $\Pi$ form a fermionic analogue of a $\mathbb Z_D$ clock-shift pair. The operator $\Pi$ is the fermionic shift operator, while $\Gamma$ and $\Gamma'$ are fermion-parity-odd clock-like operators. And, for sites labeled by $p$, we have
\begin{align}
    \{\Gamma_p,\Gamma_{p'}^\dagger\} &= 2\delta_{pp'}, \\
    \{\Gamma'_p,\Gamma_{p'}'^\dagger\} &= 2\delta_{pp'}, \\
    \{\Gamma_p,\Gamma_{p'}'^\dagger\} &= 0.
\end{align}

Lastly, to get some intuition on the fermionic clock and shift operators, let us examine the following examples.

\textit{Example 1: $D=2$.} We have $d=1$. The bosonic ``qudit'' is only one dimensional and can therefore be suppressed. The algebra reduces to the ordinary Majorana algebra
\begin{equation}
\Gamma=\gamma,\qquad \Gamma'=\gamma',\qquad \Pi=P.
\end{equation}

\textit{Example 2: $D=4$.} We have $d=2$, and the qudit is simply a qubit.  In this case, the fermionic clock and shift operators are
\begin{equation}
    \Gamma=i^{(1-Z)/2}\gamma,\qquad
    \Gamma'=i^{(Z-1)/2}\gamma',
    \qquad
    \Pi=XP^{(1-Z)/2}.
\end{equation}
They satisfy
\begin{equation}
    \Pi^2=P,\qquad \Gamma^2=\Gamma'^2=Z,
\end{equation}
and
\begin{equation}
    \Gamma\Pi=i\Pi\Gamma,\qquad
    \Gamma'\Pi=-i\Pi\Gamma'.
\end{equation}
This is the $\ZZ^F_4$ fermionic clock algebra introduced in Ref.~\cite{TV18}. A related construction in terms of ``spinful fermions" was given in Ref.~\cite{Chew18}.

\textit{Example 3: $D=8$.} We have $d=4$, so the bosonic qudit is a $\ZZ_4$ qudit and
\begin{equation}
    \omega=i,\qquad \zeta=e^{\pi i/4}.
\end{equation}
The $\ZZ^F_8$ fermionic clock and shift operators are
\begin{equation}
    \Gamma=\sqrt Z\,\gamma,\qquad
    \Gamma'=\sqrt Z^\dagger\,\gamma',
    \qquad
    \Pi=XP^{|3\rangle\langle 3|}.
\end{equation}
Equivalently, defining the projector
\begin{equation}
    \mathcal P_3=|3\rangle\langle 3|
    =\frac{1}{4}(1+iZ-Z^2-iZ^3),
\end{equation}
then
\begin{equation}
    \Pi=X(1-\mathcal P_3+\mathcal P_3P).
\end{equation}
Thus, $\Pi$ shifts the $\ZZ_4$ qudit and inserts the physical fermion parity $P$ only when the shift wraps from $|3\rangle$ to $|0\rangle$.
The algebra becomes
\begin{equation}
    \Pi^4=P,\quad \Pi^8=1,\quad
    \Gamma^2=Z,\quad \Gamma'^2=Z^\dagger,
    \quad -i\Gamma\Gamma'=P,
\end{equation}
together with
\begin{equation}
    \Gamma\Pi=e^{\pi i/4}\Pi\Gamma,\qquad
    \Gamma'\Pi=e^{-\pi i/4}\Pi\Gamma'.
\end{equation}
This is the $\ZZ^F_8$ fermionic clock and shift algebra. In the next section, we will use this to construct the stabilizer code for the $\mathbb Z_8^F$ SPT.

\subsection{Exact Bosonization for \texorpdfstring{$\ZZ_D^F$}{ZDF}}
\label{sec:ZDF_exact_Bosonization}

As an application of our $\ZZ_D$ fermionic clock and shift operators, we present an exact bosonization for the $\ZZ_D^F$ group in $(2{+}1)D$. The $\ZZ^F_2$ version was introduced in Ref.~\cite{CKR18} and its generalization to higher dimensions was discussed in Ref.~\cite{CK18,YuAn2020ExactBosAnyDim}.

On the fermionic side, there is one complex fermion and a $\ZZ_d$ qudit on each plaquette forming the fermionic clock algebra. On the bosonic side, there is one $\ZZ_D$ gauge qudit on each edge, with Pauli operators $\widetilde X_e$ and $\widetilde Z_e$. The map sends operators symmetric under $\ZZ_D^F$ symmetry generated by $\prod_p \Pi_p$ to gauge-invariant bosonic operators respecting a $\ZZ_D$ 1-form symmetry. Thus it can also be viewed as gauging $\ZZ_D^F$ symmetry. For the local operators used below, the map is
\begin{eqs}
i\times
\tikz[baseline=-0.6ex]{
  \def\L{\Ledge}
  \draw (0,0) -- (\L,0);
  \node[op, above] at (0.5*\L,0.25*\L) {$\color{red}\Gamma_{L(e)}$};
  \node[op, below] at (0.5*\L,-0.15*\L) {$\color{red}\Gamma'_{R(e)}$};
  \node[op] at (0.5*\L,0.15) {$e$};
}\quad
&
\tikz[baseline=-0.6ex]{
  \draw[<->] (0,0) -- (0.8,0);
}
\quad 
\raisebox{1.5em}{\tikz[baseline=-0.6ex]{
  \def\L{\Ledge}
  \draw (0,0) -- (\L,0);
  \draw (0,0) -- (0,-\L);
  \node[op, above] at (0.5*\L,-0.2) {$\color{magenta}\tilde X_e$};
  \node[op, left] at (0.15,-0.5*\L) {$\color{magenta}\tilde Z^d$};
}}
\quad, \\[1.4em]
i\times
\tikz[baseline=-0.6ex]{
  \def\L{\Ledge}
  \draw (0,-0.5*\L) -- (0,0.5*\L);
  \node[op, left] at (-0.1*\L,0) {$\color{red}\Gamma_{L(e)}$};
  \node[op, right] at (0.2*\L,0) {$\color{red}\Gamma'_{R(e)}$};
  \node[op, right] at (0,0.28) {$e$};
}\quad
&
\tikz[baseline=-0.6ex]{
  \draw[<->] (0,0) -- (0.8,0);
}
\quad
\raisebox{-0.6cm}{\tikz[baseline=-0.6ex]{
  \def\L{\Ledge}
  \draw (0,0) -- (\L,0);
  \draw (\L,0) -- (\L,\L);
  \node[op, above] at (0.5*\L,-0.2) {$\color{magenta}\tilde Z^d$};
  \node[op, right] at (0.82*\L,0.5*\L) {$\color{magenta}\tilde X_e$};
}}
\quad, \\[1.4em]
{\color{red} \Pi_p}
\quad
&
\tikz[baseline=-0.6ex]{
  \draw[<->] (0,0) -- (0.8,0);
}
\quad
\raisebox{-0.6cm}{\tikz[baseline=-0.6ex]{
  \def\L{\Ledge}
  \draw (0,0) rectangle (\L,\L);
  \node[op, above] at (0.5*\L,0.87*\L) {$\color{magenta}\tilde Z$};
  \node[op, below] at (0.5*\L,0.17*\L) {$\color{magenta}\tilde Z^\dagger$};
  \node[op, left] at (0.14*\L,0.5*\L) {$\color{magenta}\tilde Z$};
  \node[op, right] at (0.86*\L,0.5*\L) {$\color{magenta}\tilde Z^\dagger$};
  \node[op] at (0.5*\L,0.5*\L) {$p$};
}}
\quad .
\label{eq:bosonization_review_ZD}
\end{eqs}

Here $L(e)$ and $R(e)$ denote the two plaquettes adjacent to the oriented edge $e$. The first two lines are the images of fermionic clock hopping operators, while the last line is the image of {the local $\ZZ_D^F$ symmetry generator $\Pi_p$. Physical fermion parity is $P_p=\Pi_p^d$}.

{To reproduce the local symmetry-invariant operator algebra, we impose a $\ZZ_D$ Gauss-law constraint at each vertex}
\begin{equation}
    G_v :=
    \begin{tikzpicture}[baseline=-0.6ex, line width=0.5pt]
        \def\L{\Ledge}
        \draw (-\L,0) -- (\L,0);
        \draw (0,0) -- (0,-\L);
        \draw (0,0) -- (0,\L);
        \draw (0,\L) -- (\L,\L);
        \draw (\L,\L) -- (\L,0);
        \node[op] at (0,0) {$v$};
        \node[op] at (-0.5*\L,0) {${\color{magenta}\tilde{X}^\dagger}$};
        \node[op] at (0,-0.5*\L) {${\color{magenta}\tilde{X}^\dagger}$};
        \node[op] at (0,0.5*\L) {$\color{magenta}\tilde{X}\tilde{Z}^d$};
        \node[op] at (0.5*\L,0) {$\color{magenta}\tilde{X}\tilde{Z}^d$};
        \node[op] at (0.5*\L,\L) {$\color{magenta}\tilde{Z}^d$};
        \node[op] at (\L,0.5*\L) {$\color{magenta}\tilde{Z}^d$};
    \end{tikzpicture}
    =1 .
    \label{eq:bosonization_vertex_constraint_D}
\end{equation}
{These local gauge constraints require}
\begin{equation}
    G_v|\Psi\rangle=|\Psi\rangle,
    \qquad \forall v .
\end{equation}
{The operator $G_v$ is a small closed loop of the order-$D$ fermion in the $\ZZ_D$ toric code. On a periodic lattice, the plaquette images obey $\prod_p\Pi_p=1$, so the displayed map selects the globally neutral sector. A Hilbert-space isomorphism also requires matching the noncontractible Wilson-loop sectors to the chosen fermionic boundary conditions.}

Meanwhile we also get a $\ZZ_D$ 1-form symmetry which corresponds to the non-contractible Wilson loop of this order $D$ fermion. If we have a noncontractible loop $l$ on the lattice, the 1-form symmetry is generated by string operator
\begin{eqs}
    \begin{tikzpicture}[
  baseline=(current bounding box.center),
  scale=0.95,
  op/.style={inner sep=1pt, fill=white, text=blue},
  wire/.style={black, line width=0.8pt},
  dashline/.style={gray, dashed, line width=0.55pt}
]
\def\L{1.2}

\begin{scope}[shift={(-2.3,0)}]
  \draw[wire] (-1*\L,\L) -- (3*\L,\L);
  \draw[dashline] (-1*\L,\L) -- (-1*\L,-0.15*\L);
  \draw[dashline] (0,\L) -- (0,0);
  \draw[dashline] (\L,\L) -- (\L,0);
  \draw[dashline] (2*\L,\L) -- (2*\L, 0);
  \node[op] at (0.5*\L,\L) {$\textcolor{magenta}{\widetilde Z^d}$};
  \node[op] at (1.5*\L,\L) {$\textcolor{magenta}{\widetilde Z^d}$};
    \node[op] at (2.5*\L,\L) {$\textcolor{magenta}{\widetilde Z^d}$};
        \node[op] at (-0.5*\L,\L) {$\textcolor{magenta}{\widetilde Z^d}$};
  \node[op] at (0,0.5*\L) {$\textcolor{magenta}{\widetilde X}$};
    \node[op] at (-\L,0.5*\L) {$\textcolor{magenta}{\widetilde X}$};
  \node[op] at (\L,0.5*\L) {$\textcolor{magenta}{\widetilde X}$};
    \node[op] at (2*\L,0.5*\L) {$\textcolor{magenta}{\widetilde X}$};
\node at (-1.3*\L,0.5*\L) {$\cdots$};
\node at (2.8*\L,0.5*\L) {$\cdots$};
  \end{scope}
\end{tikzpicture}
\end{eqs}

{Starting with the product-state stabilizers $\{\Pi_p\}$ of the trivial $\ZZ_D^F$ SPT, the bosonization map gives a Pauli stabilizer group generated by their plaquette images $B_p$ and the Gauss-law operators $G_v$}, which realizes the $\mathbb Z_D$ toric code. In order to obtain a non-trivial SPT, we will need to start with an appropriate twisted quantum double, and reverse the bosonization process. We will focus our example on $D=8$.

\section{Majorana-Pauli stabilizer codes for the \texorpdfstring{$\ZZ^F_8$}{Z8F} SPT and its duality web}
\label{sec:Z8F_and_duality_web}

{The main result of this section is a Majorana-Pauli stabilizer model for the nontrivial $\ZZ_8^F$ fermionic SPT. This phase represents the nonzero element of the known $\ZZ_2$ classification and has no free-fermion representative~\cite{W16,WLG17}. It therefore tests the framework on an interacting SPT with non-split fermionic symmetry: its order-eight symmetry generator has fermion parity as its fourth power. The classification and the bosonic parent topological order are known inputs; the explicit stabilizer realization uses the $D=8$ fermionic clock operators developed in Sec.~\ref{sec:fermionicclock}.

{Sec.}~\ref{sec:ZF8_SPT_from_condensation} explains which charge subgroup to ungauge, why the resulting phase has no intrinsic topological order, and how its symmetry-flux statistics identify the SPT. {Sec.}~\ref{sec:Z8F_SPT} gives the local stabilizers and identifies the SPT through its boundary symmetry action. {Secs.}~\ref{sec:Z4enrichedTC_from_gauging_Z2F_in_ZF8SPT} and~\ref{sec:Z4entrichedTC_from_U88} then construct the same $\ZZ_4$-enriched toric-code phase by fermion-parity gauging and by a bosonic ungauging route. These comparisons verify the corresponding relations in Fig.~\ref{fig:Z_16_commutative_diagram}.}

\subsection{\texorpdfstring{$\ZZ^F_8$}{Z8F} fermionic SPT from ungauging}
\label{sec:ZF8_SPT_from_condensation}

We now construct a nontrivial fermionic SPT protected by a cyclic fermionic symmetry group $\ZZ^F_D$. For the family considered here, the first nontrivial case occurs at $D=8$. We first give the anyon-level ungauging argument for the $\ZZ^F_8$ SPT, and then implement the same ungauging at the stabilizer level in Sec.~\ref{sec:Z8F_SPT}. {{Sec.}~\ref{sec:general_constructions} develops the cyclic generalization and specifies the scope of the construction for other Abelian fermionic symmetries.}

It is well known that gauging a bosonic SPT protected by a finite group $G$ gives a twisted gauge theory of $G$~\cite{LevinGu2012Z2SPT}. Similarly, to construct a fermionic SPT protected by {$G_f$}, it is useful to start from a bosonic topological order containing an emergent fermion with the desired {$G_f$} charge, and then ungauge the corresponding charge sector by binding this emergent object to the corresponding local physical charge. Since local charges are topologically transparent, they do not affect the braiding calculation. The role of the bosonic parent topological order is to encode the correct symmetry-flux response of the resulting SPT. For the nontrivial $\ZZ^F_8$ SPT, the appropriate bosonic parent theory is the twisted $\ZZ_8$ gauge theory~\cite{W16,WLG17}
\begin{equation}
    K_{\rm std} =
    \begin{pmatrix}
        0 & 8 \\
        8 & -8
    \end{pmatrix}.
    \label{eq:Z8F_cover}
\end{equation}

An anyon is labeled by an integer vector
\begin{equation}
    \ell =
    \begin{pmatrix}
        p\\ q
    \end{pmatrix}_{\rm std},
\end{equation}
with the identification $\ell\sim \ell+K_{\rm std}\Lambda$ for $\Lambda\in\ZZ^2$~\cite{LevinStern2012AbelianKtheory}. We use
\begin{equation}{
    \theta_\ell=\exp\left(\pi i\ell^T K_{\rm std}^{-1}\ell\right)
}\end{equation}
for the topological spin.
The mutual braiding between two anyons $\ell$ and $\ell'$ is
\begin{equation}
    M_{\ell,\ell'}
    =
    \exp\!\left(2\pi i\,\ell^T K_{\rm std}^{-1}\ell'\right).
\end{equation}

The order-eight anyon
\begin{equation}
    f:=
    \begin{pmatrix}
        4\\-1
    \end{pmatrix}_{\rm std},
    \qquad
    h_f=\frac{1}{2},
    \label{eq:order_8_fermion}
\end{equation}
is a fermion. We identify this emergent fermion with the unit $\ZZ^F_8$ symmetry charge. In the ungauging construction, $f$ is made local by binding it to the corresponding physical local charge $\psi$. The bosonic composite used to implement ungauging is $f\psi$.

Let us first check that this ungauging removes all intrinsic topological order. Since
\begin{equation}
    K_{\rm std}^{-1}f
    =
    \begin{pmatrix}
        3/8\\
        1/2
    \end{pmatrix},
\end{equation}
the mutual braiding between a general anyon $\ell=(p,q)^T_{\rm std}$ and $f$ is
\begin{equation}
    M_{\ell,f}
    =
    \exp\!\left(2\pi i\,\ell^T K_{\rm std}^{-1}f\right)
    =
    \exp\!\left[
        2\pi i
        \left(
            \frac{3p}{8}+\frac{q}{2}
        \right)
    \right].
\end{equation}
An anyon remains deconfined after ungauging only if it braids trivially with $f$. Therefore it must satisfy
\begin{equation}
    3p+4q=0 \pmod{8}.
\end{equation}
The solutions are precisely the anyons generated by powers of $f$. After ungauging, these powers are identified with local physical charge excitations, and hence do not represent nontrivial deconfined topological anyons. Therefore no intrinsic topological order remains, as expected for an SPT phase.

It remains to determine which $\ZZ^F_8$ SPT is obtained. This can be diagnosed from the statistics of a unit symmetry flux in the gauged theory.
Since $f$ is identified with the unit $\ZZ_8$ charge, a unit $\ZZ_8$ flux $\phi=(p',q')^T_{\rm std}$ is characterized by the charge-flux braiding condition
\begin{equation}
    M_{\phi,f}
    =
    \exp\left(2\pi i\,\phi^T K_{\rm std}^{-1}f\right)
    =
    \exp\left(\frac{2\pi i}{8}\right),
\end{equation}
which is equivalent to
\begin{equation}
    3p'+4q'=1 \pmod{8}.
\end{equation}
One convenient representative is
\begin{equation}
    \phi=
    \begin{pmatrix}
        3\\4
    \end{pmatrix}_{\rm std}.
\end{equation}
Indeed, this anyon satisfies
\begin{equation}
    M_{\phi,f}
    =
    \exp\left(2\pi i\,\phi^T K_{\rm std}^{-1}f\right)
    =
    \exp\left(\frac{2\pi i}{8}\right),
\end{equation}
as required for the mutual braiding between a unit $\ZZ_8$ flux and a
unit $\ZZ_8$ charge. Its topological spin phase is
\begin{equation}
    \theta_\phi
    =
    \exp\left(i\pi\,\phi^T K_{\rm std}^{-1}\phi\right)
    =
    {\exp\left(2\pi i\,\frac{1}{16}\right)}.
\end{equation}
The phase $\theta_\phi$ itself is not a well-defined invariant, because
a symmetry flux can be redefined by attaching symmetry charge:
\begin{equation}
    \phi \longrightarrow \phi+kf,
    \qquad k\in\ZZ_8.
\end{equation}
Under this redefinition, the spin phase changes as
\begin{equation}
    \theta_{\phi+kf}
    =
    \theta_\phi\,M_{\phi,f}^{\,k}\,\theta_f^{\,k^2}.
\end{equation}
Since $M_{\phi,f}=\exp(2\pi i/8)$ and $\theta_f=-1$, we have
\begin{equation}
    \theta_{\phi+kf}
    =
    \theta_\phi\,
    \exp\left(\frac{2\pi i k}{8}\right)
    (-1)^{k^2}.
\end{equation}
Therefore, we define the invariant as $\theta_\phi^8$ since
\begin{equation}
    \left(\theta_{\phi+kf}\right)^8
    =
    \theta_\phi^8\,
    \exp(2\pi i k)\,
    (-1)^{8k^2}
    =
    \theta_\phi^8,
\end{equation}
which is invariant under charge attachment and is independent
of the choice of flux representative. For the representative
{$\phi=(3,4)^T_{\rm std}$}, we find
\begin{equation}
    \theta_\phi^8
    =
    {\exp\left(2\pi i\,\frac{1}{2}\right)}
    =
    -1.
\end{equation}
This nontrivial symmetry-flux spin invariant distinguishes the nontrivial
$\ZZ^F_8$ fermionic SPT from the trivial one~\cite{W16,WLG17}. Hence,
ungauging the charge subgroup generated by the order-eight fermion $f=(4,-1)^T_{\rm std}$ in the twisted
$\ZZ_8$ gauge theory of Eq.~\eqref{eq:Z8F_cover} produces the unique
nontrivial $\ZZ^F_8$ fermionic SPT.

Finally, let us clarify the basis convention used here and in Fig.~\ref{fig:Z_16_commutative_diagram}. The calculation above used the standard twisted-gauge-theory basis of Eq.~\eqref{eq:Z8F_cover}. The diagonal basis for the same anyon theory is obtained by the $GL(2,\ZZ)$ transformation
\begin{equation}
    K_{\rm diag}
    =
    W^T K_{\rm std}W
    =
    \begin{pmatrix}
        8&0\\
        0&-8
    \end{pmatrix},
    \qquad
    W=
    \begin{pmatrix}
        1&0\\
        1&1
    \end{pmatrix}.
\end{equation}
Anyon vectors transform as
\begin{equation}
    \ell_{\rm diag}=W^T\ell_{\rm std},
    \qquad
    \ell_{\rm std}=W^{-T}\ell_{\rm diag}.
\end{equation}
Therefore the fermionic charge to be ungauged and the unit flux become
\begin{equation}
    f=
    \begin{pmatrix}
        4\\-1
    \end{pmatrix}_{\rm std}
    \longmapsto
    \begin{pmatrix}
        3\\-1
    \end{pmatrix}_{\rm diag}
    \equiv
    \begin{pmatrix}
        3\\7
    \end{pmatrix}_{\rm diag},
\end{equation}
and
\begin{equation}
    \phi=
    \begin{pmatrix}
        3\\4
    \end{pmatrix}_{\rm std}
    \longmapsto
    \begin{pmatrix}
        7\\4
    \end{pmatrix}_{\rm diag}.
\end{equation}
In Sec.~\ref{sec:Z8F_SPT}, we implement this ungauging explicitly as a Majorana-Pauli stabilizer code for the $\ZZ^F_8$ SPT.

\subsection{Majorana-Pauli stabilizer code for \texorpdfstring{$\mathbb{Z}_8^F$}{Z8F} SPT}
\label{sec:Z8F_SPT}

Here, we construct the generalized Majorana-Pauli stabilizer code for the $\ZZ^F_8$ SPT. We follow the strategy discussed above in Sec.~\ref{sec:ZF8_SPT_from_condensation}. We first present the result and then discuss the technical details to construct the model on the lattice.

\Needspace{8\baselineskip}
Our $\ZZ^F_8$ SPT Majorana-Pauli stabilizer model has stabilizer generators
\begin{eqs}
A_v &=
\begin{tikzpicture}[baseline=-0.6ex, line width=0.5pt]
  \def\L{1.2}
  \draw (-\L,0) -- (\L,0);
  \draw (0,0) -- (0,-\L);
  \draw (0,0) -- (0,\L);
  \draw (0,\L) -- (\L,\L);
  \draw (\L,\L) -- (\L,0);

  \node[op] at (0,0) {$v$};
  \node[op] at (-0.5*\L,0) {$\textcolor{blue}{X^\dagger}$};
  \node[op] at (0,-0.5*\L) {$\textcolor{blue}{X^\dagger}$};
  \node[op] at (0,0.5*\L) {$\textcolor{blue}{XZ}$};
  \node[op] at (0.5*\L,0) {$\textcolor{blue}{XZ^\dagger}$};
  \node[op] at (0.5*\L,\L) {$\textcolor{blue}{Z}$};
  \node[op] at (\L,0.5*\L) {$\textcolor{blue}{Z^\dagger}$};
  \node[op] at (0.5*\L,0.5*\L) {$\textcolor{red}{\Pi^\dagger}$};
\end{tikzpicture}
,\qquad
B_p =
\raisebox{-0.6cm}{
\begin{tikzpicture}[baseline=-0.6ex, line width=0.5pt]
  \def\L{1.2}
  \draw (0,0) rectangle (\L,\L);

  \node[op] at (0.5*\L,\L) {$\textcolor{blue}{Z^{-2}}$};
  \node[op] at (0,0.5*\L) {$\textcolor{blue}{Z^{-2}}$};
  \node[op] at (\L,0.5*\L) {$\textcolor{blue}{Z^2}$};
  \node[op] at (0.5*\L,0) {$\textcolor{blue}{Z^2}$};
  \node[op] at (0.5*\L,0.5*\L) {$\textcolor{red}{P_p}$};
\end{tikzpicture}}
,
\\
C_e &=
\raisebox{-0.6cm}{
\begin{tikzpicture}[baseline=-0.6ex, line width=0.5pt]
  \def\L{1.2}
  \draw (-\L,0) -- (0,0);
  \draw (0,0) -- (0,\L);

  \node[op] at (-0.5*\L,0) {$\textcolor{blue}{Z^8}$};
  \node[op] at (0,0.5*\L) {$\textcolor{blue}{X_e^8}$};
\end{tikzpicture}}
,\qquad
\raisebox{0.6cm}{
\begin{tikzpicture}[baseline=-0.6ex, line width=0.5pt]
  \def\L{1.2}
  \draw (0,0) -- (\L,0);
  \draw (0,0) -- (0,-\L);

  \node[op] at (0.5*\L,0) {$\textcolor{blue}{X_e^8}$};
  \node[op] at (0,-0.5*\L) {$\textcolor{blue}{Z^8}$};
\end{tikzpicture}}
,
\\
(-i)\cdot D_e &=
\raisebox{-0.6cm}{
\begin{tikzpicture}[baseline=-0.6ex, line width=0.5pt]
  \def\L{1.2}
  \draw (-\L,0) -- (0,0);
  \draw (0,0) -- (0,\L);

  \node[op] at (-0.5*\L,0) {$\textcolor{blue}{Z^{-2}}$};
  \node[op] at (-0.5*\L,0.5*\L) {$\textcolor{red}{\Gamma}$};
  \node[op] at (0,0.5*\L) {$\textcolor{blue}{X_e^4}$};
  \node[op] at (0.5*\L,0.53*\L) {$\textcolor{red}{\Gamma'}$};
\end{tikzpicture}}
,\qquad
\raisebox{0.6cm}{
\begin{tikzpicture}[baseline=-0.6ex, line width=0.5pt]
  \def\L{1.2}
  \draw (0,0) -- (\L,0);
  \draw (0,0) -- (0,-\L);

  \node[op] at (0.5*\L,0) {$\textcolor{blue}{X_e^4}$};
  \node[op] at (0.5*\L,0.5*\L) {$\textcolor{red}{\Gamma}$};
  \node[op] at (0,-0.5*\L) {$\textcolor{blue}{Z^2}$};
  \node[op] at (0.53*\L,-0.5*\L) {$\textcolor{red}{\Gamma'}$};
\end{tikzpicture}}
\quad,
\label{eq:Z8F_SPT_stabilizers}
\end{eqs}
where Pauli $\color{blue} X$ and $\color{blue} Z$ act on $\mathbb{Z}_{16}$ qudits, and $\color{red} \Pi$, $\color{red} P$, $\color{red} \Gamma$, $\color{red} \Gamma'$ are fermionic clock operators with $D=8$ defined in Sec.~\ref{sec:fermionicclock}. We write the stabilizer group for the $\ZZ^F_8$ SPT as
\begin{equation}
    \mathcal{S}_{\ZZ^F_8\text{-SPT}} = 
    \big\langle
        \{A_v\},\{B_p\},\{C_e\},\{D_e\}
    \big\rangle \, .
    \label{eq:Z8F_SPT_stabilizer_group}
\end{equation}
Each $A_v$ is an order-sixteen operator. $B_p$ and $D_e$ are of order-eight. $C_e$ is of order-two. The groundspace is the simultaneous $+1$ eigenspace of all the stabilizers. The commutativity of these stabilizers follows straightforwardly from our lattice-level construction, which we discuss next. The Hamiltonian has a $\ZZ^F_8$ symmetry represented by
\begin{equation}
    U\left((g,a)\right):= \left(\prod_p {\color{red}\Pi_p}\right)^{g+4a}, \quad g \in G_b =\ZZ_4,~~a \in \ZZ_2^F,
\end{equation}
where the group law of elements $(g,a)\in \ZZ^F_8$ is defined in Eq.~\eqref{eq:group_law_Gf}. The fermion parity symmetry is
\begin{equation}
    U^F=U\left((0,1)\right) = \left(\prod_p {\color{red}\Pi_p}\right)^{4} = \prod_p {\color{red}P_p},
\end{equation}
as expected.

We now derive {the stabilizers in Eq.}~\eqref{eq:Z8F_SPT_stabilizers} following the ungauging construction described in Sec.~\ref{sec:ZF8_SPT_from_condensation}. The first step is to construct a Pauli stabilizer model for the bosonic $\ZZ_8$ twisted gauge theory of Eq.~\eqref{eq:Z8F_cover}. We follow the condensation procedure of Ref.~\cite{Ellison22}, reviewed in Sec.~\ref{sec:fermionic_toric_code}. We begin with the $\ZZ_{16}$ toric code on the square lattice, whose stabilizer group is
\begin{equation}
    \mathcal{S}_1 = \left\langle
        \raisebox{-0.5\height}{\includegraphics[scale=0.8]{Figures/AvTC.pdf}},
        \quad
        \raisebox{-0.5\height}{\includegraphics[scale=0.8]{Figures/BpTC.pdf}}
    \right\rangle  = \mathcal{S}_{\ZZ_{16}\text{ toric code}}\, .
\end{equation}
Anyons in the $\ZZ_{16}$ toric code are generated by the electric charge $e$ and the magnetic flux $m$, both of which have order sixteen. The generators are both bosons, and their mutual braiding is
\begin{equation}
M_{e,m}=\exp\left(\frac{i\pi}{8}\right).
\end{equation}
To get the $\ZZ_8$ twisted gauge theory, we then condense the boson
\begin{equation}
    e^8m^8,
\end{equation}
{whose lattice hopping operators are the $\{C_e\}$ terms} in Eq.~\eqref{eq:Z8F_SPT_stabilizers}. The deconfined anyons after this condensation are generated by
\begin{equation}
    em,\qquad e^2 .
\end{equation}
These two generators have the same fusion and braiding data as the anyons of the twisted $\ZZ_8$ gauge theory described by Eq.~\eqref{eq:Z8F_cover}. Indeed, under the identification of
\begin{equation}
    \begin{pmatrix}1\\0\end{pmatrix}_{\rm std} \longleftrightarrow em,
    \qquad
    \begin{pmatrix}0\\1\end{pmatrix}_{\rm std} \longleftrightarrow e^2,
    \label{eq:Z8_twisted_gauge_theory_identification}
\end{equation}
the mutual and self-statistics reproduce those computed from $K^{-1}$. Thus, condensing $e^8m^8$ in the $\ZZ_{16}$ toric code gives a Pauli stabilizer realization of the bosonic twisted $\ZZ_8$ gauge theory appearing in Eq.~\eqref{eq:Z8F_cover}. After condensation of $e^8m^8$ using $\{C_e\}$, our stabilizer group is generated by the $\{C_e\}$ terms together with elements in $\mathcal{S}_1$ that commute with the $\{C_e\}$ terms, i.e.
\begin{equation}
    \mathcal{S}_2 = 
    \big\langle
        \{C_e\},~\mathcal{Z}_{\mathcal{S}_1}(\{C_e\})
    \big\rangle \, ,
\end{equation}
where $\mathcal Z_{\mathcal S_1}(\{C_e\})$ is the centralizer of the condensation terms $\{C_e\}$ inside the initial stabilizer group $\mathcal S_1$, namely
\begin{equation}
    \mathcal Z_{\mathcal S_1}(\{C_e\}) := \{s\in \mathcal S_1\mid [s,C_e]=0\ \text{for all }e\}.
\end{equation}
Graphically, $\mathcal Z_{\mathcal S_1}(\{C_e\})$ is generated by
\begin{equation}
    A^{\mathcal{S}_2}_v =
    \begin{tikzpicture}[baseline=-0.6ex, line width=0.5pt]
      \def\L{1.2}
      \draw (-\L,0) -- (\L,0);
      \draw (0,0) -- (0,-\L);
      \draw (0,0) -- (0,\L);
      \draw (0,\L) -- (\L,\L);
      \draw (\L,\L) -- (\L,0);
      \node[op] at (0,0) {$v$};
      \node[op] at (-0.5*\L,0) {$\textcolor{blue}{X^\dagger}$};
      \node[op] at (0,-0.5*\L) {$\textcolor{blue}{X^\dagger}$};
      \node[op] at (0,0.5*\L) {$\textcolor{blue}{XZ}$};
      \node[op] at (0.5*\L,0) {$\textcolor{blue}{XZ^\dagger}$};
      \node[op] at (0.5*\L,\L) {$\textcolor{blue}{Z}$};
      \node[op] at (\L,0.5*\L) {$\textcolor{blue}{Z^\dagger}$};
    \end{tikzpicture}
    ,\quad
    B^{\mathcal{S}_2}_p =
    \raisebox{-0.6cm}{\begin{tikzpicture}[baseline=-0.6ex, line width=0.5pt]
      \def\L{1.2}
      \draw (0,0) rectangle (\L,\L);
      \node[op] at (0.5*\L,\L) {$\textcolor{blue}{Z^{-2}}$};
      \node[op] at (0,0.5*\L) {$\textcolor{blue}{Z^{-2}}$};
      \node[op] at (\L,0.5*\L) {$\textcolor{blue}{Z^2}$};
      \node[op] at (0.5*\L,0) {$\textcolor{blue}{Z^2}$};
    \end{tikzpicture}}
    \label{eq:Z8F_twisted_gauge_theory_AB_terms}
\end{equation}

Next, to obtain the $\ZZ_8^F$ SPT, we must take the order-eight emergent fermion
\begin{equation}
f=\begin{pmatrix}4\\-1\end{pmatrix}_{\rm std}
\end{equation}
in the twisted $\ZZ_8$ gauge theory and bind it to the physical transparent fermion $\psi$. Under the identification in Eq.~\eqref{eq:Z8_twisted_gauge_theory_identification}, this anyon is represented in terms of anyons of the initial $\ZZ_{16}$ toric code as
\begin{equation}
    f=\begin{pmatrix}4\\-1\end{pmatrix}_{\rm std}
    \quad\longleftrightarrow\quad
    (em)^4(e^2)^{-1}=e^2m^4 .
\end{equation}
Thus, the second step is ungauging the charge subgroup generated by $f$, implemented using the bosonic composite $e^2m^4\psi$. To implement this ungauging on the lattice, we introduce the $\ZZ^F_8$ fermionic clock and shift operators at each plaquette. That is, the stabilizer group before ungauging is generated by the generators in $\mathcal{S}_2$ as well as single $\ZZ^F_8$ fermionic clock operators ${\color{red}\Pi_p}$ on each plaquette. Then, the short-string operators corresponding to $e^2m^4\psi$ are precisely the $\{D_e\}$ terms in Eq.~\eqref{eq:Z8F_SPT_stabilizers}. The blue generalized Pauli part creates the short string of $e^2m^4$, while the red fermionic clock operators hop the physical fermion $\psi$. Note that we have chosen $\{D_e\}$ so that they commute with $\{C_e\}$. After imposing the ungauging terms for $e^2 m^4 \psi$, we again look for elements commuting with both $\{D_e\}$ and $\{C_e\}$. The final stabilizer group we get is then Eq.~\eqref{eq:Z8F_SPT_stabilizer_group}.

Having presented our Majorana-Pauli stabilizer code for the $\ZZ_8^F$ SPT, we now show that the model indeed realizes the nontrivial $\ZZ_8^F$ SPT. We do this by extracting the boundary symmetry action and identifying the corresponding supercohomology data. Following the procedure of Sec.~\ref{sec:Z2Z2F_SPT}, consider acting the symmetry on a half-plane $R$ with an upper boundary,
\begin{equation}
U_R((1,0))=\prod_{p\in R}{\color{red}\Pi_p}.
\end{equation}
{
As before, multiplying by bulk vertex stabilizers cancels the plaquette clock factors and the Pauli operators on interior edges, leaving the effective boundary operator
\begin{eqs}
&\vcenter{\hbox{
{\color{black}\begin{tikzpicture}[scale=1.2,baseline=5pt]
\foreach \x in {0,1,2,3,4}{
    \draw[dashed] (\x,0) -- (\x,4);
}
\foreach \y in {0,1,2,3}{
    \draw[dashed] (-0.3,\y) -- (4.3,\y);
}
\draw[line width=0.8pt] (-0.3,4) -- (4.3,4);
\foreach \x in {0.5,1.5,2.5,3.5}{
    \foreach \y in {0.5,1.5,2.5,3.5}{
        \node at (\x,\y) {$\textcolor{red}{\Pi}$};
    }
}
\end{tikzpicture}}
}} \\[1em]
\cong\quad
&\vcenter{\hbox{
{\color{black}\begin{tikzpicture}[scale=1.2,baseline=5pt]
\foreach \x in {0,1,2,3,4}{
    \draw[dashed] (\x,0) -- (\x,-4);
}
\foreach \y in {-1,-2,-3,-4}{
    \draw[dashed] (-0.3,\y) -- (4.3,\y);
}
\draw[line width=0.8pt] (-0.3,0) -- (4.3,0);
\foreach \x in {0,1,2,3,4}{
    \node[op] at (\x,-0.5) {$\textcolor{blue}{X}$};
}
\foreach \x in {0.5,1.5,2.5,3.5}{
    \node[op] at (\x,0) {$\textcolor{blue}{Z}$};
}
\end{tikzpicture}}
}} \,
:= \widetilde{U}((1,0))~.
\label{eq:Z8F_boundary_effective_symmetry}
\end{eqs}
The $\mathbb{Z}_8^F$ group law implies $\widetilde{U}((1,0))^8\sim \mathbf{1}$ up to bulk stabilizers. Restricting $\widetilde{U}((1,0))$ to a finite boundary interval $l=[a,b]$, denoted $\widetilde{U}_l((1,0))$, this relation no longer holds strictly. Instead, $\widetilde{U}_l((1,0))^8$ reduces to bulk stabilizers multiplied by residual operators localized near the endpoints $a$ and $b$. These endpoint terms encode the supercohomology data.
We choose the following truncation, with the same endpoint convention as in Sec.~\ref{sec:Z2Z2F_SPT}:
\begin{eqs}
   \widetilde{U}_l ((1,0))=\;& \!\!
\raisebox{0.8em}{{\color{black}\begin{tikzpicture}[
  baseline=(current bounding box.center),
  scale=0.85,
  op/.style={inner sep=1pt, fill=white, text=blue},
  redop/.style={inner sep=1pt, fill=white, text=red},
  orangeop/.style={inner sep=1pt, fill=white, text=orange},
  wire/.style={black, line width=0.8pt},
  dashline/.style={gray, dashed, line width=0.55pt}
]
\def\L{1.2}

\begin{scope}[shift={(-2.3,0)}]
  \draw[wire] (-0.7*\L,\L) -- (1.9*\L,\L);
  \draw[dashline] (0,\L) -- (0,-0.15*\L);
  \draw[dashline] (\L,\L) -- (\L,-0.15*\L);
  \draw[dashline] (-0.7*\L,0) -- (1.9*\L,0);
  \node at (0,1.3*\L) {$a$};
  \fill (0,\L) circle (2pt);
  \node[op] at (0.5*\L,\L) {$Z$};
  \node[op] at (1.5*\L,\L) {$Z$};
  \node[op] at (0,0.5*\L) {$X$};
  \node[op] at (\L,0.5*\L) {$X$};
\end{scope}

\node at (0.6,0.5*\L) {$\cdots$};

\begin{scope}[shift={(2.0,0)}]
  \draw[wire] (-0.7*\L,\L) -- (2.0*\L,\L);
  \draw[dashline] (0,\L) -- (0,-0.15*\L);
  \draw[dashline] (\L,\L) -- (\L,-0.15*\L);
  \draw[dashline] (-0.7*\L,0) -- (1.9*\L,0);
  \node at (2*\L,1.3*\L) {$b$};
  \fill (2*\L,\L) circle (2pt);
  \node[op] at (0.5*\L,\L) {$Z$};
  \node[op] at (1.5*\L,\L) {$Z$};
  \node[op] at (0,0.5*\L) {$X$};
  \node[op] at (\L,0.5*\L) {$X$};
\end{scope}
\end{tikzpicture}}}
\label{eq:truncated_Z8f_boundary_symmetry}
\end{eqs}
For the non-split fermionic symmetry, composing the truncated quotient-symmetry operators also produces a fermion-parity factor:
\begin{eqs}
    &\widetilde{U}_l((g,0))\,\widetilde{U}_l((h,0)) \\
    \sim&~ \Omega(g,h)\,
    \widetilde{U}_l\left(([g+h]_4,0)\right)\,
    \left(\widetilde{U}_l^F\right)^{w_2(g,h)},
\label{eq:fermionic_defect}
\end{eqs}
where $\widetilde{U}_l^F$ denotes the truncated effective fermion parity symmetry,
$g, h, [g+h]_4\in \ZZ_4$ take values in $\{0,1,2,3\}$, and
\begin{equation}
    w_2(g, h)
    =
    \left\lfloor \frac{g+h}{4}\right\rfloor
    \in \{0,1\}
\end{equation}
is the extension cocycle, denoted $\omega_2$ in the supercohomology data, for
\[
    1\rightarrow \ZZ_2^F
    \rightarrow \ZZ_8^F
    \rightarrow \ZZ_4
    \rightarrow 1 .
\]
Before carrying out the
calculation, we emphasize that the effective fermion parity
$\widetilde{U}_l^F$ must itself be non-anomalous. More precisely, after
eliminating the middle terms in $(\widetilde{U}_l^F)^2$ using
stabilizers, the remaining endpoint operators near the two ends of $l$ must
commute with effective boundary symmetry $\widetilde{U}_l((1,0))$.
We first multiply $\widetilde{U}^F=\widetilde{U}((1,0))^4$ by plaquette stabilizers, moving the horizontal $Z^4$ factors to the lower row:
\begin{eqs}
    \widetilde{U}^F
    &=\raisebox{-0.3\height}{\includegraphics[scale=0.2]{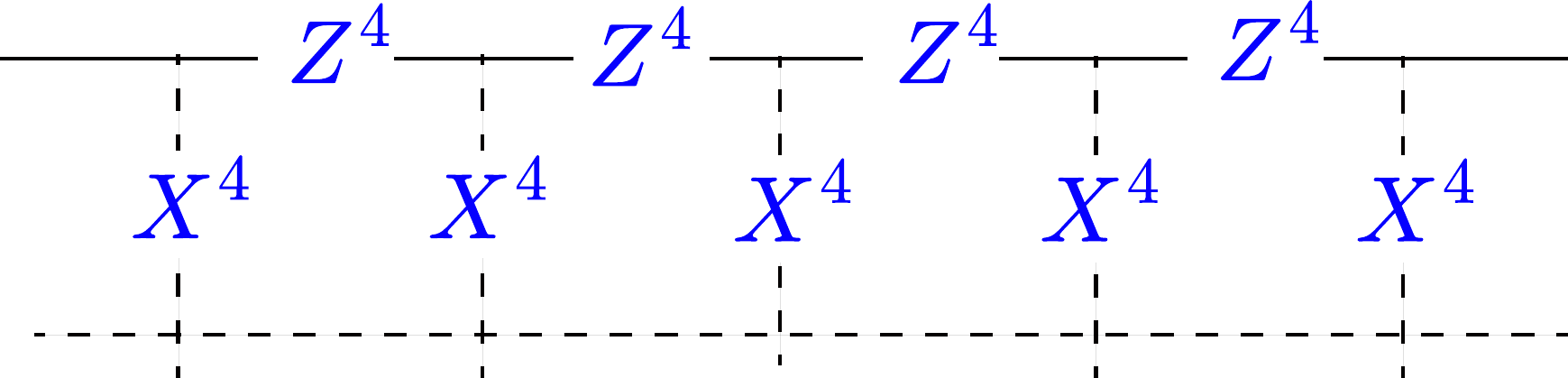}}\\[1em]
    &\sim
    \raisebox{-0.5\height}{\includegraphics[scale=0.2]{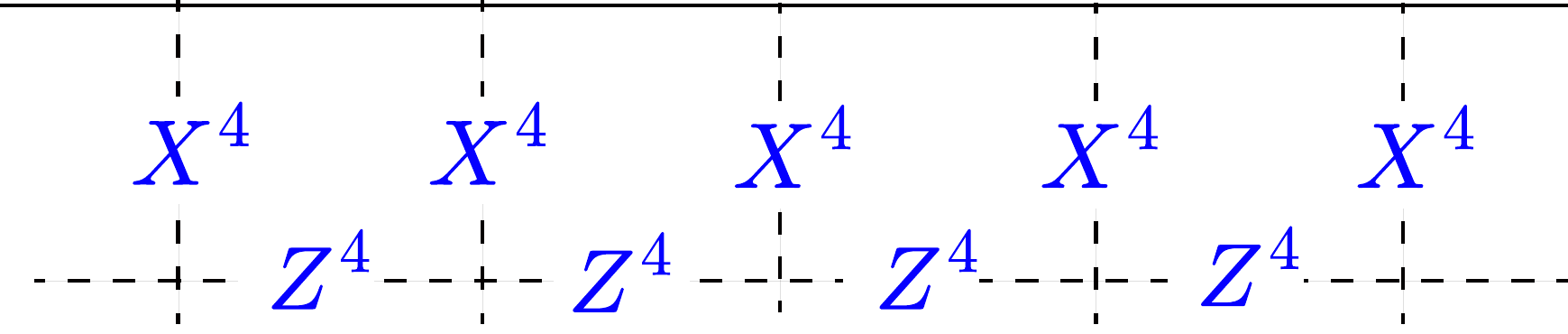}} .
\end{eqs}
We then truncate this representative:
\begin{equation}
    \raisebox{-0.8em}{$\widetilde{U}^F_l=$}~~
    {\color{black}\begin{tikzpicture}[
    baseline=(current bounding box.center),
    scale=0.85,
    op/.style={inner sep=1pt, fill=white, text=blue},
    redop/.style={inner sep=1pt, fill=white, text=red},
    orangeop/.style={inner sep=1pt, fill=white, text=orange},
    wire/.style={black, line width=0.8pt},
    dashline/.style={gray, dashed, line width=0.55pt}
    ]
\def\L{1.2}

\begin{scope}[shift={(-2.3,0)}]
  \draw[wire] (-1*\L,\L) -- (1.9*\L,\L);
  \draw[dashline] (0,\L) -- (0,-0.15*\L);
  \draw[dashline] (\L,\L) -- (\L,-0.15*\L);
  \draw[dashline] (-1*\L,0) -- (1.9*\L,0);
  \node at (0,1.3*\L) {$a$};
  \fill (0,\L) circle (2pt);
  \node[op] at (0.5*\L,0) {$Z^{4}$};
  \node[op] at (-0.5*\L,0) {$Z^{4}$};
  \node[op] at (0*\L,0.5*\L) {$X^4$};
  \node[op] at (1*\L,0.5*\L) {$X^4$};
\end{scope}

\node at (0.6,0.5*\L) {$\cdots$};

\begin{scope}[shift={(2.0,0)}]
  \draw[wire] (-0.7*\L,\L) -- (2.0*\L,\L);
  \draw[dashline] (0,\L) -- (0,-0.15*\L);
  \draw[dashline] (\L,\L) -- (\L,-0.15*\L);
  \draw[dashline] (-0.7*\L,0) -- (2*\L,0);
  \node at (2*\L,1.3*\L) {$b$};
  \fill (2*\L,\L) circle (2pt);
  \node[op] at (0.5*\L,0) {$Z^{4}$};
  \node[op] at (-0.5*\L,0) {$Z^{4}$};
  \node[op] at (0*\L,0.5*\L) {$X^4$};
  \node[op] at (1*\L,0.5*\L) {$X^4$};
\end{scope}
\end{tikzpicture}}
\label{eq:truncated_Z8F_boundary_fermion_parity}
\end{equation}
Squaring the displayed string doubles every exponent, giving a product of the vertical-edge $C_e$ stabilizers. Therefore, no nontrivial operator remains at the endpoints,
and $\widetilde{U}^F_l$ is indeed non-anomalous. We emphasize that the naive
choice
$\widetilde{U}^F_l=\big(\widetilde{U}_l((1,0))\big)^4$
does not work, since its endpoint operators do not commute with the effective symmetry $\widetilde{U}((1,0))$ on the boundary.

Using Eqs.~\eqref{eq:truncated_Z8f_boundary_symmetry}, \eqref{eq:fermionic_defect}, and~\eqref{eq:truncated_Z8F_boundary_fermion_parity}, we compare four quotient-symmetry strings with the chosen parity string. Multiplication by squared plaquette stabilizers removes the interior, leaving the endpoint operators shown below. As before, we extract the cocycle locally at the left endpoint, independently of the dressing near the right endpoint.
\begin{equation}
\begin{aligned}
&\Omega(1,3)=\Omega(2,2)
\\
\raisebox{-0.5em}{$=$}\;&\quad
{\color{black}\begin{tikzpicture}[
  baseline=(current bounding box.center),
  scale=0.95,
  op/.style={inner sep=1pt, fill=white, text=blue},
  wire/.style={black, line width=0.8pt},
  dashline/.style={gray, dashed, line width=0.55pt}
]
\def\L{1.2}

\begin{scope}[shift={(-2.3,0)}]
  \draw[wire] (-1*\L,\L) -- (1.9*\L,\L);
  \draw[dashline] (-1*\L,\L) -- (-1*\L,-0.15*\L);
  \draw[dashline] (0,\L) -- (0,-0.15*\L);
  \draw[dashline] (\L,\L) -- (\L,-0.15*\L);
  \draw[dashline] (-1*\L,0) -- (1.9*\L,0);
  \node at (0,1.3*\L) {$a$};
  \fill (0,\L) circle (2pt);
  \node[op] at (0.5*\L,\L) {$Z^4$};
  \node[op] at (1.5*\L,\L) {$Z^4$};
  \node[op] at (-0.5*\L,0) {$Z^{-4}$};
  \node[op] at (0.5*\L,0) {$Z^{-4}$};
  \node[op] at (1.5*\L,0) {$Z^{-4}$};
\end{scope}

\node at (0.3,0.5*\L) {$\cdots$};

\begin{scope}[shift={(1.4,0)}]
  \draw[wire] (-0.6*\L,\L) -- (2.0*\L,\L);
  \draw[dashline] (0,\L) -- (0,-0.15*\L);
  \draw[dashline] (\L,\L) -- (\L,-0.15*\L);
  \draw[dashline] (2*\L,\L) -- (2*\L,-0.15*\L);
  \draw[dashline] (-0.6*\L,0) -- (2*\L,0);
  \node at (2*\L,1.3*\L) {$b$};
  \fill (2*\L,\L) circle (2pt);
  \node[op] at (-0.5*\L,\L) {$Z^4$};
  \node[op] at (0.5*\L,\L) {$Z^4$};
  \node[op] at (1.5*\L,\L) {$Z^4$};
  \node[op] at (-0.5*\L,0) {$Z^{-4}$};
  \node[op] at (0.5*\L,0) {$Z^{-4}$};
\end{scope}
\end{tikzpicture}}
\\
\raisebox{-1em}{$\sim$}\;&\quad
{\color{black}\begin{tikzpicture}[
  baseline=(current bounding box.center),
  scale=0.95,
  op/.style={inner sep=1pt, fill=white, text=blue},
  wire/.style={black, line width=0.8pt},
  dashline/.style={gray, dashed, line width=0.55pt}
]
\def\L{1.2}

\begin{scope}[shift={(-2.3,0)}]
  \draw[wire] (-1*\L,\L) -- (1.9*\L,\L);
  \draw[dashline] (-1*\L,\L) -- (-1*\L,-0.15*\L);
  \draw[dashline] (0,\L) -- (0,-0.15*\L);
  \draw[dashline] (\L,\L) -- (\L,-0.15*\L);
  \draw[dashline] (-1*\L,0) -- (1.9*\L,0);
  \node at (0,1.3*\L) {$a$};
  \fill (0,\L) circle (2pt);
  \node[op] at (-0.5*\L,0) {$Z^{-4}$};
  \node[op] at (0.1*\L,0.5*\L) {$Z^{-4}$};
\end{scope}

\node at (0.3,0.5*\L) {$\cdots$};

\begin{scope}[shift={(1.4,0)}]
  \draw[wire] (-0.6*\L,\L) -- (2.0*\L,\L);
  \draw[dashline] (0,\L) -- (0,-0.15*\L);
  \draw[dashline] (\L,\L) -- (\L,-0.15*\L);
  \draw[dashline] (2*\L,\L) -- (2*\L,-0.15*\L);
  \draw[dashline] (-0.6*\L,0) -- (2*\L,0);
  \node at (2*\L,1.3*\L) {$b$};
  \fill (2*\L,\L) circle (2pt);
  \node[op] at (2*\L,0.5*\L) {$Z^4$};
  \node[op] at (1.5*\L,0) {$Z^4$};
\end{scope}
\end{tikzpicture}}
\; .
\end{aligned}
\end{equation}
$\Omega$ can now be decomposed into two operators $\Omega_a$ and $\Omega_b$ localized near $a$ and $b$ respectively. We take $\Omega_a$ to be
\begin{equation}
\raisebox{-0.8em}{$\Omega_a(1,3)=\Omega_a(2,2)=$}
{\color{black}\begin{tikzpicture}[
  baseline=(current bounding box.center),
  scale=0.85,
  op/.style={inner sep=1pt, fill=white, text=blue},
  redop/.style={inner sep=1pt, fill=white, text=red},
  orangeop/.style={inner sep=1pt, fill=white, text=orange},
  dashline/.style={gray, dashed, line width=0.55pt}
]
\def\L{1.6}

\coordinate (c) at (1.0*\L,0);

\draw[dashline] (c) -- ++(0,\L);

\draw[dashline] (c) -- ++(-\L,0);

\node at (1.0*\L,1.2*\L) {$a$};
\fill (\L,\L) circle (2pt);

\node[op] at (1.1*\L,0.55*\L) {$Z^{-4}$};

\node[op] at (0.5*\L,0.0*\L) {$Z^{-4}$};
\end{tikzpicture}}~.
\label{eq:Z8f_edge_defect}
\end{equation}
The left endpoint operator $\Omega_a$ is purely bosonic, so $n_2=0$. The overlapping $X$ and $Z^{-4}$ contribute a conjugation phase $i$. {Eq.}~\eqref{eq:Else_Nayak_nu3} therefore gives
\begin{eqs}
    \exp\bigl(2\pi i\nu_3(g,h,k)\bigr) &= \exp\left(\frac{2\pi i}{16} g\bigl(h+k-[h+k]_4\bigr)\right) \\
    &= i^{\,g\left\lfloor \frac{h+k}{4}\right\rfloor} \, .
\end{eqs}
Using the canonical integer lift $a(g)=g$ of the $\mathbb Z_4$ coordinate, we can write the additive cocycle as
\begin{equation}
    \nu_3=\frac{1}{16}\,a\cup\delta a \, ,
\end{equation}
which generates the non-trivial $\mathbb{Z}_8^F$ SPT invariant realized by our model.
Indeed, the data
\begin{equation}
    \nu_3=\frac{1}{16}\,a\cup \delta a, \quad
    n_2= 0, \quad
    \omega_2 = \frac{\delta a}{4},
    \label{eq:Z8f_v3_n2}
\end{equation}
with $\omega_2$ understood modulo two, satisfy the supercohomology equation with nontrivial group extension:
\begin{eqs}
    \delta \nu_3 &= \frac{1}{2} n_2 \cup n_2 + \frac{1}{2} \omega_2 \cup n_2~, \\
    \delta n_2 &= 0~.
    \label{eq:general_supercohomology_equation}
\end{eqs}
We note that the choice of $\widetilde{U}_l^F$ is not unique. Different choices can lead to a nontrivial representative of $n_2$. For example, up to a fixed overall phase, we may choose the endpoint-dressed truncation
\begin{equation}
    \raisebox{-0.8em}{$\widetilde{U}'^{F}_l:=$}~~
    {\color{black}\begin{tikzpicture}[
    baseline=(current bounding box.center),
    scale=0.85,
    op/.style={inner sep=1pt, fill=white, text=blue},
    redop/.style={inner sep=1pt, fill=white, text=red},
    orangeop/.style={inner sep=1pt, fill=white, text=orange},
    wire/.style={black, line width=0.8pt},
    dashline/.style={gray, dashed, line width=0.55pt}
    ]
\def\L{1.2}

\begin{scope}[shift={(-2.3,0)}]
  \draw[wire] (-1*\L,\L) -- (1.9*\L,\L);
  \draw[dashline] (0,\L) -- (0,-0.15*\L);
  \draw[dashline] (\L,\L) -- (\L,-0.15*\L);
  \draw[dashline] (-1*\L,0) -- (1.9*\L,0);
  \node at (0,1.3*\L) {$a$};
  \fill (0,\L) circle (2pt);
  \node[op] at (0.5*\L,0) {$Z^{4}$};
  \node[op] at (-0.5*\L,0) {$Z^{4}$};
  \node[op] at (0*\L,0.5*\L) {$X^4$};
  \node[op] at (1*\L,0.5*\L) {$X^4$};
  \node[op] at (-0.5*\L,0.5*\L) {\color{red}$\Gamma$};
\end{scope}

\node at (0.6,0.5*\L) {$\cdots$};

\begin{scope}[shift={(2.0,0)}]
  \draw[wire] (-0.7*\L,\L) -- (2.0*\L,\L);
  \draw[dashline] (0,\L) -- (0,-0.15*\L);
  \draw[dashline] (\L,\L) -- (\L,-0.15*\L);
  \draw[dashline] (-0.7*\L,0) -- (2*\L,0);
  \node at (2*\L,1.3*\L) {$b$};
  \fill (2*\L,\L) circle (2pt);
  \node[op] at (0.5*\L,0) {$Z^{4}$};
  \node[op] at (-0.5*\L,0) {$Z^{4}$};
  \node[op] at (0*\L,0.5*\L) {$X^4$};
  \node[op] at (1*\L,0.5*\L) {$X^4$};
  \node[op] at (1.6*\L,0.53*\L) {\color{red}$\Gamma^{-1}$};
\end{scope}
\end{tikzpicture}}~.
\end{equation}
Compared with Eq.~\eqref{eq:truncated_Z8F_boundary_fermion_parity}, the only
difference is the insertion of the local operators ${\color{red}\Gamma}$ and
${\color{red}\Gamma^{-1}}$ at the two endpoints. This choice is also
non-anomalous, since ${\color{red}\Gamma_p^2=Z_p}$ commutes with the boundary
symmetry $\widetilde{U}((1,0))$ in
Eq.~\eqref{eq:truncated_Z8f_boundary_symmetry}. The endpoint
${\color{red}\Gamma}$ operators contribute a nontrivial $n_2$:
\begin{equation}
    n_2(g,h)=\left\lfloor \frac{g+h}{4}\right\rfloor,
    \quad \text{or equivalently} \quad
    n_2=\frac{\delta a}{4} \, .
\end{equation}
Nevertheless, this choice represents the same supercohomology class since the supercohomology data have the following equivalence relation~\cite{GW14, Barkeshli2022Classification}:
\begin{equation}
    n_2 \;\to\; n_2+\omega_2,
    \qquad
    \nu_3 \;\to\; \nu_3+\frac{1}{2}\omega_2\cup_1 n_2 \, .
\end{equation}
}

\subsection{\texorpdfstring{$\mathbb{Z}_4$}{Z4}-enriched toric code from gauging fermion parity in \texorpdfstring{$\ZZ^F_8$}{Z8F} SPT}
\label{sec:Z4enrichedTC_from_gauging_Z2F_in_ZF8SPT}

Instead of gauging the full symmetry group, we may choose to gauge only the central fermion parity subgroup $\mathbb{Z}_2^F\subset \mathbb{Z}_8^F$. The remaining global bosonic symmetry is then the quotient
\begin{equation}
    G_b=\mathbb{Z}_8^F/\mathbb{Z}_2^F\cong \mathbb{Z}_4 .
\end{equation}
We expect that gauging fermion parity in the $\ZZ^F_8$ SPT above leads to a bosonic shadow whose intrinsic topological order is the $\mathbb{Z}_2$ toric code. Since the $\ZZ_4$ symmetry group is untouched, the resulting phase is expected to be a $\mathbb{Z}_4$-symmetry-enriched $\mathbb{Z}_2$ toric code. Next, we explain how the resulting phase is indeed the $\ZZ_4$ enriched toric code.

After gauging $\ZZ_2^F$, the physical fermion line and the fermion-parity defect line become anyons in the $\ZZ_2$ toric code. We use the convention in which the physical fermion is identified with the toric-code fermion $f$, while the fermion-parity defect line becomes the toric-code flux $m$:
\begin{equation}
    \psi \longmapsto f,
    \qquad
    \widetilde U_l^F \longmapsto W_m(l).
\end{equation}
Therefore, Eq.~\eqref{eq:fermionic_defect} becomes, in the bosonic shadow,
\begin{eqs}
    &\widetilde{U}_l(r)\,\widetilde{U}_l(s)  
    \sim
    \Omega(r,s)\,
    \widetilde{U}_l(r+s)\,
    W_m(l)^{w_2(r,s)}~,
\end{eqs}
where $\widetilde U_l(g)$ is taken as the truncated boundary symmetry of $G_b\cong \ZZ_4$, identified with $\widetilde U_l((g,0))$ defined in Eq.~\eqref{eq:truncated_Z8f_boundary_symmetry}. For the representative chosen in Eq.~\eqref{eq:truncated_Z8F_boundary_fermion_parity}, the endpoint operator $\Omega(r,s)$ is bosonic. Thus $n_2=0$, and no additional toric code fermion string is produced by $\Omega(r,s)$. The only anyon-valued contribution comes from the fermion-parity defect line. Hence, the symmetry fractionalization cocycle of the resulting $\ZZ_4$-enriched toric code is
\begin{equation}
    t(r,s)
    =
    m^{w_2(r,s)}
    =
    m^{\left\lfloor (r+s)/4\right\rfloor}.
    \label{eq:Z4_SET_fractionalization_from_Z8F}
\end{equation}
For example,
\begin{equation}
    t(1,3)=m,
    \qquad
    t(2,2)=m .
\end{equation}
Equivalently, at the level of defect lines, applying the generator $g=1\in G_b=\ZZ_4$ four times leaves behind an $m$ string
\begin{equation}
    {\widetilde U_l(1)^4}\sim W_m(l).
\end{equation}
On a closed, contractible $l$, the Wilson operator $W_m(l)$ is topologically trivial and acts trivially in the ground-state sector. Thus, the symmetry operator realizes a $G_b=\mathbb{Z}_4$ global symmetry.

The same fractionalization data can be expressed in terms of the projective phases $\eta_x(r,s)$ for the toric-code anyons $x\in\{e,m,f\}$. Since the symmetry does not permute the anyons, the relation between the anyon-valued cocycle $t(r,s)$ and the phases $\eta_x(r,s)$ is
\begin{equation}
    \eta_x(r,s)
    =
    M_{x,t(r,s)}
    =
    M_{x,m}^{\,w_2(r,s)} .
\end{equation}
Using the toric-code braiding data
\begin{equation}
    M_{e,m}=-1,
    \qquad
    M_{m,m}=+1,
    \qquad
    M_{f,m}=-1,
\end{equation}
we obtain
\begin{equation}
    \eta_e(r,s)
    =\eta_f(r,s)=
    (-1)^{\left\lfloor (r+s)/4\right\rfloor},
    \quad
    \eta_m(r,s)
    =
    1.
\end{equation}
Therefore, $e$ and $f=e\times m$ carry the nontrivial $\ZZ_4$ fractionalization, while $m$ carries the trivial one. This characterizes the $\ZZ_4$-enriched $\ZZ_2$ toric code obtained by gauging fermion parity in the nontrivial $\ZZ_8^F$ fermionic SPT.

In addition, the nontrivial $\mathbb{Z}_8^F$ fSPT carries defect associator data inherited from the boundary calculation of
\begin{equation}
    \nu_3=\frac{1}{16}\,a\cup\delta a ,
    \label{eq:v3_Z4_SET_from_Z8F}
\end{equation}
where $a$ is the canonical $\mathbb{Z}_4$ one-cochain, lifted to integer values.  Thus the bosonic shadow of the nontrivial $\mathbb{Z}_8^F$ fSPT is a $\mathbb{Z}_4$-enriched $\mathbb{Z}_2$ toric code characterized by Eq.~\eqref{eq:Z4_SET_fractionalization_from_Z8F} and Eq.~\eqref{eq:v3_Z4_SET_from_Z8F}.

We now implement this gauging procedure on the lattice.  One could directly use the bosonization map described in Sec.~\ref{sec:Z2_enriched_TC}; however, the resulting lattice model would no longer be a Pauli Hamiltonian.  To keep a Pauli description, we realize the bosonization in an enlarged Pauli algebra.  More precisely, we introduce a $\ZZ_8$ Pauli group generated by $\{\textcolor{magenta}{\widetilde{X},\widetilde{Z}}\}$ and use the following local mapping
\begin{eqs}
i\times
\tikz[baseline=-0.6ex]{
  \def\L{\Ledge}
  \draw (0,0) -- (\L,0);
  \node[op, above] at (0.5*\L,0.25*\L) {$\color{red}\Gamma_{L(e)}$};
  \node[op, below] at (0.5*\L,-0.15*\L) {$\color{red}\Gamma'_{R(e)}$};
  \node[op] at (0.5*\L,0.15) {$e$};
}\quad
&
\tikz[baseline=-0.6ex]{
  \draw[<->] (0,0) -- (0.8,0);
}
\quad 
\raisebox{1.5em}{\tikz[baseline=-0.6ex]{
  \def\L{\Ledge}
  \draw (0,0) -- (\L,0);
  \draw (0,0) -- (0,-\L);
  \node[op, above] at (0.5*\L,-0.2) {$\color{magenta}\widetilde X^{-1}_e$};
  \node[op, left] at (0.15,-0.5*\L) {$\color{magenta}\widetilde Z^4$};
}}
\quad, \\[1.4em]
i\times
\tikz[baseline=-0.6ex]{
  \def\L{\Ledge}
  \draw (0,-0.5*\L) -- (0,0.5*\L);
  \node[op, left] at (-0.1*\L,0) {$\color{red}\Gamma_{L(e)}$};
  \node[op, right] at (0.2*\L,0) {$\color{red}\Gamma'_{R(e)}$};
  \node[op, right] at (0,0.28) {$e$};
}\quad
&
\tikz[baseline=-0.6ex]{
  \draw[<->] (0,0) -- (0.8,0);
}
\quad
\raisebox{-0.6cm}{\tikz[baseline=-0.6ex]{
  \def\L{\Ledge}
  \draw (0,0) -- (\L,0);
  \draw (\L,0) -- (\L,\L);
  \node[op, above] at (0.5*\L,-0.2) {$\color{magenta}\widetilde Z^4$};
  \node[op, right] at (0.82*\L,0.5*\L) {$\color{magenta}\widetilde X^{-1}_e$};
}}
\quad, \\[1.4em]
{\color{red} P_p=-i\Gamma_p\Gamma_p'}
\quad
&
\tikz[baseline=-0.6ex]{
  \draw[<->] (0,0) -- (0.8,0);
}
\quad
\raisebox{-0.6cm}{\tikz[baseline=-0.6ex]{
  \def\L{\Ledge}
  \draw (0,0) rectangle (\L,\L);
  \node[op, above] at (0.5*\L,0.87*\L) {$\color{magenta}\widetilde Z^4$};
  \node[op, below] at (0.5*\L,0.17*\L) {$\color{magenta}\widetilde Z^4$};
  \node[op, left] at (0.14*\L,0.5*\L) {$\color{magenta}\widetilde Z^4$};
  \node[op, right] at (0.86*\L,0.5*\L) {$\color{magenta}\widetilde Z^4$};
  \node[op] at (0.5*\L,0.5*\L) {$p$};
}}
\quad , \\[1.4em]
{\color{red} \Pi_p}
\quad
&
\tikz[baseline=-0.6ex]{
  \draw[<->] (0,0) -- (0.8,0);
}
\quad
\raisebox{-0.6cm}{\tikz[baseline=-0.6ex]{
  \def\L{\Ledge}
  \draw (0,0) rectangle (\L,\L);
  \node[op, above] at (0.6*\L,0.87*\L) {$\color{magenta}\widetilde Z^{-1}$};
  \node[op, below] at (0.5*\L,0.17*\L) {$\color{magenta}\widetilde Z$};
  \node[op, left] at (0.25*\L,0.5*\L) {$\color{magenta}\widetilde Z^{-1}$};
  \node[op, right] at (0.86*\L,0.5*\L) {$\color{magenta}\widetilde Z$};
  \node[op] at (0.5*\L,0.5*\L) {$\textcolor{red}{X_p}$};
}}
\label{eq:bosonization_Z8f}
\end{eqs}
Here $\textcolor{red}{X}$ belongs to the $\ZZ_4$ Pauli group used to define the fermionic clock operators.  Since we gauge only the $\ZZ_2^F$ subgroup, these $\ZZ_4$ Pauli operators remain as physical operators. Combining this fact with $\textcolor{red}{\Gamma^2}=\textcolor{red}{Z}$ and $\textcolor{red}{\Gamma'^2}=\textcolor{red}{Z^\dagger}$, we must impose the additional constraint
\begin{eqs}
\tikz[baseline=-0.6ex]{
  \def\L{\Ledge}
  \draw (0,0) -- (\L,0);
  \node[op, above] at (0.5*\L,0.4*\L) {$\color{red}Z$};
  \node[op, below] at (0.5*\L,-0.3*\L) {$\color{red}Z^\dagger$};
  \node[op,above] at (0.5*\L,-0.2) {$\textcolor{magenta}{\widetilde X^{2}_e}$};
}\quad,
\quad
\tikz[baseline=-0.6ex]{
  \def\L{\Ledge}
  \draw (0,-0.5*\L) -- (0,0.5*\L);
  \node[op, left] at (-0.25*\L,0) {$\color{red}Z$};
  \node[op, right] at (0.4*\L,0) {$\color{red}Z^\dagger$};
  \node[op, right] at (-0.1*\L,0) {$\textcolor{magenta}{\widetilde X^2_e}$};
}\quad
= \quad 1
\label{bosonization_Z8F_constraintX}
\end{eqs}
On the fermionic side, the physical Hilbert space is restricted to the total fermion-parity even sector
\begin{equation}
    P_F=\prod_p P_p = \prod_p(-i\Gamma_p\Gamma'_p) = 1 .
\end{equation}
{The bosonic side must also obey the following local Gauss-law constraint at each vertex}
\begin{equation}
    G_v :=
    \begin{tikzpicture}[baseline=-0.6ex, line width=0.5pt]
        \def\L{\Ledge}
        \draw (-\L,0) -- (\L,0);
        \draw (0,0) -- (0,-\L);
        \draw (0,0) -- (0,\L);
        \draw (0,\L) -- (\L,\L);
        \draw (\L,\L) -- (\L,0);
        \node[op] at (0,0) {$v$};
        \node[op] at (-0.5*\L,0) {${\color{magenta}\tilde{X}^\dagger}$};
        \node[op] at (0,-0.5*\L) {${\color{magenta}\tilde{X}^\dagger}$};
        \node[op] at (0,0.5*\L) {$\color{magenta}\tilde{X}\tilde{Z}^4$};
        \node[op] at (0.5*\L,0) {$\color{magenta}\tilde{X}\tilde{Z}^4$};
        \node[op] at (0.5*\L,\L) {$\color{magenta}\tilde{Z}^4$};
        \node[op] at (\L,0.5*\L) {$\color{magenta}\tilde{Z}^4$};
    \end{tikzpicture}
    =1 .
    \label{eq:bosonization_vertex_constraint_Z8F}
\end{equation}
Using the constraint in Eq.~\eqref{bosonization_Z8F_constraintX}, one checks that $G_v$ remains as an order-two operator.

Substituting the above mapping into the Hamiltonian of the $\ZZ_8^F$ SPT in Eq.~\eqref{eq:Z8F_SPT_stabilizers}, we obtain a commuting Pauli Hamiltonian for the $\ZZ_4$-enriched toric code
\begin{eqs}
{A'_v} &{=
\begin{tikzpicture}[baseline=-0.6ex, line width=0.5pt]
  \def\L{1.2}
  \draw (-\L,0) -- (\L,0);
  \draw (0,0) -- (0,-\L);
  \draw (0,0) -- (0,\L);
  \draw (0,\L) -- (\L,\L);
  \draw (\L,\L) -- (\L,0);
  \node[op] at (0,0) {$v$};
  \node[op] at (-0.5*\L,0) {$\textcolor{blue}{X^\dagger}$};
  \node[op] at (0,-0.5*\L) {$\textcolor{blue}{X^\dagger}$};
  \node[op] at (-0.15*\L,0.5*\L) {$\textcolor{blue}{XZ}\textcolor{magenta}{\widetilde Z}$};
  \node[op] at (0.6*\L,0) {$\textcolor{blue}{XZ^\dagger}\textcolor{magenta}{\widetilde Z^\dagger}$};
  \node[op] at (0.5*\L,\L) {$\textcolor{blue}{Z}\textcolor{magenta}{\widetilde Z}$};
  \node[op] at (1.1*\L,0.5*\L) {$\textcolor{blue}{Z^\dagger}\textcolor{magenta}{\widetilde Z^\dagger}$};
  \node[op] at (0.5*\L,0.5*\L) {$\textcolor{red}{X^\dagger}$};
\end{tikzpicture}
,}\quad
B'_p =
\raisebox{-0.6cm}{\begin{tikzpicture}[baseline=-0.6ex, line width=0.5pt]
  \def\L{1.2}
  \draw (0,0) rectangle (\L,\L);
  \node[op] at (0.58*\L,1.2*\L) {$\textcolor{magenta}{\tilde{Z}^4}\textcolor{blue}{Z^{-2}}$};
  \node[op] at (-0.1*\L,0.5*\L) {$\textcolor{magenta}{\tilde{Z}^4}\textcolor{blue}{Z^{-2}}$};
  \node[op] at (1.15*\L,0.5*\L) {$\textcolor{magenta}{\tilde{Z}^4}\textcolor{blue}{Z^2}$};
  \node[op] at (0.5*\L,-0.17*\L) {$\textcolor{magenta}{\tilde{Z}^4}\textcolor{blue}{Z^2}$};
  \node[op] at (0.5*\L,0.5*\L) {${p}$};
\end{tikzpicture}}
~~, \\
C_e &= ~~
\raisebox{-0.6cm}{\begin{tikzpicture}[baseline=-0.6ex, line width=0.5pt]
  \def\L{1.2}
  \draw (-\L,0) -- (0,0);
  \draw (0,0) -- (0,\L);
  \node[op] at (-0.5*\L,0.05*\L) {$\textcolor{blue}{Z^8}$};
  \node[op] at (0.15*\L,0.5*\L) {$\textcolor{blue}{X_e^{8}}$};
\end{tikzpicture}}
~~,\quad 
\raisebox{0.6cm}{\begin{tikzpicture}[baseline=-0.6ex, line width=0.5pt]
  \def\L{1.2}
  \draw (0,0) -- (\L,0);
  \draw (0,0) -- (0,-\L);
  \node[op] at (0.5*\L,0) {$\textcolor{blue}{X_e^8}$};
  \node[op] at (0,-0.5*\L) {$\textcolor{blue}{Z^8}$};
\end{tikzpicture}}
\quad ,\\
D'_e &= ~~
\raisebox{-0.6cm}{\begin{tikzpicture}[baseline=-0.6ex, line width=0.5pt]
  \def\L{1.2}
  \draw (-1.1*\L,0) -- (0,0);
  \draw (0,0) -- (0,\L);
  \node[op] at (-0.55*\L,0.07*\L) {$\textcolor{magenta}{\tilde{Z}^4}\textcolor{blue}{Z^{-2}}$};
  \node[op] at (0.*\L,0.5*\L) {$\textcolor{magenta}{\tilde{X}^{-1}}\textcolor{blue}{X_e^{4}}$};
\end{tikzpicture}}
~~,\qquad 
\raisebox{0.6cm}{\begin{tikzpicture}[baseline=-0.6ex, line width=0.5pt]
  \def\L{1.2}
  \draw (0,0) -- (\L,0);
  \draw (0,0) -- (0,-\L);
  \node[op] at (0.5*\L,0.25*\L) {$\textcolor{magenta}{\tilde{X}^{-1}}\textcolor{blue}{X_e^4}$};
  \node[op] at (0,-0.5*\L) {$\textcolor{magenta}{\tilde{Z}^4}\textcolor{blue}{Z^2}$};
\end{tikzpicture}}
~~ ,\\
E_e &= ~~
\tikz[baseline=-0.6ex]{
  \def\L{\Ledge}
  \draw (0,-0.5*\L) -- (0,0.5*\L);
  \node[op] at (-0.5*\L,0) {$\color{red}Z$};
  \node[op] at (0,0) {$\color{magenta}\widetilde X_e^2$};
  \node[op] at (0.5*\L,0) {$\color{red}Z^\dagger$};
}
~~,\quad
\tikz[baseline=-0.6ex]{
  \def\L{\Ledge}
  \draw (-0.5*\L,0) -- (0.5*\L,0);
  \node[op] at (0,0.5*\L) {$\color{red}Z$};
  \node[op] at (0,0) {$\color{magenta}\widetilde X_e^2$};
  \node[op] at (0,-0.5*\L) {$\color{red}Z^\dagger$};
}
\quad ,\\
F_v &= 
    \begin{tikzpicture}[baseline=-0.6ex, line width=0.5pt]
        \def\L{\Ledge}
        \draw (-\L,0) -- (\L,0);
        \draw (0,0) -- (0,-\L);
        \draw (0,0) -- (0,\L);
        \draw (0,\L) -- (\L,\L);
        \draw (\L,\L) -- (\L,0);
        \node[op] at (0,0) {$v$};
        \node[op] at (-0.5*\L,0) {${\color{magenta}\tilde{X}^\dagger}$};
        \node[op] at (0,-0.5*\L) {${\color{magenta}\tilde{X}^\dagger}$};
        \node[op] at (0,0.5*\L) {$\color{magenta}\tilde{X}\tilde{Z}^4$};
        \node[op] at (0.5*\L,0) {$\color{magenta}\tilde{X}\tilde{Z}^4$};
        \node[op] at (0.5*\L,\L) {$\color{magenta}\tilde{Z}^4$};
        \node[op] at (\L,0.5*\L) {$\color{magenta}\tilde{Z}^4$};
    \end{tikzpicture}~,
\label{eq:Z8F_SET_stabilizers}
\end{eqs}
where $\color{blue}X,Z$ denote Pauli operators on the $\mathbb{Z}_{16}$ qudits, $\color{red}X,Z$ denote Pauli operators on the $\mathbb{Z}_{4}$ qudits, and $\color{magenta}\widetilde X,\widetilde Z$ denote Pauli operators on the auxiliary $\mathbb{Z}_{8}$ qudits. The terms $E_e$ impose the compatibility condition between the physical $\mathbb{Z}_4$ variables and the enlarged Pauli algebra used for bosonization, while the terms $F_v$ impose the bosonized fermion-parity Gauss law. Together, these constraints define the physical Hilbert space obtained by gauging only the central $\mathbb{Z}_2^F$ subgroup. The remaining $\mathbb{Z}_4$ degrees of freedom are kept as physical symmetry variables, and the resulting commuting Pauli Hamiltonian realizes the $\mathbb{Z}_4$-enriched $\mathbb{Z}_2$ toric code described above.

One can verify that the intrinsic topological order of this theory is the same as that of the toric code, with anyon content $\mathcal A=\{1,e,m,f\}$.  Representative string operators for the anyon excitations are
{\begin{eqs}
    m&=
\raisebox{-0.6cm}{\tikz[baseline=-0.6ex]{
  \def\L{\Ledge}
  \draw (0,0) -- (\L,0);
  \draw (\L,0) -- (\L,\L);
  \draw[dashed, magenta, -{Latex[length=2mm]}] (0.5*\L,0.7*\L) -- (0.5*\L, -0.4*\L);
  \node[op, above] at (0.5*\L,-0.2) {$\color{magenta}\widetilde X_e$};
  \node[op, right] at (0.82*\L,0.5*\L) {$\color{blue}Z^2$};
}},\quad
\raisebox{-0.6cm}{\tikz[baseline=-0.6ex]{
  \def\L{\Ledge}
  \draw (0,0) -- (0,\L);
  \draw (0,\L) -- (\L,\L);
  \draw[dashed, magenta, -{Latex[length=2mm]}] (-0.4*\L,0.5*\L) -- (0.7*\L, 0.5*\L);
  \node[op, above] at (0,0.3*\L) {$\color{magenta}\widetilde X_e$};
  \node[op, right] at (0.3*\L,\L) {$\color{blue}Z^2$};
}},
\\
    e&=
\raisebox{0cm}{\tikz[baseline=-0.6ex]{
  \def\L{\Ledge}
  \draw (0,0) -- (\L,0);
  \draw[dashed, magenta, -{Latex[length=2mm]}] (0,-0.1*\L) -- (1.3*\L, -0.1*\L);
  \node[op] at (0.5*\L,0) {$\color{blue}{Z^2}\color{magenta}\widetilde Z^4_e$};
}},\quad
\raisebox{-0.6cm}{\tikz[baseline=-0.6ex]{
  \def\L{\Ledge}
  \draw (0,0) -- (0,\L);
  \draw[dashed, magenta, -{Latex[length=2mm]}] (0.1*\L,1*\L) -- (0.1*\L, -0.2*\L);
  \node[op, above] at (0,0.3*\L) {$\color{blue}{Z^2}\color{magenta}\widetilde Z^4_e$};
}},\\
f&=\raisebox{-0.6cm}{\tikz[baseline=-0.6ex]{
  \def\L{\Ledge}
  \draw (0,0) -- (\L,0);
  \draw (\L,0) -- (\L,\L);
  \draw[dashed, magenta, -{Latex[length=2mm]}] (0.5*\L,0.7*\L) -- (0.5*\L, -0.4*\L);
  \node[op, above] at (0.5*\L,-0.2) {$\color{magenta}\widetilde X_e$};
  \node[op, right] at (0.82*\L,0.5*\L) {$\color{magenta}\widetilde{Z}^4$};
}},\quad
\raisebox{-0.6cm}{\tikz[baseline=-0.6ex]{
  \def\L{\Ledge}
  \draw (0,0) -- (0,\L);
  \draw (0,\L) -- (\L,\L);
  \draw[dashed, magenta, -{Latex[length=2mm]}] (-0.4*\L,0.5*\L) -- (0.7*\L, 0.5*\L);
  \node[op, above] at (0,0.3*\L) {$\color{magenta}\widetilde X_e$};
  \node[op, right] at (0.3*\L,\L) {$\color{magenta}\widetilde{Z}^4$};
}}.
\end{eqs}}\noindent
The system also has a $\ZZ_4$ global symmetry generated by
\begin{align}
    U(g)=\left(\prod_p \textcolor{red}{X_p}\right)^g ~, \quad g\in \ZZ_4 =\{0, 1, 2, 3\}~.
\end{align}
As in Sec.~\ref{sec:Z2_enriched_TC}, we extract the symmetry fractionalization data $t(g,h)\in \mathcal A$ from the effective boundary symmetry operator
\begin{eqs}
&\vcenter{\hbox{
\begin{tikzpicture}[scale=1.2,baseline=5pt]
\foreach \x in {0,1,2,3,4}{
    \draw[dashed] (\x,0) -- (\x,4);
}
\foreach \y in {0,1,2,3}{
    \draw[dashed] (-0.3,\y) -- (4.3,\y);
}
\draw[line width=0.8pt] (-0.3,4) -- (4.3,4);
\foreach \x in {0.5,1.5,2.5,3.5}{
    \foreach \y in {0.5,1.5,2.5,3.5}{
        \node at (\x,\y) {$\textcolor{red}{X}$};
    }
}
\end{tikzpicture}
}} \\[1em]
\cong\quad
&\vcenter{\hbox{
\begin{tikzpicture}[scale=1.2,baseline=5pt]
\foreach \x in {0,1,2,3,4}{
    \draw[dashed] (\x,0) -- (\x,-4);
}
\foreach \y in {-1,-2,-3,-4}{
    \draw[dashed] (-0.3,\y) -- (4.3,\y);
}
\draw[line width=0.8pt] (-0.3,0) -- (4.3,0);
\foreach \x in {0,1,2,3,4}{
    \node[op] at (\x,-0.5) {$\textcolor{blue}{X}$};
}
\foreach \x in {0.5,1.5,2.5,3.5}{
    \node[op] at (\x,0) {$\textcolor{blue}{Z}\textcolor{magenta}{\widetilde Z}$};
}
\end{tikzpicture}
}} \,
:= \widetilde{U}(1)~.
\label{eq:symmetry_Z4}
\end{eqs}
We then truncate the $\ZZ_4$ symmetry to an interval $l=[a,b]$ along the boundary and denote the resulting operator by $\widetilde U_l(g)=\widetilde U_l(1)^g,g\in\{0,1,2,3\}$. We choose $\widetilde U_l(1)$ to be
{ 
\begin{eqs}
&\widetilde U_l(1) = \\
&\quad
\begin{tikzpicture}[
  baseline=(current bounding box.center),
  scale=0.95,
  op/.style={inner sep=1pt, fill=white, text=blue},
  wire/.style={black, line width=0.8pt},
  dashline/.style={gray, dashed, line width=0.55pt}
]
\def\L{1.2}

\begin{scope}[shift={(-2.3,0)}]
  \draw[wire] (-1*\L,\L) -- (1.9*\L,\L);
  \draw[dashline] (-1*\L,\L) -- (-1*\L,-0.15*\L);
  \draw[dashline] (0,\L) -- (0,-0.15*\L);
  \draw[dashline] (\L,\L) -- (\L,-0.15*\L);
  \draw[dashline] (-1*\L,0) -- (1.9*\L,0);

  \node at (0,1.3*\L) {$a$};
  \fill (0*\L,\L) circle (2pt);
  \node[op] at (0.5*\L,\L) {$Z\textcolor{magenta}{\widetilde Z}$};
  \node[op] at (1.5*\L,\L) {$Z\textcolor{magenta}{\widetilde Z}$};
  \node[op] at (0,0.5*\L) {$X$};
  \node[op] at (\L,0.5*\L) {$X$};
\end{scope}

\node at (0.3,0.5*\L) {$\cdots$};

\begin{scope}[shift={(1.4,0)}]
  \draw[wire] (-0.6*\L,\L) -- (2.0*\L,\L);
  \draw[dashline] (0,\L) -- (0,-0.15*\L);
  \draw[dashline] (\L,\L) -- (\L,-0.15*\L);
  \draw[dashline] (-0.6*\L,0) -- (2*\L,0);

  \node at (2*\L,1.3*\L) {$b$};
  \fill (2*\L,\L) circle (2pt);
  \node[op] at (-0.5*\L,\L) {$Z\textcolor{magenta}{\widetilde Z}$};
  \node[op] at (0.5*\L,\L) {$Z\textcolor{magenta}{\widetilde Z}$};
  \node[op] at (1.5*\L,\L) {$Z\textcolor{magenta}{\widetilde Z}$};
  \node[op] at (0,0.5*\L) {$X$};
  \node[op] at (\L,0.5*\L) {$X$};
\end{scope}
\end{tikzpicture}
~.
\end{eqs}
}
The relevant endpoint operator is $\Omega(1,3)=\Omega(2,2)=\widetilde U_l(4)$.  Up to multiplication by bulk stabilizers, it can be written as
\begin{eqs}
    &\Omega(1,3)=\Omega(2,2)
\\
\raisebox{-0.5em}{$=$}\;&\quad\quad~
\begin{tikzpicture}[
  baseline=(current bounding box.center),
  scale=0.95,
  op/.style={inner sep=1pt, fill=white, text=blue},
  wire/.style={black, line width=0.8pt},
  dashline/.style={gray, dashed, line width=0.55pt}
]
\def\L{1.2}

\begin{scope}[shift={(-2.3,0)}]
  \draw[wire] (-1*\L,\L) -- (1.9*\L,\L);
  \draw[dashline] (-1*\L,\L) -- (-1*\L,-0.15*\L);
  \draw[dashline] (0,\L) -- (0,-0.15*\L);
  \draw[dashline] (\L,\L) -- (\L,-0.15*\L);
  \draw[dashline] (-1*\L,0) -- (1.9*\L,0);

  \node at (0,1.3*\L) {$a$};
  \fill (0,\L) circle (2pt);
  \node[op] at (0.5*\L,\L) {$Z^4\textcolor{magenta}{\widetilde Z^4}$};
  \node[op] at (1.5*\L,\L) {$Z^4\textcolor{magenta}{\widetilde Z^4}$};
  \node[op] at (0,0.5*\L) {$X^4$};
  \node[op] at (\L,0.5*\L) {$X^4$};
\end{scope}

\node at (0.3,0.5*\L) {$\cdots$};

\begin{scope}[shift={(1.4,0)}]
  \draw[wire] (-0.6*\L,\L) -- (2.0*\L,\L);
  \draw[dashline] (0,\L) -- (0,-0.15*\L);
  \draw[dashline] (\L,\L) -- (\L,-0.15*\L);
  \draw[dashline] (-0.6*\L,0) -- (2*\L,0);

  \node at (2*\L,1.3*\L) {$b$};
  \fill (2*\L,\L) circle (2pt);
  \node[op] at (-0.5*\L,\L) {$Z^4\textcolor{magenta}{\widetilde Z^4}$};
  \node[op] at (0.5*\L,\L) {$Z^4\textcolor{magenta}{\widetilde Z^4}$};
  \node[op] at (1.5*\L,\L) {$Z^4\textcolor{magenta}{\widetilde Z^4}$};
  \node[op] at (0,0.5*\L) {$X^4$};
  \node[op] at (\L,0.5*\L) {$X^4$};
\end{scope}
\end{tikzpicture}
\\
\raisebox{-1em}{$\sim$}\;&
\begin{tikzpicture}[
  baseline=(current bounding box.center),
  >=Latex,
  scale=0.95,
  wire/.style={black, line width=0.8pt},
  dashline/.style={gray, dashed, line width=0.55pt},
  blueop/.style={inner sep=1pt, fill=white, text=blue},
  pinkop/.style={inner sep=1pt, fill=white, text=magenta},
  orangeop/.style={inner sep=1pt, fill=white, text=orange},
  redarrow/.style={red, -{Latex[length=2mm]}, line width=0.6pt}
]
\def\L{1.2}

\begin{scope}[shift={(-2.3,0)}]
  \draw[wire] (-1*\L,\L) -- (1.9*\L,\L);
  \draw[dashline] (-\L,\L) -- (-\L,-0.2*\L);
  \draw[dashline] (0,\L) -- (0,-0.2*\L);
  \draw[dashline] (\L,\L) -- (\L,-0.2*\L);
  \draw[dashline] (-1*\L,0) -- (1.9*\L,0);

  \node at (0,1.3*\L) {$a$};
  \fill (0*\L,\L) circle (2pt);
  \coordinate (Aend) at (0.25*\L,1.3*\L);

  \node[op] at (0.5*\L,1*\L) {$\color{blue} Z^2$};
  \node[op] at (1.5*\L,1*\L) {$\color{blue} Z^2$};

  \node[op] at (-0.5*\L,0.5*\L) {$\color{red} Z^\dagger$};
  \node[op] at (0*\L,0.5*\L) {${\color{magenta}\widetilde{X}}$};
  \node[pinkop] at (1*\L,0.55*\L) {$\widetilde{X}$};

  \node[op]   at (-0.56*\L,1.1*\L) {${\color{magenta}\widetilde{Z}^4}{\color{blue}Z^{-2}}$};
  \node[op]   at (-1.2*\L,0.55*\L) {${\color{magenta}\widetilde{Z}^4}{\color{blue}Z^{-2}}$};
\end{scope}

\node at (0.3,0.5*\L) {$\cdots$};

\begin{scope}[shift={(1.4,0)}]
  \draw[wire] (-0.6*\L,\L) -- (2.0*\L,\L);
  \draw[dashline] (0,\L) -- (0,-0.2*\L);
  \draw[dashline] (\L,\L) -- (\L,-0.2*\L);
  \draw[dashline] (-0.6*\L,0) -- (1.9*\L,0);

  \node at (2*\L,1.3*\L) {$b$};
  \fill (2*\L,\L) circle (2pt);
  \coordinate (Bend) at (1.75*\L,1.3*\L);

  \node[op] at (-0.5*\L,1*\L) {$\color{blue}Z^2$};
  \node[op] at (0.5*\L,1*\L) {$\color{blue}Z^2$};
  \node[op] at (1.5*\L,1*\L) {$\color{blue}Z^4\color{magenta}\widetilde Z^4$};

  \node[pinkop]   at (0,0.55*\L) {$\widetilde{X}$};
  \node[op]   at (\L,0.55*\L) {${\color{magenta}\widetilde X \widetilde Z^4}{\color{blue}Z^2}$};
  \node[op] at (1.7*\L,0.5*\L) {$\color{red} Z$};
\end{scope}

\draw[redarrow] (Aend) -- (Bend)
  node[midway, above=2pt, fill=white, inner sep=1pt, text=red] {$m$};

\end{tikzpicture}
~.
\end{eqs}
Thus, upon removing the local endpoint operators near $a$ and $b$, the fusion of the truncated $\ZZ_4$ symmetry operators leaves behind the worldline of the boson $m$.  In terms of symmetry fractionalization, this yields
\begin{eqs}
    t(1,3)=t(2,2)=m. 
\end{eqs}

\subsection{\texorpdfstring{$\ZZ_4$}{Z4}-enriched toric code from \texorpdfstring{$U(1)_8 \times U(1)_{-8}$}{U(1)8 x U(1)-8}}
\label{sec:Z4entrichedTC_from_U88}

We now give an equivalent realization of the same
$\ZZ_4$-enriched toric code from a different microscopic route. Instead of starting from the $\ZZ_8^F$ fSPT and gauging fermion parity, we directly ungauge the charge subgroup generated by the boson represented by the vector
\begin{equation}
    {\begin{pmatrix}4\\-2\end{pmatrix}_{\rm std}}
\end{equation}
in the Abelian topological order of
\begin{equation}
    K=\begin{pmatrix}
        0&8\\
        8&-8
    \end{pmatrix}_{\rm std}.
\end{equation}
We note that the above $K$-matrix is equivalent to the {$U(1)_{8} \times U(1)_{-8}$} by the following $GL_2(\ZZ)$ transformation
\begin{equation}
    \begin{pmatrix}
        1&0\\
        1&1
    \end{pmatrix}^T
    \begin{pmatrix}
        0&8\\
        8&-8
    \end{pmatrix}
    \begin{pmatrix}
        1&0\\
        1&1
    \end{pmatrix}
    =
    \begin{pmatrix}
        8&0\\
        0&-8
    \end{pmatrix} \, ,
\end{equation}
{The same change of basis sends the bosonic charge $(4,-2)^T_{\rm std}$ to $(2,-2)^T_{\rm diag}$. Under the identification in Eq.~\eqref{eq:Z8_twisted_gauge_theory_identification}, it corresponds to $(em)^4(e^2)^{-2}=m^4$ in the parent $\ZZ_{16}$ toric code. Thus this route first condenses $e^8m^8$ and then ungauges the charge subgroup generated by $m^4$. Following the lattice construction in Sec.~\ref{sec:Z8F_SPT}, the stabilizer generators of the resulting SET are}
\begin{eqs}
A''_v &=
\begin{tikzpicture}[baseline=-0.6ex, line width=0.5pt]
  \def\L{1.2}
  \draw (-\L,0) -- (\L,0);
  \draw (0,0) -- (0,-\L);
  \draw (0,0) -- (0,\L);
  \draw (0,\L) -- (\L,\L);
  \draw (\L,\L) -- (\L,0);
  \node[op] at (0,0) {$v$};
  \node[op] at (-0.5*\L,0) {$\textcolor{blue}{X^\dagger}$};
  \node[op] at (0,-0.5*\L) {$\textcolor{blue}{X^\dagger}$};
  \node[op] at (-0.05*\L,0.5*\L) {$\textcolor{blue}{XZ}$};
  \node[op] at (0.5*\L,0) {$\textcolor{blue}{XZ^\dagger}$};
  \node[op] at (0.5*\L,\L) {$\textcolor{blue}{Z}$};
  \node[op] at (\L,0.5*\L) {$\textcolor{blue}{Z^\dagger}$};
  \node[op] at (0.5*\L,0.5*\L) {$\textcolor{red}{X^\dagger}$};
\end{tikzpicture}
,\quad
B''_p =
\raisebox{-0.6cm}{\begin{tikzpicture}[baseline=-0.6ex, line width=0.5pt]
  \def\L{1.2}
  \draw (0,0) rectangle (\L,\L);
  \node[op] at (0.58*\L,1*\L) {$\textcolor{blue}{Z^{-2}}$};
  \node[op] at (-0.1*\L,0.5*\L) {$\textcolor{blue}{Z^{-2}}$};
  \node[op] at (1.15*\L,0.5*\L) {$\textcolor{blue}{Z^2}$};
  \node[op] at (0.5*\L,0) {$\textcolor{blue}{Z^2}$};
  \node[op] at (0.5*\L,0.5*\L) {$\textcolor{red}{X^2_p}$};
\end{tikzpicture}}
, \\
C_e &= ~~
\raisebox{-0.6cm}{\begin{tikzpicture}[baseline=-0.6ex, line width=0.5pt]
  \def\L{1.2}
  \draw (-\L,0) -- (0,0);
  \draw (0,0) -- (0,\L);
  \node[op] at (-0.5*\L,0.05*\L) {$\textcolor{blue}{Z^8}$};
  \node[op] at (0.15*\L,0.5*\L) {$\textcolor{blue}{X_e^{8}}$};
\end{tikzpicture}}
,\qquad 
\raisebox{0.6cm}{\begin{tikzpicture}[baseline=-0.6ex, line width=0.5pt]
  \def\L{1.2}
  \draw (0,0) -- (\L,0);
  \draw (0,0) -- (0,-\L);
  \node[op] at (0.5*\L,0) {$\textcolor{blue}{X_e^8}$};
  \node[op] at (0,-0.5*\L) {$\textcolor{blue}{Z^8}$};
\end{tikzpicture}}
\quad ,\\
E'_e &= ~~
\tikz[baseline=-0.6ex]{
  \def\L{\Ledge}
  \draw (0,-0.5*\L) -- (0,0.5*\L);
  \node[op] at (-0.5*\L,0) {$\color{red}Z$};
  \node[op] at (0,0) {$\color{blue} X_e^4$};
  \node[op] at (0.5*\L,0) {$\color{red}Z^\dagger$};
}
\;,\quad
\tikz[baseline=-0.6ex]{
  \def\L{\Ledge}
  \draw (-0.5*\L,0) -- (0.5*\L,0);
  \node[op] at (0,0.5*\L) {$\color{red}Z$};
  \node[op] at (0,0) {$\color{blue}X_e^4$};
  \node[op] at (0,-0.5*\L) {$\color{red}Z^\dagger$};
}
\quad ,
\label{eq:Z4SET_from_U1_8_stabilizers}
\end{eqs}
where Pauli $\textcolor{blue}{X}$ and $\textcolor{blue}{Z}$ act on the $\ZZ_{16}$ edge qudits, while Pauli $\textcolor{red}{X}$ and $\textcolor{red}{Z}$ act on the $\ZZ_4$ plaquette qudits. The stabilizers $C_e$ implement the first condensation, namely the order-two boson $e^8m^8$ of the parent $\ZZ_{16}$ toric code. This condensation gives the twisted $\ZZ_8$ gauge theory, equivalently $U(1)_8\times U(1)_{-8}$. The stabilizers $E'_e$ then implement ungauging of the charge subgroup generated by the boson {$(2,-2)^T_{\rm diag}$}, or equivalently $m^4$ in the parent $\ZZ_{16}$ toric code. The endpoint $\ZZ_4$ operators in $E'_e$ dress its hopping operators so that the dual $\ZZ_4$ zero-form symmetry remains manifest. The stabilizers $A''_v$ and $B''_p$ are the corresponding centralizer terms that survive after imposing these constraints. Modulo the bosons identified by these constraints, the remaining deconfined anyons have the fusion and braiding data of the ordinary toric code, with anyon content $\{1,e,m,f=em\}$. A convenient choice of representative string operators is
\begin{eqs}
    e&=
\raisebox{-0.6cm}{\tikz[baseline=-0.6ex]{
  \def\L{\Ledge}
  \draw (0,0) -- (\L,0);
  \draw[dashed, magenta, -{Latex[length=2mm]}] (0.5*\L,0.7*\L) -- (0.5*\L, -0.4*\L);
  \node[op, above] at (0.5*\L,-0.2) {$\color{blue}X^2_e$};
}},\quad
\raisebox{-0.6cm}{\tikz[baseline=-0.6ex]{
  \def\L{\Ledge}
  \draw (0,0) -- (0,\L);
  \draw[dashed, magenta, -{Latex[length=2mm]}] (-0.4*\L,0.5*\L) -- (0.7*\L, 0.5*\L);
  \node[op, above] at (0,0.3*\L) {$\color{blue}X^2_e$};
}},
\\
    m&=
\raisebox{0cm}{\tikz[baseline=-0.6ex]{
  \def\L{\Ledge}
  \draw (0,0) -- (\L,0);
  \draw[dashed, magenta, -{Latex[length=2mm]}] (0,-0.1*\L) -- (1.3*\L, -0.1*\L);
  \node[op] at (0.5*\L,0) {$\color{blue}{Z^4_e}$};
}},\quad
\raisebox{-0.6cm}{\tikz[baseline=-0.6ex]{
  \def\L{\Ledge}
  \draw (0,0) -- (0,\L);
  \draw[dashed, magenta, -{Latex[length=2mm]}] (0.1*\L,1*\L) -- (0.1*\L, -0.2*\L);
  \node[op, above] at (0,0.3*\L) {$\color{blue}{Z^4_e}$};
}},\\
f&=\raisebox{-0.6cm}{\tikz[baseline=-0.6ex]{
  \def\L{\Ledge}
  \draw (0,0) -- (\L,0);
  \draw (\L,0) -- (\L,\L);
  \draw[dashed, magenta, -{Latex[length=2mm]}] (0.5*\L,0.7*\L) -- (0.5*\L, -0.4*\L);
  \node[op, above] at (0.5*\L,-0.2) {$\color{blue}X^2_e$};
  \node[op, right] at (0.82*\L,0.5*\L) {$\color{blue}Z^4$};
}},\quad
\raisebox{-0.6cm}{\tikz[baseline=-0.6ex]{
  \def\L{\Ledge}
  \draw (0,0) -- (0,\L);
  \draw (0,\L) -- (\L,\L);
  \draw[dashed, magenta, -{Latex[length=2mm]}] (-0.4*\L,0.5*\L) -- (0.7*\L, 0.5*\L);
  \node[op, above] at (0,0.3*\L) {$\color{blue}X^2_e$};
  \node[op, right] at (0.3*\L,\L) {$\color{blue}Z^4$};
}}.
\end{eqs}

We next examine the $\ZZ_4$ symmetry fractionalization.  The onsite symmetry is again generated by
\begin{equation}
    U(g)=\left(\prod_p {\color{red}X_p} \right)^g ~, \quad g\in \ZZ_4 =\{0, 1, 2, 3\}~.
\end{equation}
Because the boundary action of this symmetry is unchanged by these condensation and ungauging steps, the effective boundary symmetry operator $\widetilde U(g)$ is the same as the one in Eq.~\eqref{eq:symmetry_Z4}. Therefore, the fusion rule of truncated symmetry operators can be computed in exactly the same way. For an interval $l=[a,b]$ along the boundary, the nontrivial endpoint operator is
\begin{eqs}
    &\Omega(1,3)=\Omega(2,2) \\
\raisebox{-0.5em}{$=$}\;&\quad~~
\begin{tikzpicture}[
  baseline=(current bounding box.center),
  scale=0.95,
  op/.style={inner sep=1pt, fill=white, text=blue},
  wire/.style={black, line width=0.8pt},
  dashline/.style={gray, dashed, line width=0.55pt}
]
\def\L{1.2}

\begin{scope}[shift={(-2.3,0)}]
  \draw[wire] (-1*\L,\L) -- (1.9*\L,\L);
  \draw[dashline] (-1*\L,\L) -- (-1*\L,-0.15*\L);
  \draw[dashline] (0,\L) -- (0,-0.15*\L);
  \draw[dashline] (\L,\L) -- (\L,-0.15*\L);
  \draw[dashline] (-1*\L,0) -- (1.9*\L,0);

  \node at (0,1.3*\L) {$a$};
  \fill (0*\L,\L) circle (2pt);
  \node[op] at (0.5*\L,\L) {$Z^4$};
  \node[op] at (1.5*\L,\L) {$Z^4$};
  \node[op] at (0,0.5*\L) {$X^4$};
  \node[op] at (\L,0.5*\L) {$X^4$};
\end{scope}

\node at (0.3,0.5*\L) {$\cdots$};

\begin{scope}[shift={(1.4,0)}]
  \draw[wire] (-0.6*\L,\L) -- (2.0*\L,\L);
  \draw[dashline] (0,\L) -- (0,-0.15*\L);
  \draw[dashline] (\L,\L) -- (\L,-0.15*\L);
  \draw[dashline] (-0.6*\L,0) -- (2*\L,0);

  \node at (2*\L,1.3*\L) {$b$};
  \fill (2*\L,\L) circle (2pt);
  \node[op] at (-0.5*\L,\L) {$Z^4$};
  \node[op] at (0.5*\L,\L) {$Z^4$};
  \node[op] at (1.5*\L,\L) {$Z^4$};
  \node[op] at (0,0.5*\L) {$X^4$};
  \node[op] at (\L,0.5*\L) {$X^4$};
\end{scope}
\end{tikzpicture}
\\
\raisebox{-1em}{$\sim$}&\quad~~
\begin{tikzpicture}[
  baseline=(current bounding box.center),
  >=Latex,
  scale=0.95,
  wire/.style={black, line width=0.8pt},
  dashline/.style={gray, dashed, line width=0.55pt},
  blueop/.style={inner sep=1pt, fill=white, text=blue},
  pinkop/.style={inner sep=1pt, fill=white, text=magenta},
  orangeop/.style={inner sep=1pt, fill=white, text=orange},
  redarrow/.style={red, -{Latex[length=2mm]}, line width=0.6pt}
]
\def\L{1.2}

\begin{scope}[shift={(-2.3,0)}]
  \draw[wire] (-1*\L,\L) -- (1.9*\L,\L);
  \draw[dashline] (-\L,\L) -- (-\L,-0.2*\L);
  \draw[dashline] (0,\L) -- (0,-0.2*\L);
  \draw[dashline] (\L,\L) -- (\L,-0.2*\L);
  \draw[dashline] (-1*\L,0) -- (1.9*\L,0);

  \node at (0,1.3*\L) {$a$};
  \fill (0*\L,\L) circle (2pt);
  \coordinate (Aend) at (0.25*\L,1.3*\L);

  \node[op] at (0.5*\L,1*\L) {$\color{blue} Z^4$};
  \node[op] at (1.5*\L,1*\L) {$\color{blue} Z^4$};
  \node[op] at (-0.5*\L,0.5*\L) {$\color{red} Z^\dagger$};
\end{scope}

\node at (0.3,0.5*\L) {$\cdots$};

\begin{scope}[shift={(1.4,0)}]
  \draw[wire] (-0.6*\L,\L) -- (2.0*\L,\L);
  \draw[dashline] (0,\L) -- (0,-0.2*\L);
  \draw[dashline] (\L,\L) -- (\L,-0.2*\L);
  \draw[dashline] (-0.6*\L,0) -- (1.9*\L,0);

  \node at (2*\L,1.3*\L) {$b$};
  \fill (2*\L,\L) circle (2pt);
  \coordinate (Bend) at (1.75*\L,1.3*\L);

  \node[op] at (-0.5*\L,1*\L) {$\color{blue}Z^4$};
  \node[op] at (0.5*\L,1*\L) {$\color{blue}Z^4$};
  \node[op] at (1.5*\L,1*\L) {$\color{blue}Z^4$};

  \node[op] at (1.5*\L,0.5*\L) {$\color{red} Z$};
\end{scope}

\draw[redarrow] (Aend) -- (Bend)
  node[midway, above=2pt, fill=white, inner sep=1pt, text=red] {$m$};
\end{tikzpicture}
\; .
\end{eqs}
Thus, after removing local endpoint operators near $a$ and $b$, the fusion of truncated symmetry operators leaves behind an $m$ worldline. Equivalently, the anyon-valued fractionalization cocycle is
\begin{equation}
    t(r,s)=m^{\left\lfloor (r+s)/4\right\rfloor},
    \qquad r,s\in\{0,1,2,3\}.
\end{equation}
and
\begin{equation}
    t(1,3)=t(2,2)=m .
\end{equation}

This shows that the present ungauging construction realizes the same $\ZZ_4$ SET phase as the bosonic shadow obtained above from the $\ZZ_8^F$ fSPT for the following reasons. First, the intrinsic topological order is the toric code. Second, the $\ZZ_4$ symmetry does not permute the anyons in either realization. Third, under the natural anyon identification induced by the ungauged theory, the fractionalization class agrees with the previous result
\begin{equation}
    t(r,s)=m^{\left\lfloor (r+s)/4\right\rfloor}.
\end{equation}
{Consequently, the projective symmetry quantum numbers for the generator $g=1$ are also identical}
\begin{equation}
    {U_e(1)^4=-1,\qquad U_m(1)^4=+1,\qquad U_f(1)^4=-1} .
\end{equation}
Moreover, because the effective boundary symmetry operator $\widetilde U(g)$ is the same as in Eq.~\eqref{eq:symmetry_Z4}, the defect associator data are unchanged
\begin{equation}
    \nu_3=\frac{1}{16}\,a\cup\delta a .
\end{equation}
The two Hamiltonians therefore share the same {anyon content}, symmetry fractionalization, and defect-associator data, so they belong to the same $\ZZ_4$-enriched toric-code phase.

\section{General stabilizer construction of Abelian fermionic topological phases}
\label{sec:general_constructions}

{
This section develops general Majorana-Pauli stabilizer constructions for Abelian fermionic topological orders and fermionic SPT phases. Their scope is summarized in Fig.~\ref{fig:general_construction_flow}. Starting from Abelian bosonic topological orders realized by the Pauli stabilizer models~\cite{Ellison22}, we bind an emergent fermion to a local physical fermion and condense the resulting boson. This procedure yields stabilizer realizations of all $(2+1)$-dimensional Abelian fermionic topological orders admitting a fully gapped boundary.

Section~\ref{sec:general_fermionic_topological_orders} formulates this condensation in the $K$-matrix formalism and implements it on the lattice. We then turn to fermionic SPT phases. Section~\ref{sec:split_spt_scope} identifies the split fSPTs obtained by ungauging the electric charges, while Sec.~\ref{sec:nonsplit_spt_scope} constructs non-split fSPTs directly from their gauged bosonic shadows. Section~\ref{sec:general_fermionic_SPTs} presents an explicit family of lattice models with cyclic symmetry. Throughout the SPT discussion, we focus on finite Abelian unitary onsite symmetries and restrict attention to the supercohomology sector without Kitaev-chain decorations~\cite{WangGu2018Toward, WangGu2020Construction}.
}

\begin{figure*}[t]
\begingroup

\centering

\pgfmathsetlengthmacro{\flowwidth}{0.98\linewidth}
\pgfmathsetlengthmacro{\topboxwidth}{0.24*\flowwidth}
\pgfmathsetlengthmacro{\bottomboxwidth}{0.32*\flowwidth}
\pgfmathsetlengthmacro{\toplabelwidth}{0.12*\flowwidth}
\pgfmathsetlengthmacro{\bottomlabelwidth}{0.27*\flowwidth}

\begin{tikzpicture}[
    x=\flowwidth,
    y=1cm,
    every node/.style={text=black},
    phasebox/.style={
        draw=black,
        fill=white,
        line width=0.5pt,
        align=center,
        minimum height=1.85cm,
        inner xsep=4pt,
        inner ysep=5pt,
        outer sep=0pt,
        font=\small
    },
    topbox/.style={
        phasebox,
        minimum width=\topboxwidth,
        text width=\dimexpr\topboxwidth-8pt\relax
    },
    bottombox/.style={
        phasebox,
        minimum height=1.65cm,
        minimum width=\bottomboxwidth,
        text width=\dimexpr\bottomboxwidth-8pt\relax
    },
    paneltitle/.style={
        anchor=west,
        inner sep=0pt,
        outer sep=0pt,
        font=\small
    },
    actionlabel/.style={
        draw=none,
        fill=none,
        align=center,
        inner sep=0pt,
        outer sep=0pt,
        font=\footnotesize
    },
    flowarrow/.style={
        draw=black,
        -{Latex[length=2mm]},
        line width=0.6pt,
        shorten <=0.5pt,
        shorten >=0.5pt
    }
]

\node[paneltitle] at (0,1.45)
    {(a) Split symmetry: $G_f=\ZZ_2^f\times G_b$};

\node[topbox] (a-source) at (0.12,0)
    {Abelian bosonic topological order\\
     with a gapped boundary\\
     and an order-$2$\\
     emergent fermion $f$};

\node[topbox] (a-middle) at (0.50,0)
    {Abelian fermionic\\
     topological order\\
     with a gapped boundary};

\node[topbox] (a-result) at (0.88,0)
    {Split fermionic SPT\\
     type-I $n_2$\\
     type-I or type-II $\nu_3$};

\draw[flowarrow] (a-source.east) --
    node[actionlabel, midway, above=4pt,
         text width=\toplabelwidth]
    {Condense $f\psi$}
    (a-middle.west);

\draw[flowarrow] (a-middle.east) --
    node[actionlabel, midway, above=4pt,
         text width=\toplabelwidth]
    {Ungauge\\electric charges}
    (a-result.west);

\node[paneltitle] at (0,-1.55)
    {(b) Non-split symmetry:
     $G_f=\ZZ_2^f\times_{\omega_2}G_b$};

\node[bottombox] (b-source) at (0.19,-2.90)
    {Gauged bosonic shadow\\
     with an order-$2N$ fermion $f$};

\node[bottombox] (b-result) at (0.81,-2.90)
    {Non-split fermionic SPT\\
     type-I $\omega_2,n_2$\\
     type-I or type-II $\nu_3$};

\draw[flowarrow] (b-source.east) --
    node[actionlabel, midway, above=5pt,
         text width=\bottomlabelwidth]
    {Ungauge $f$ and\\
      the bosonic charges}
    (b-result.west);

\end{tikzpicture}

\caption{Overview of the general constructions in
Sec.~\ref{sec:general_constructions}.
(a) For split symmetry, $G_f=\ZZ_2^f\times G_b$.
A preliminary purely bosonic condensation~\cite{Ellison22}
prepares the Abelian bosonic topological order, which contains an order-$2$ emergent fermion $f$. Condensing $f\psi$, where $\psi$ is
the physical fermion, yields the target Abelian
fermionic topological order. Ungauging the electric charges
associated with a chosen gapped boundary then gives a split fermionic SPT phase.
(b) For non-split symmetry, $G_f$ is a central extension of
$G_b$ by $\ZZ_2^f$ with nonzero extension class
$[\omega_2]\in H^2(G_b,\ZZ_2)$.
Starting from the gauged bosonic shadow, we ungauge $f$
and the remaining bosonic gauge charges to obtain
the non-split fSPT, binding them to the corresponding local clock charges.
In particular, $f$ is bound to the local fermionic-clock charge $\psi$.
The SPT constructions assume finite Abelian unitary onsite
symmetries and are restricted to the supercohomology sector
without Kitaev-chain decorations.
\textbf{Type-I} denotes components supported on a single cyclic
symmetry factor, whereas \textbf{type-II} denotes homogeneous
$\nu_3$ responses coupling two factors.
The explicit cyclic lattice family in
Sec.~\ref{sec:general_fermionic_SPTs} uses $n_2=0$.
The general non-split construction and its cochain dictionary
are presented in Sec.~\ref{sec:nonsplit_spt_scope} and
Appendix~\ref{app:nonsplit_bosonic_shadow}.}
\label{fig:general_construction_flow}
\endgroup
\end{figure*}

\subsection{General construction of Abelian fermionic twisted gauge theories}
\label{sec:general_fermionic_topological_orders}

{We generalize the fermionic toric code of Sec.~\ref{sec:fermionic_toric_code} by specifying a bosonic parent, choosing mutually local condensates, and verifying the resulting topological order.}

\subsubsection{K-matrix formalism construction}
\label{sec:general_fermionic_topological_order_K_matrix}

{A gapped boundary selects an electric Lagrangian subgroup. In a basis adapted to this subgroup, the target Abelian fermionic order has the form established in Appendix~\ref{sec:boundary_adapted_normal_form},}
\begin{equation}
    K_{\text{fTQD}}
    =
    \begin{pmatrix}
        0_{M\times M} & N\\
        N & -S_{\mathcal I}
    \end{pmatrix} .
    \label{eq:K_fTQD}
\end{equation}
{Here we use the invariant-factor presentation
\begin{equation}
    N=\operatorname{diag}(n_1,\ldots,n_M),
    \qquad
    1\leq n_1\mid n_2\mid\cdots\mid n_M .
    \label{eq:N_defintion_general}
\end{equation}
The cyclic orders $n_i$ may be even or odd, and $S_{\mathcal I}$ is a symmetric integer twist matrix}
\begin{equation}
    S_{\mathcal I}
    =
    \begin{pmatrix}
        s_1 & s_{12} & \cdots & s_{1M}\\
        s_{12} & s_2 & \cdots & s_{2M}\\
        \vdots & \vdots & \ddots & \vdots\\
        s_{1M} & s_{2M} & \cdots & s_M
    \end{pmatrix} .
    \label{eq:S_matrix_entries}
\end{equation}
{Odd diagonal entries signal a fermionic $K$-matrix presentation (Appendix~\ref{sec:K_matrix_fermionic_condition}). Stable equivalence removes those attached to odd-order factors and combines the remaining odd entries into one (Appendix~\ref{app: Assumptions on the fermionic twisted quantum double}). In the intrinsically fermionic branch, we can therefore choose a distinguished index $i_\star$ with even $n_{i_\star}$ such that
\begin{equation}
    s_i \in
    \begin{cases}
        \ZZ_{\text{odd}}, & i=i_\star,\\
        \ZZ_{\text{even}}, & i\neq i_\star.
    \end{cases}
    \label{eq:single_odd_diagonal_condition}
\end{equation}
The basis changes preserve the ordered invariant factors in $N$, and $i_\star$ need not be the first index. If all $s_i$ become even, the gauge block is bosonic and has a stabilizer realization by Ref.~\cite{Ellison22}.

After identifying excitations that differ by any local excitation, including the physical fermion, the number of topological sectors is
\begin{equation}
    |\det K_\text{fTQD}| = |(-1)^M (\det N)^2| = (\det N)^2 \, ,
\end{equation}
independent of the matrix $S_\mathcal{I}$.}

{The invariant-factor ordering automatically gives the integrality condition
\begin{equation}
    \frac{2n_i s_{ij}}{n_j}\in \ZZ,
    \qquad
    i>j ,
    \label{eq:integrality_condition_S}
\end{equation}
since $n_j\mid n_i$ for $i>j$. We use this below to ensure integer exponents in the deconfined generators and lattice operators; it places no additional restriction on $S_{\mathcal I}$.}

{To realize Eq.~\eqref{eq:K_fTQD}, choose a parent consisting of $\mathbb{Z}_{2n_i^2}$ toric-code layers and a transparent physical fermion:}
\begin{equation}
    \begin{aligned}
    K_\text{parent} &=
        \begin{pmatrix}
            0_{M\times M} & 2N^2 &  & \\
            2N^2 & 0_{M\times M} &  & \\
             &  & 1 & 0\\
             &  & 0 & -1
        \end{pmatrix} \\
    &= K_{\text{TCs}} \oplus K_f \, ,
    \end{aligned}
    \label{eq:K_parent_general}
\end{equation}
where $N^2=\operatorname{diag}(n_1^2,\ldots,n_M^2)$ and the lower right block $K_f$ describes the transparent physical fermion.

{
We condense $M$ mutually local bosons, identifying them with local excitations and confining anyons that braid nontrivially with any of them. The integer vectors representing these bosons form the columns of}
\begin{equation}
    Q
    :=
    \begin{pmatrix}
        -N\\
        2NU_{\mathcal{I}}\\
        0_{1\times M}\\
        \bar{s}^{T}
    \end{pmatrix},
    \label{eq:general_condensate_Q}
\end{equation}
where $U_{\mathcal{I}}$ is an upper-triangular matrix {with entries in $\tfrac{1}{2}\mathbb{Z}$} satisfying
\begin{equation}
    U_{\mathcal{I}}+U_{\mathcal{I}}^{T}=S_{\mathcal{I}}.
\end{equation}
Explicitly,
\begin{equation}
    (U_{\mathcal{I}})_{ij}
    :=
    \begin{cases}
        s_i/2, & i=j,\\
        s_{ij}, & i<j,\\
        0, & i>j.
    \end{cases}
\end{equation}
The vector $\bar{s}$ in Eq.~\eqref{eq:general_condensate_Q} records the parity of each $s_i$:
\begin{equation}
    \bar{s}=(\bar{s}_1,\ldots,\bar{s}_M)^T,
    \qquad
    \bar{s}_i=s_i \bmod 2.
\end{equation}
{
The rows of $Q$ follow the same basis ordering as $K_{\text{parent}}$ in Eq.~\eqref{eq:K_parent_general}. The first block specifies the reduction of the parent toric-code anyon labels to the target cyclic orders, while the second encodes the twist $S_{\mathcal I}$. The last two rows attach a physical fermion to the $i$th condensate whenever $s_i$ is odd. Although $U_{\mathcal I}$ has half-integer diagonal entries, $2NU_{\mathcal I}$ is integral. Thus, $Q$ is an integer matrix.
}

{
Let $q_i$ denote the $i$th column of $Q$. The mutual braiding statistics of the corresponding anyons are determined by
\begin{equation}
    2\pi Q^T K_{\text{parent}}^{-1}Q
    =
    -2\pi\left(S_{\mathcal{I}}+\bar{s}\bar{s}^{T}\right).
\end{equation}
Every entry is an integer multiple of $2\pi$, so these anyons have trivial mutual braiding. Their topological spins are
\begin{equation}
    h_{q_i}
    =
    \frac12 q_i^T K_{\text{parent}}^{-1}q_i
    =
    -\frac12\left(s_i+\bar{s}_i^2\right)
    \equiv 0 \pmod{1},
\end{equation}
where the last congruence follows because $\bar{s}_i$ and $s_i$ have the same parity. Thus, every $q_i$ represents a boson.}

{
In the intrinsically fermionic case, the condensation specified by $Q$ can be performed in two stages. Write the condensate corresponding to the unique odd diagonal entry $s_{i_\star}$ as $f\psi$, where $f$ is an emergent fermion in the toric-code layers. First, condense the bosons represented by $q_i$ for $i\ne i_\star$, together with $f^2$, using only Pauli operators and leaving the physical fermions decoupled. This produces the Abelian bosonic order with a gapped boundary shown in the leftmost box of Fig.~\ref{fig:general_construction_flow}(a). Since $f$ remains deconfined while its square is condensed, it descends to an order-$2$ emergent fermion. Condensing $f\psi$ then implements the first arrow in Fig.~\ref{fig:general_construction_flow}(a) and yields the target fermionic topological order.
}

{To identify the resulting order, we first find all deconfined excitations. Write a parent-theory anyon as}
\begin{equation}
    \ell
    =
    \begin{pmatrix}
        x\\
        y\\
        u\\
        v
    \end{pmatrix},
    \qquad
    x,y\in \ZZ^M,
    \qquad
    u,v\in \ZZ .
\end{equation}
The condition that $\ell$ braid trivially with every condensed anyon is
\begin{equation}
    Q^T K_{\text{parent}}^{-1}\ell \in \ZZ^M .
\end{equation}
Since
\begin{equation}
    K_{\text{parent}}^{-1}
    =
    \begin{pmatrix}
        0_{M\times M} & \frac12 N^{-2} &  & \\
        \frac12 N^{-2} & 0_{M\times M} &  & \\
         &  & 1 & 0\\
         &  & 0 & -1
    \end{pmatrix},
\end{equation}
we obtain
\begin{equation}
    Q^T K_{\text{parent}}^{-1}\ell
    =
    U_{\mathcal{I}}^{T}N^{-1}x
    -
    \frac12 N^{-1}y
    -
    v\bar{s}.
\end{equation}
The term $v\bar{s}$ is already an integer vector, so it does not affect the integrality condition. Hence, $\ell$ is deconfined when
\begin{equation}
    U_{\mathcal{I}}^{T}N^{-1}x
    -
    \frac12 N^{-1}y
    \in \ZZ^M .
\end{equation}
Equivalently, for some $m\in \ZZ^M$,
\begin{equation}
    U_{\mathcal{I}}^{T}N^{-1}x
    -
    \frac12 N^{-1}y
    =
    -m .
\end{equation}
Solving for $y$ gives
\begin{equation}
    y
    =
    2NU_{\mathcal{I}}^{T}N^{-1}x
    +
    2Nm .
    \label{eq:y_deconf_solution}
\end{equation}
{Thus $x$, $m$, $u$, and $v$ parametrize the deconfined excitations. Their unit vectors will give a basis, provided the resulting $y$ is integral.}

{This requires
\begin{equation}
    2NU_{\mathcal{I}}^{T}N^{-1}
    \in
    \operatorname{Mat}_{M}(\ZZ).
    \label{eq:L_integrality_condition}
\end{equation}
Its diagonal entries are $s_i$, and for $i>j$ its only potentially nonintegral entries are}
\begin{equation}
    \left(2NU_{\mathcal{I}}^{T}N^{-1}\right)_{ij}
    =
    \frac{2n_i s_{ji}}{n_j}
    =
    \frac{2n_i s_{ij}}{n_j}.
\end{equation}
{Their integrality follows from Eq.~\eqref{eq:integrality_condition_S}.}

Let $r_i$ be the $i$-th standard basis vector of $\ZZ^M$. Setting $x=r_i$ and $m=0$ in Eq.~\eqref{eq:y_deconf_solution} gives the first type of generator,
\begin{equation}
    \begin{pmatrix}
        r_i\\
        2NU_{\mathcal{I}}^{T}N^{-1}r_i\\
        0\\
        0
    \end{pmatrix}.
\end{equation}
Setting $x=0$ and $m=r_i$ gives the second type,
\begin{equation}
    \begin{pmatrix}
        0\\
        2Nr_i\\
        0\\
        0
    \end{pmatrix}.
\end{equation}
Together with the two generators from the local fermion block, these vectors form the columns of
\begin{equation}
    L
    =
    \begin{pmatrix}
        \mathbb{I}_M & 0_{M\times M} &  & \\
        2NU_{\mathcal{I}}^{T}N^{-1} & 2N &  & \\
         &  & 1 & 0\\
         &  & 0 & -1
    \end{pmatrix}.
    \label{eq:general_deconfined_excitations}
\end{equation}

The columns of $L$ braid trivially with the condensed anyons. Indeed,
\begin{equation}
    L^T K_{\text{parent}}^{-1}Q
    =
    \begin{pmatrix}
        0_{M\times M}\\
        -\mathbb{I}_M\\
        0_{1\times M}\\
        \bar{s}^{T}
    \end{pmatrix},
\end{equation}
{whose entries are integers. {Eq.}~\eqref{eq:y_deconf_solution} also shows that these columns span every deconfined vector over the integers.}

{An excitation labeled by $\lambda\in\ZZ^{2M+2}$ has parent representative $L\lambda$. Restricting the parent braiding form to this basis gives}
\begin{equation}
    K_{\text{final}}^{-1}
    =
    L^T K_{\text{parent}}^{-1}L,
\end{equation}
where a direct computation yields
\begin{equation}
    L^T K_{\text{parent}}^{-1}L
    =
    \begin{pmatrix}
        N^{-1}S_{\mathcal{I}}N^{-1} & N^{-1} &  & \\
        N^{-1} & 0_{M\times M} &  & \\
         &  & 1 & 0\\
         &  & 0 & -1
    \end{pmatrix}.
\end{equation}
In other words,
\begin{equation}
    \begin{aligned}
    K_{\text{final}} &=
        \begin{pmatrix}
            0_{M\times M} & N &  & \\
            N & -S_{\mathcal{I}} &  & \\
             &  & 1 & 0\\
             &  & 0 & -1
        \end{pmatrix}  \\
    &= K_\text{fTQD} \oplus K_f
    \end{aligned}
\end{equation}

\subsubsection{Lattice-level construction}
\label{sec:general_fermionic_topological_order_lattice_model}

{We implement the parent theory in Eq.~\eqref{eq:K_parent_general} on a square lattice, with $M$ qudits of dimensions $2n_i^2$ on each edge and a physical fermion on each plaquette. The parent stabilizers are}
\begin{equation}
    A_{v,i}
    =
    \begin{tikzpicture}[baseline={(v.base)}, scale=1]
        \draw[line width=0.45pt] (0,-1.2) -- (0,1.2);
        \draw[line width=0.45pt] (-1.2,0) -- (1.2,0);

        \node[inner sep=1pt, fill=white] (v) at (0,0) {$v$};

        \node[blue, inner sep=1pt, fill=white]
            at (0,0.6) {$X_i^{\dagger}$};
        \node[blue, inner sep=1pt, fill=white]
            at (0,-0.6) {$X_i$};
        \node[blue, inner sep=1pt, fill=white]
            at (-0.6,0) {$X_i$};
        \node[blue, inner sep=1pt, fill=white]
            at (0.6,0) {$X_i^{\dagger}$};
    \end{tikzpicture},
    \quad
    B_{p,i}
    =
    \begin{tikzpicture}[baseline={(p.base)}, scale=1]
        \draw[line width=0.45pt] (-0.6,-0.6) rectangle (0.6,0.6);

        \node[font=\scriptsize, inner sep=1pt, fill=white] (p) at (0,0) {$p$};

        \node[blue, inner sep=1pt, fill=white]
            at (0,0.6) {$Z_i^{\dagger}$};
        \node[blue, inner sep=1pt, fill=white]
            at (-0.6,0) {$Z_i^{\dagger}$};
        \node[blue, inner sep=1pt, fill=white]
            at (0.6,0) {$Z_i$};
        \node[blue, inner sep=1pt, fill=white]
            at (0,-0.6) {$Z_i$};
    \end{tikzpicture},
    \quad
    \textcolor{red}{P_p} \, ,
    \label{eq:parent_lattice_stabilizers_general}
\end{equation}
{where $P_p=-i\gamma_p\gamma_p'$, with Majorana conventions as in Sec.~\ref{sec:fermionic_toric_code}.}

{The condensates are specified by Eq.~\eqref{eq:general_condensate_Q},}
\begin{equation*}
    Q
    =
    \begin{pmatrix}
        -N\\
        2NU_{\mathcal{I}}\\
        0_{1\times M}\\
        \bar{s}^{T}
    \end{pmatrix}.
\end{equation*}
{If $e_i$ and $m_i$ generate the anyons of the $i$-th toric-code layer, the $i$-th column of $Q$ represents}
\begin{equation}
    q_i
    =
    \psi^{\bar{s}_i}
    m_i^{-n_i}
    e_i^{n_i s_i}
    \prod_{j<i} e_j^{2n_j s_{ji}}
    \label{eq:lattice_condensed_anyon_bi}
\end{equation}
for $i \in \{1,\ldots,M\}$. Here, $\psi$ denotes a physical fermion.

{The corresponding condensation terms and surviving parent stabilizers generate}
\begin{equation}
    \mathcal{S}_\text{fTQD} = 
    \big\langle
    \{A'_{v,i}\},\{B'_{p,i}\},\{C_{e,i}\}
    \big\rangle_{i \in \{1,...,M\}} \, ,
\end{equation}
where the generators are defined as follows. At each vertex, we have
\begin{equation}
    A'_{v,i} =
    \begin{tikzpicture}[baseline=-0.6ex, line width=0.5pt]
      \def\L{1.5}
      \draw (-\L,0) -- (\L,0);
      \draw (0,0) -- (0,-\L);
      \draw (0,0) -- (0,\L);
      \draw (0,\L) -- (\L,\L);
      \draw (\L,\L) -- (\L,0);
      \node[op] at (0,0) {$v$};
      \node[op] at (-0.5*\L,0) {$\textcolor{blue}{X_i^\dagger}$};
      \node[op] at (0,-0.5*\L) {$\textcolor{blue}{X_i^\dagger}$};
      \node[op] at (0,0.5*\L) {$\textcolor{blue}{X_i \hat{Z}_i}$};
      \node[op] at (0.5*\L,0) {$\textcolor{blue}{X_i{\hat{Z}}^{\dagger}_i}$};
      \node[op] at (0.5*\L,\L) {$\textcolor{blue}{\hat{Z}_i}$};
      \node[op] at (\L,0.5*\L) {$\textcolor{blue}{\hat{Z}_i^{\dagger}}$};
    \end{tikzpicture}
\end{equation}
{where $\hat Z_{i}=Z_{i}^{s_i}\mathop{\Pi}\limits_{j>i}Z_{j}^{2 n_{j}s_{ij}/n_{i}}$.} 

The plaquette terms are
{
\begin{equation}
\begin{aligned}
    B'_{p,i_\star} &=
    \vcenter{\hbox{%
\begin{tikzpicture}[baseline=-0.6ex, line width=0.5pt]
      \def\L{1.5}
      \draw (0,0) rectangle (\L,\L);
      \node[op] at (0.5*\L,\L) {$\textcolor{blue}{Z_{i_\star}^{-n_{i_\star}}}$};
      \node[op] at (0,0.5*\L) {$\textcolor{blue}{Z_{i_\star}^{-n_{i_\star}}}$};
      \node[op] at (\L,0.5*\L) {$\textcolor{blue}{Z_{i_\star}^{n_{i_\star}}}$};
      \node[op] at (0.5*\L,0) {$\textcolor{blue}{Z_{i_\star}^{n_{i_\star}}}$};
      \node[op] at (0.5*\L,0.5*\L) {$\textcolor{red}{P_p}$};
    \end{tikzpicture}}},
    \\[8pt]
    B'_{p,i} &=
    \vcenter{\hbox{%
\begin{tikzpicture}[baseline=-0.6ex, line width=0.5pt]
      \def\L{1.5}
      \draw (0,0) rectangle (\L,\L);
      \node[op] at (0.5*\L,\L) {$\textcolor{blue}{Z_i^{-2n_i}}$};
      \node[op] at (0,0.5*\L) {$\textcolor{blue}{Z_i^{-2n_i}}$};
      \node[op] at (\L,0.5*\L) {$\textcolor{blue}{Z_i^{2n_i}}$};
      \node[op] at (0.5*\L,0) {$\textcolor{blue}{Z_i^{2n_i}}$};
    \end{tikzpicture}}},
    \qquad i\neq i_\star.
\end{aligned}
\label{eq:general_plaquette_stabilizers}
\end{equation}}
Finally, the condensation terms corresponding to Eq.~\eqref{eq:lattice_condensed_anyon_bi} are
{\begin{equation}
    (-i)\cdot C_{e,i_\star} =
    \raisebox{-0.6cm}{\begin{tikzpicture}[baseline=-0.6ex, line width=0.5pt]
      \def\L{1.5}
      \draw (-\L,0) -- (0,0);
      \draw (0,0) -- (0,\L);
      \node[op] at (-0.5*\L,0) {$\textcolor{blue}{\check Z_{i_\star}}$};
      \node[op] at (-0.5*\L,0.5*\L) {$\textcolor{red}{\gamma}$};
      \node[op] at (0.0*\L,0.5*\L) {$\textcolor{blue}{X_{e,i_\star}^{-n_{i_\star}}}$};
      \node[op] at (0.5*\L, 0.53*\L) {$\textcolor{red}{\gamma'}$};
    \end{tikzpicture}}
    ,\qquad 
    \raisebox{0.6cm}{\begin{tikzpicture}[baseline=-0.6ex, line width=0.5pt]
      \def\L{1.5}
      \draw (0,0) -- (\L,0);
      \draw (0,0) -- (0,-\L);
      \node[op] at (0.5*\L,0) {$\textcolor{blue}{X_{e,i_\star}^{n_{i_\star}}}$};
      \node[op] at (0.5*\L,0.5*\L) {$\textcolor{red}{\gamma}$};
      \node[op] at (0,-0.5*\L) {$\textcolor{blue}{\check Z_{i_\star}}$};
      \node[op] at (0.53*\L,-0.5*\L) {$\textcolor{red}{\gamma'}$};
    \end{tikzpicture}} \, ,
\end{equation}
and for $i\neq i_\star$}
\begin{equation}
    C_{e,i} =
    \raisebox{-0.6cm}{\begin{tikzpicture}[baseline=-0.6ex, line width=0.5pt]
      \def\L{1.2}
      \draw (-\L,0) -- (0,0);
      \draw (0,0) -- (0,\L);
      \node[op] at (-0.5*\L,0) {$\textcolor{blue}{\check{Z}_i}$};
      \node[op] at (0.0*\L,0.5*\L) {$\textcolor{blue}{X_{e,i}^{-n_i}}$};
    \end{tikzpicture}}
    ,\qquad 
    \raisebox{0.6cm}{\begin{tikzpicture}[baseline=-0.6ex, line width=0.5pt]
      \def\L{1.2}
      \draw (0,0) -- (\L,0);
      \draw (0,0) -- (0,-\L);
      \node[op] at (0.5*\L,0) {$\textcolor{blue}{X_{e,i}^{n_i}}$};
      \node[op] at (0,-0.5*\L) {$\textcolor{blue}{\check{Z}_i}$};
    \end{tikzpicture}} \, ,
\end{equation}
{where $\check{Z}_i= Z_{i}^{n_{i}s_{i}}\prod_{j<i}Z_{j}^{2n_{j}s_{ij}}$ for every layer, including $i=i_\star$. The earlier-layer factors in $\check Z_{i_\star}$ are required when the distinguished index is not first: they implement the corresponding column of $Q$ and ensure commutation with the vertex terms in those layers.

If every diagonal entry of $S_{\mathcal I}$ is even, the special $i_\star$ terms are absent: one uses the ordinary $B'_{p,i}$ and $C_{e,i}$ terms for every layer, while the physical-fermion parity stabilizers $P_p$ remain decoupled.}

The deconfined excitations in the basis of Eq.~\eqref{eq:general_deconfined_excitations} are given by
{
\newcommand{\generalstringdiagram}[1]{%
  \makebox[2.2cm][c]{$\vcenter{\hbox{%
    \begin{tikzpicture}[x=\Ledge cm,y=\Ledge cm,line width=.5pt,
      stringhop/.style={dashed,magenta,-{Latex[length=1.6mm]}},
      stringop/.style={inner sep=1pt,fill=white,text=blue}]
      #1
    \end{tikzpicture}}}$}}
\begin{alignat}{2}
    c_i &={}\generalstringdiagram{
      \draw (0,-.5)--(0,.5);
      \draw[stringhop] (.36,-.5)--(.36,.5);
      \node[stringop] at (0,0) {$Z_i^{2n_i}$};
    },\quad &&\generalstringdiagram{
      \draw (-.5,0)--(.5,0);
      \draw[stringhop] (-.5,-.24)--(.5,-.24);
      \node[stringop] at (0,0) {$Z_i^{2n_i}$};
    },
    \label{eq:general_c_string_operator}
\\[5pt]
    \phi_i &={}\generalstringdiagram{
      \draw (-.5,0)--(.5,0)--(.5,1);
      \draw[stringhop] (0,.5)--(0,-.5);
      \node[stringop] at (0,0) {$X_{e,i}$};
      \node[stringop] at (.5,.5) {$\hat Z_i^{\dagger}$};
    },\quad &&\generalstringdiagram{
      \draw (0,-.5)--(0,.5)--(1,.5);
      \draw[stringhop] (-.5,0)--(.5,0);
      \node[stringop] at (0,0) {$X_{e,i}$};
      \node[stringop] at (.5,.5) {$\hat Z_i$};
    }.
    \label{eq:general_phi_string_operator}
\end{alignat}
The spins and mutual braiding phases of these strings reproduce $K_{\mathrm{fTQD}}^{-1}$ in Eq.~\eqref{eq:K_fTQD}, as derived in Appendix~\ref{app:general_lattice_topological_data}.}

\subsection{{Split fSPTs from the Abelian \texorpdfstring{$K$}{K} matrix}}
\label{sec:split_spt_scope}

{For a \emph{split} fermionic symmetry, fermion parity is an independent direct factor,
\begin{equation}
G_f=G\times \mathbb Z_2^F,
\qquad
G=\prod_{i=1}^r\mathbb Z_{N_i},
\end{equation}
so the extension cocycle is $\omega_2=0$. We first separate the fermion-decoration data from the independent bosonic responses, then determine which of them are encoded by the Abelian $K$ matrix.}

\paragraph{{General solution of the Gu--Wen equation.}}
{The supercohomology data are represented by
\begin{equation}
n_2\in Z^2(G,\mathbb Z_2),
\qquad
\nu_3\in C^3(G,\mathbb R/\mathbb Z),
\end{equation}
subject to the Gu--Wen equation~\cite{GW14}
\begin{equation}
\delta n_2=0,
\qquad
\delta\nu_3=\frac12 n_2\cup n_2 .
\label{eq:GW-general}
\end{equation}
The second equality takes values in $\mathbb R/\mathbb Z$, with $\mathbb Z_2$ embedded by multiplication by $1/2$. We fix a representative of $n_2$ and solve for $\nu_3$; pairs related by the standard supercohomology cochain equivalences describe the same phase.

Let $a_i\in Z^1(G,\mathbb Z_{N_i})$ be the projection onto the $i$-th cyclic factor, and let
\begin{equation}
\widetilde a_i(g)\in\{0,1,\ldots,N_i-1\}
\end{equation}
denote its canonical integer lift. Since $a_i$ is flat modulo $N_i$,
\begin{equation}
B_i:=\frac{\delta\widetilde a_i}{N_i}\in Z^2(G,\mathbb Z)
\end{equation}
records the integer carry when two elements are added modulo $N_i$.

Only an even-order cyclic factor contributes to $H^1(G,\mathbb Z_2)$ or to the single-factor part of $H^2(G,\mathbb Z_2)$. For every even $N_i$, define
\begin{align}
x_i&:=a_i\pmod 2\in Z^1(G,\mathbb Z_2),
\nonumber\\
b_i&:=B_i\pmod 2\in Z^2(G,\mathbb Z_2),
\end{align}
and let
\begin{equation}
E=\{i\mid N_i\ {\rm is\ even}\}.
\end{equation}
A representative of every class in $H^2(G,\mathbb Z_2)$ can then be written as
\begin{equation}
n_2
=
\sum_{i\in E}r_i b_i
+
\sum_{\substack{i<j\\i,j\in E}}
r_{ij}\,x_i\cup x_j,
\qquad
r_i,r_{ij}\in\mathbb Z_2 .
\label{eq:general-n2}
\end{equation}
The single-factor cocycle $b_i$ is a \emph{type-I decoration}, also called a \emph{self-decoration}; $x_i\cup x_j$ couples two factors and is a \emph{mixed-decoration}.

We solve the Gu--Wen equation on each generator before combining them. For a type-I generator $b_i$, define
\begin{equation}
q_i=\frac{1}{2N_i}\widetilde a_i\cup B_i .
\end{equation}
Using $\delta\widetilde a_i=N_iB_i$ and $\delta B_i=0$ gives
\begin{equation}
\delta q_i=\frac12 b_i\cup b_i .
\end{equation}
For a mixed generator
\begin{equation}
m_{ij}=x_i\cup x_j,
\end{equation}
one may use
\begin{align}
q_{ij}(g,h,k)
={}&
\frac14 x_i(g)x_j(h)x_j(k)
\nonumber\\
&+
\frac12 x_i(g)x_i(h)x_j(h)x_j(k).
\end{align}
Here $x_i$ and $x_j$ take their integer values $0$ or $1$. Direct substitution gives
\begin{equation}
\delta q_{ij}
=
\frac12 m_{ij}\cup m_{ij}.
\end{equation}

Collect these generators as $e_A\in\{b_i,m_{ij}\}$ and their primitives as $q_A$. For
\begin{equation}
n_2=\sum_A r_Ae_A,
\end{equation}
a particular solution is
\begin{equation}
Q(n_2)
=
\sum_A r_Aq_A
+
\frac12\sum_{A<B}
r_Ar_B\,e_A\cup_1e_B .
\label{eq:Q-n2}
\end{equation}
The cup-one term supplies the cross terms because
\begin{equation}
\delta(e_A\cup_1e_B)
=
e_A\cup e_B+e_B\cup e_A
\pmod 2,
\end{equation}
and hence
\begin{equation}
\delta Q(n_2)
=
\frac12n_2\cup n_2 .
\end{equation}
Thus every class in $H^2(G,\mathbb Z_2)$ is unobstructed in this finite Abelian, unitary, split-symmetry setting.

Any other solution for the same $n_2$ differs by a closed $3$-cocycle:
\begin{equation}
\nu_3
=
Q(n_2)+\omega_3+\delta\lambda_2,
\qquad
[\omega_3]\in H^3(G,\mathbb R/\mathbb Z).
\label{eq:general-nu3}
\end{equation}
Here $Q(n_2)$ is fixed by the decoration, and $\omega_3$ is an independent bosonic SPT response.

For $G=\prod_i\mathbb Z_{N_i}$,
\begin{equation}
H^3(G,\mathbb R/\mathbb Z)
\simeq
\prod_i\mathbb Z_{N_i}
\times
\prod_{i<j}\mathbb Z_{N_{ij}}
\times
\prod_{i<j<k}\mathbb Z_{N_{ijk}},
\end{equation}
where
\begin{equation}
N_{ij}=\gcd(N_i,N_j),
\qquad
N_{ijk}=\gcd(N_i,N_j,N_k).
\end{equation}
The three factors give the homogeneous type-I, type-II, and type-III responses, involving one, two, and three cyclic factors, respectively. Representatives are
\begin{align}
\omega_3^{\rm I}
&=
\sum_i
\frac{p_i}{N_i}\,
\widetilde a_i\cup B_i,
\\
\omega_3^{\rm II}
&=
\sum_{i<j}
\frac{p_{ij}}{N_i}\,
\widetilde a_i\cup B_j,
\\
\omega_3^{\rm III}
&=
\sum_{i<j<k}
\frac{p_{ijk}}{N_{ijk}}\,
\widetilde a_i\cup\widetilde a_j\cup\widetilde a_k .
\end{align}
The coefficients are integer representatives of
\begin{equation}
p_i\in\mathbb Z_{N_i},
\qquad
p_{ij}\in\mathbb Z_{N_{ij}},
\qquad
p_{ijk}\in\mathbb Z_{N_{ijk}}.
\end{equation}
}

\paragraph{{Data encoded by the Abelian $K$ matrix.}}
{We now match the decoration and response data to the matrix used in the construction. A nonchiral Abelian twisted gauge theory with an electric gapped boundary admits the stable integral presentation
\begin{equation}
K_{\rm Abelian}
=
\begin{pmatrix}
0 & N\\
N & -S
\end{pmatrix},
\qquad
N=\operatorname{diag}(N_1,\ldots,N_r),
\label{eq:abelian-K-SPT}
\end{equation}
where the symmetric integer matrix $S$ encodes the twist. We choose the standard stable representative with $S_{ii}$ even for odd $N_i$, since an odd-order factor has no nontrivial $\mathbb Z_2$ decoration.

The quadratic cochain obtained from this matrix is
\begin{equation}
\nu^{(0)}_{3,S}
=
\sum_{i,j}
\frac{S_{ij}}{2N_i}\,
\widetilde a_i\cup B_j .
\label{eq:bare-K-nu}
\end{equation}
Define
\begin{equation}
k_i=
\begin{cases}
S_{ii}\pmod 2, & N_i\ {\rm even},\\
0, & N_i\ {\rm odd}.
\end{cases}
\end{equation}
The associated fermion decoration is
\begin{equation}
n_{2,S}
=
\sum_{i\in E}k_i b_i
=
\sum_{i\in E}
S_{ii}\frac{\delta\widetilde a_i}{N_i}
\pmod 2.
\label{eq:K-n2}
\end{equation}
Thus the parity of each diagonal entry selects a type-I decoration.

To determine the accompanying response, compare the coboundary of Eq.~\eqref{eq:bare-K-nu},
\begin{align}
\delta\nu^{(0)}_{3,S}
={}&
\frac12\sum_i S_{ii}B_i\cup B_i
\nonumber\\
&+
\frac12\sum_{i<j}
S_{ij}
\left(
B_i\cup B_j+B_j\cup B_i
\right).
\end{align}
with the required Gu--Wen term,
\begin{align}
\frac12 n_{2,S}\cup n_{2,S}
={}&
\frac12\sum_i k_iB_i\cup B_i
\nonumber\\
&+
\frac12\sum_{i<j}
k_ik_j B_i\cup B_j
\nonumber\\
&+
\frac12\sum_{i<j}
k_ik_j B_j\cup B_i
\pmod 1.
\end{align}
The diagonal terms agree because $S_{ii}=k_i\pmod 2$. The off-diagonal coefficients agree only when
\begin{equation}
S_{ij}=k_ik_j\pmod 2.
\label{eq:S-parity-condition}
\end{equation}
For general $S$, the cup-one correction gives
\begin{equation}
\nu_{3,S}
=
\sum_{i,j}
\frac{S_{ij}}{2N_i}\,
\widetilde a_i\cup B_j
+
\frac12
\sum_{i<j}
(k_ik_j-S_{ij})\,B_i\cup_1B_j .
\label{eq:corrected-K-nu}
\end{equation}
The cup-one identity then ensures
\begin{equation}
\delta\nu_{3,S}
=
\frac12n_{2,S}\cup n_{2,S}.
\end{equation}
This correction is fixed by $S$ and $n_{2,S}$; it introduces no additional invariant.

To identify the independent homogeneous responses, write
\begin{equation}
S_{ii}=k_i+2p_i.
\end{equation}
Then each diagonal contribution splits as
\begin{equation}
\frac{S_{ii}}{2N_i}\widetilde a_i\cup B_i
=
k_iq_i+
\frac{p_i}{N_i}\widetilde a_i\cup B_i.
\end{equation}
The first term contributes to $Q(n_{2,S})$ and the second is a type-I bosonic response. After subtracting $Q(n_{2,S})$ from Eq.~\eqref{eq:corrected-K-nu}, the term proportional to each off-diagonal entry $S_{ij}$ is
\begin{equation}
T_{ij}=\frac{\widetilde a_i\cup B_j}{2N_i}
+\frac{\widetilde a_j\cup B_i}{2N_j}
-\frac12 B_i\cup_1B_j.
\label{eq:split_typeII_generator}
\end{equation}
This closed cocycle vanishes on either individual factor. On the cyclic subgroup generated by $(1,1)$, of order $\ell_{ij}=\operatorname{lcm}(N_i,N_j)$, its invariant is $\sum_{t=0}^{\ell_{ij}-1}T_{ij}(g,tg,g)=1/N_{ij}$ modulo one, so it generates the full type-II factor $\mathbb Z_{N_{ij}}$. 
The resulting sector is therefore
\begin{equation}
\boxed{
\begin{aligned}
n_2
&=
\sum_{i\in E}r_i
\frac{\delta\widetilde a_i}{N_i}
\pmod 2,
\\
\nu_3
&=
Q(n_2)
+
\omega_3^{\rm I}
+
\omega_3^{\rm II}
+
\delta\lambda_2 .
\end{aligned}}
\label{eq:K-accessible-supercohomology}
\end{equation}
All data in Eq.~\eqref{eq:K-accessible-supercohomology} are realized, modulo supercohomology equivalences. Mixed $n_2$ decorations and independent type-III responses require a more general construction. Ungauging the specified electric charges gives the split fSPT in Fig.~\ref{fig:general_construction_flow}(a), using ordinary clock variables for $G$ and retaining the physical fermions. The charge labels fix the symmetry action and must be tracked through changes of cyclic basis. The invariant-factor presentation of Sec.~\ref{sec:general_fermionic_topological_orders} guarantees lattice integrality without restricting the intrinsic target orders.}

\subsection{{Non-split fSPTs from bosonic shadows}}
\label{sec:nonsplit_spt_scope}

{We next consider a non-split fermion-parity extension. For a finite Abelian fermionic symmetry, a change of cyclic generators gives
\begin{equation}
\begin{aligned}
G_f&=\mathbb Z_{2N_0}^{F}\times\prod_{i=1}^{M}\mathbb Z_{N_i},\\
G_b&=G_f/\mathbb Z_2^F=\mathbb Z_{N_0}\times\prod_{i=1}^{M}\mathbb Z_{N_i},
\end{aligned}
\end{equation}
with $N_0$ even. This reduction, proved in Appendix~\ref{sec:sufficiency_non_split_ZF2n}, isolates the non-split extension on one cyclic factor without eliminating SPT response data involving the other factors. With $B_\alpha=\delta\widetilde a_\alpha/N_\alpha$ and $b_\alpha=B_\alpha\pmod 2$, we can write $\omega_2=b_0$. The non-split supercohomology consistency equation is~\cite{Barkeshli2022Classification}
\begin{equation}
\delta\nu_3=\frac12\bigl(n_2\cup n_2+\omega_2\cup n_2\bigr).
\label{eq:nonsplit_GW_scope}
\end{equation}
Consider the equivalence
\begin{equation}
n_2\mapsto n_2+\omega_2,
\qquad
\nu_3\mapsto\nu_3+\frac12\omega_2\cup_1 n_2~.
\end{equation}
For the type-I fermion decorations $n_2$, this equivalence
removes the component on factor $0$. We may therefore choose
\begin{equation}
n_2=\sum_{i=1}^{M}k_i b_i,
\qquad k_i\in\mathbb Z_2,
\label{eq:nonsplit_n2_choice}
\end{equation}
with $k_i=0$ for odd $N_i$. Thus the remaining split factors can still carry fermion decorations. Their coupling to the extension contributes the additional particular solution
\begin{equation}
R_{\omega,n}=\sum_{i=1}^{M}\frac{k_i}{2N}\,
\widetilde a_0\cup B_i,
\qquad
\delta R_{\omega,n}=\frac12\omega_2\cup n_2.
\label{eq:nonsplit_cross_term}
\end{equation}
Consequently,
\begin{equation}
\nu_3=Q(n_2)+R_{\omega,n}
+\omega_3^{\rm I}+\omega_3^{\rm II}+\delta\lambda_2
\label{eq:nonsplit_cochain_family}
\end{equation}
solves Eq.~\eqref{eq:nonsplit_GW_scope} for type-I components of $\omega_2,n_2$ and type-I/type-II $\nu_3$. The homogeneous cocycles can involve any quotient factor, including factor $0$, and their labels remain subject to fermionic cochain equivalences. {Eq.}~\eqref{eq:nonsplit_cochain_family} therefore specifies admissible representatives rather than a count of inequivalent phases.

In the case of a non-split extension, $G_b$ has no lift to a complementary subgroup of $G_f$, so one cannot obtain the desired intermediate fermionic topological order by gauging $G_b$ as in the split construction. Instead, we use the bosonic $G_b$-enriched toric-code shadow obtained by gauging fermion parity~\cite{BGK17,Barkeshli2022Classification}. Gauging its genuine bosonic symmetry $G_b$ gives the Abelian bosonic parent below. We then recover the fSPT by fermionic ungauging directly from this parent.

Here we derive the explicit K-matrix representation of its bosonic parent from the supercohomology data. Write $r=M+1$ and introduce
\begin{equation}
\begin{gathered}
\mathsf N=\operatorname{diag}(N_0,\ldots,N_M),\qquad N_0=N,\\
\boldsymbol w=(1,0,\ldots,0)^T,\qquad
\boldsymbol k=(0,k_1,\ldots,k_M)^T,\\
\boldsymbol v=\boldsymbol w+\boldsymbol k .
\end{gathered}
\label{eq:nonsplit_shadow_parameters}
\end{equation}
The components of these vectors are integer lifts, with $k_i=0,1$. A useful ordered representative of the particular Gu--Wen solution is
\begin{equation}
\nu^{\rm can}_3
=\frac12\sum_{\alpha,\beta=0}^{M}
v_\alpha k_\beta\,
\frac{\widetilde a_\alpha}{N_\alpha}\cup B_\beta
+\widetilde\nu_3 .
\label{eq:nonsplit_shadow_canonical_nu}
\end{equation}
Here $\widetilde\nu_3$ is an arbitrary type-I/type-II bosonic cocycle. Appendix~\ref{app:nonsplit_bosonic_shadow} derives this representative from the existing solution $Q(n_2)+R_{\omega,n}$ in Eqs.~\eqref{eq:Q-n2} and~\eqref{eq:nonsplit_cross_term}, including the shift of homogeneous type-II labels. It also fixes the correspondence between $\widetilde\nu_3$ and the matrix $P$ below, using the sign convention of Eq.~\eqref{eq:abelian-K-SPT}.

In the ordered basis $(x,\boldsymbol b,y,\boldsymbol a)$, take
\begin{equation}
K_B=
\begin{pmatrix}
0&0&-2&\boldsymbol v^T\\
0&0&0&\mathsf N\\
-2&0&0&\boldsymbol k^T\\
\boldsymbol v&\mathsf N&\boldsymbol k&P
\end{pmatrix},
\qquad P=P^T,\quad P_{\alpha\alpha}\in2\ZZ .
\label{eq:nonsplit_shadow_KB}
\end{equation}

Appendix~\ref{app:nonsplit_bosonic_shadow} provides a transformation to the twisted-quantum-double form. Consequently, it admits a Pauli stabilizer realization by Ref.~\cite{Ellison22}.

Then we identify the order-$2N$ emergent fermion in this model and ungauge it together with the remaining bosonic charges. 
The columns of $K_B$ give the anyon relations
\begin{equation}
\begin{gathered}
N_\alpha a_\alpha=0,\qquad
2y=\sum_\alpha v_\alpha a_\alpha,\qquad
2x=\sum_\alpha k_\alpha a_\alpha .
\end{gathered}
\label{eq:nonsplit_shadow_charge_relations}
\end{equation}
Define, in additive anyon notation,
\begin{equation}
f=x+y-\sum_\alpha k_\alpha a_\alpha .
\label{eq:nonsplit_shadow_fermion}
\end{equation}
Then
\begin{equation}
2f=a_0,\qquad
\theta_f=-1,\qquad
\operatorname{ord}(f)=2N_0
.
\label{eq:nonsplit_shadow_fermion_order}
\end{equation}
Moreover, every $a_\alpha$ is a boson and all mutual braiding phases among $f$ and the $a_\alpha$ are trivial. Thus
\begin{equation}
\begin{aligned}
\mathcal L&=\langle f,a_1,\ldots,a_M\rangle
\simeq\ZZ_{2N_0}\times\prod_{i=1}^{M}\ZZ_{N_i},\\
|\mathcal L|&=2\det\mathsf N=\sqrt{|\det K_B|},\qquad
\mathcal L^\perp=\mathcal L .
\end{aligned}
\label{eq:nonsplit_shadow_Lagrangian}
\end{equation}
Maximality of $\mathcal L$ implies that the ungauged phase has no intrinsic topological order modulo the local physical fermion. On the lattice level, we introduce one $\ZZ_{2N}^F$ fermionic clock from Sec.~\ref{sec:fermionicclock} and ordinary $\ZZ_{N_i}$ clocks for $i>0$ at each plaquette. Bind $f$ to the local fermionic-clock charge, and bind $a_i$ to the $i$-th bosonic-clock charge. The relation $2f=a_0$ is then realized by $\Gamma_0^2=Z_0$. The symmetry generators are then
\begin{equation}
\begin{gathered}
U_0=\prod_p\Pi_{0,p},\qquad
U_i=\prod_p X_{i,p}\quad(i>0),\\
U_0^N=\prod_p P_p=(-1)^F,\qquad
U_0^{2N}=U_i^{N_i}=1 .
\end{gathered}
\label{eq:nonsplit_shadow_onsite_symmetry}
\end{equation}
These relations realize the desired onsite symmetry group of the non-split fermionic SPT phase.
}

\Needspace{6\baselineskip}
\subsection{{Explicit lattice example: \texorpdfstring{$\ZZ_{2^n}^F$}{Z(2 to the n)-F} fermionic SPT phases}}
\label{sec:general_fermionic_SPTs}

{For the explicit cyclic construction, take
\begin{equation}
    G_f=\ZZ_{2^n}^F,
    \qquad
    G_b=G_f/\ZZ_2^F\simeq\ZZ_{2^{n-1}}.
\end{equation}
The case $n=3$, namely $\ZZ_8^F$, was constructed in Sec.~\ref{sec:Z8F_SPT}; here we extend the construction uniformly to $n>3$. We first identify a bosonic condensation followed by fermionic ungauging that remove the intrinsic topological order, then give their stabilizer realization and extract the fSPT response from its boundary symmetry.}

{Begin with the $\ZZ_{2^{n+1}}$ toric code, whose anyons are generated by $e$ and $m$, and condense the boson}
\begin{eqs}
    a=e^{2^n}m^{2^n}.
\end{eqs}
The resulting topological order is generated by
\begin{eqs}
    \{ em^{-1},\,m^2\},
\end{eqs}
and admits the $K$-matrix description
\begin{eqs}
    K=
    \begin{pmatrix}
        0 & 2^n\\
        2^n & 2^n
    \end{pmatrix}.
\end{eqs}
In this basis, consider the anyon represented by
\begin{eqs}
    \begin{pmatrix}
        2\\
        1-2^{n-2}
    \end{pmatrix},
\end{eqs}
which, in terms of the original generators of the $\ZZ_{2^{n+1}}$ toric code, is
\begin{eqs}
    b=(em^{-1})^2(m^2)^{1-2^{n-2}}
      =e^2m^{-2^{n-1}}.
\end{eqs}
Its topological spin is
\begin{eqs}
    \theta_b
    =
    \exp\left(
        \frac{2\pi i}{2^{n+1}}\,
        2(-2^{n-1})
    \right)
    =-1,
\end{eqs}
so $b$ is a fermionic anyon. To ungauge the fermionic charge sector generated by $b$, we bind $b$ to a local fermionic-clock charge $\psi$ and retain the resulting onsite fermionic symmetry.

{To verify that the resulting phase has no intrinsic topological order, consider the subgroup}
\begin{eqs}
    H=\langle a,b\rangle
\end{eqs}
of the anyon group of the original $\ZZ_{2^{n+1}}$ toric code, with the understanding that $b$ is identified with the local physical fermion in the intrinsic topological sector after ungauging. {The parent anyon group $\mathcal A=\mathbb Z_{2^{n+1}}\times\mathbb Z_{2^{n+1}}$ has a nondegenerate braiding pairing. The generators obey
\begin{equation}
\begin{gathered}
\operatorname{ord}(a)=2,\qquad\operatorname{ord}(b)=2^n,\\
B(a,a)=B(a,b)=B(b,b)=1.
\end{gathered}
\end{equation}
where $B$ denotes the mutual-braiding phase. Moreover, $a\notin\langle b\rangle$: matching the electric exponent would require a power $b^{2^{n-1}}$, whose magnetic exponent vanishes modulo $2^{n+1}$ for $n>3$, whereas that of $a$ is $2^n$. Thus
\begin{equation}
|H|=2^{n+1}=\sqrt{|\mathcal A|},
\qquad H=H^\perp.
\end{equation}
Binding every odd power of $b$ to $\psi$ makes the entire bound-state subgroup bosonic. Maximality of $H$ then shows that the resulting phase has no nontrivial intrinsic excitations modulo the local physical fermion. The order of $b$ is unchanged by the first condensation because $\langle a\rangle\cap\langle b\rangle=\{1\}$.

It remains to implement this fermionic ungauging on the lattice, retaining the required onsite fermionic symmetry.}

{The lattice realization places $\ZZ_{2^{n+1}}$ Pauli degrees of freedom on edges and $\ZZ_{2^n}^F$ fermionic clock variables $\{\Gamma,\Gamma',\Pi\}$ on plaquettes (Sec.~\ref{sec:fermionicclock}). Its stabilizer generators are}
\begin{align}
A_v' &=
\begin{tikzpicture}[baseline=-0.6ex, line width=0.5pt]
  \def\L{1.2}
  \draw (-\L,0) -- (\L,0);
  \draw (0,0) -- (0,-\L);
  \draw (0,0) -- (0,\L);
  \draw (0,\L) -- (\L,\L);
  \draw (\L,\L) -- (\L,0);
  \node[op] at (0,0) {$v$};
  \node[op] at (-0.5*\L,0) {$\textcolor{blue}{X^\dagger}$};
  \node[op] at (0,-0.5*\L) {$\textcolor{blue}{X^\dagger}$};
  \node[op] at (0,0.5*\L) {$\textcolor{blue}{XZ}$};
  \node[op] at (0.5*\L,0) {$\textcolor{blue}{XZ^\dagger}$};
  \node[op] at (0.5*\L,\L) {$\textcolor{blue}{Z}$};
  \node[op] at (\L,0.5*\L) {$\textcolor{blue}{Z^\dagger}$};
  \node[op] at (0.5*\L,0.5*\L) {$\textcolor{red}{\Pi^\alpha}$};
\end{tikzpicture}
,\quad
B_p =
\raisebox{-0.6cm}{\begin{tikzpicture}[baseline=-0.6ex, line width=0.5pt]
  \def\L{1.2}
  \draw (0,0) rectangle (\L,\L);
  \node[op] at (0.5*\L,\L) {$\textcolor{blue}{Z^{-2}}$};
  \node[op] at (0,0.5*\L) {$\textcolor{blue}{Z^{-2}}$};
  \node[op] at (\L,0.5*\L) {$\textcolor{blue}{Z^2}$};
  \node[op] at (0.5*\L,0) {$\textcolor{blue}{Z^2}$};
  \node[op] at (0.5*\L,0.5*\L) {$\textcolor{red}{P_p}$};
\end{tikzpicture}}
, \\
(-i)\cdot C_e &= ~~
\raisebox{-0.6cm}{\begin{tikzpicture}[baseline=-0.6ex, line width=0.5pt]
  \def\L{1.2}
  \draw (-\L,0) -- (0,0);
  \draw (0,0) -- (0,\L);
  \node[op] at (-0.5*\L,0) {$\textcolor{blue}{Z^{-2}}$};
  \node[op] at (-0.5*\L,0.5*\L) {$\textcolor{red}{\Gamma}$};
  \node[op] at (0.0*\L,0.5*\L) {$\textcolor{blue}{X_e^{2^{n-1}}}$};
  \node[op] at (0.5*\L, 0.53*\L) {$\textcolor{red}{\Gamma'}$};
\end{tikzpicture}}
,\qquad 
\raisebox{0.6cm}{\begin{tikzpicture}[baseline=-0.6ex, line width=0.5pt]
  \def\L{1.2}
  \draw (0,0) -- (\L,0);
  \draw (0,0) -- (0,-\L);
  \node[op] at (0.5*\L,0) {$\textcolor{blue}{X_e^{2^{n-1}}}$};
  \node[op] at (0.5*\L,0.5*\L) {$\textcolor{red}{\Gamma}$};
  \node[op] at (0,-0.5*\L) {$\textcolor{blue}{Z^2}$};
  \node[op] at (0.53*\L,-0.5*\L) {$\textcolor{red}{\Gamma'}$};
\end{tikzpicture}}
, \\
 D_e &= ~~
\raisebox{-0.6cm}{\begin{tikzpicture}[baseline=-0.6ex, line width=0.5pt]
  \def\L{1.2}
  \draw (-\L,0) -- (0,0);
  \draw (0,0) -- (0,\L);
  \node[op] at (-0.5*\L,0) {$\textcolor{blue}{Z^{2^n}}$};
  \node[op] at (0.0*\L,0.5*\L) {$\textcolor{blue}{X_e^{2^{n}}}$};
\end{tikzpicture}}
,\qquad 
\raisebox{0.6cm}{\begin{tikzpicture}[baseline=-0.6ex, line width=0.5pt]
  \def\L{1.2}
  \draw (0,0) -- (\L,0);
  \draw (0,0) -- (0,-\L);
  \node[op] at (0.5*\L,0) {$\textcolor{blue}{X_e^{2^{n}}}$};
  \node[op] at (0,-0.5*\L) {$\textcolor{blue}{Z^{2^n}}$};
\end{tikzpicture}}
\quad .
\end{align}
Here $\alpha=1-2^{n-2}$. {The onsite symmetry is generated by
\begin{equation}
U=\prod_p\Pi_p,\qquad
U^{2^{n-1}}=\prod_p P_p=(-1)^F.
\end{equation}
The clock relations show that $U$ commutes with the displayed stabilizers: in each ungauging term the two endpoint clock factors acquire opposite phases. For $n>3$, $\alpha$ is odd and we choose its inverse modulo the full clock order $2^n$,
\begin{equation}
\alpha^{-1}\equiv1+2^{n-2}\pmod{2^n}.
\end{equation}
This modulus ensures that $A_v'^{-\alpha^{-1}}$ cancels a bulk factor of $\Pi_p$ exactly.}

{To extract the supercohomology data, restrict $U$ to a region and multiply by $A_v'^{-\alpha^{-1}}$ in its interior. This cancels the bulk $\Pi_p$ factors and leaves the effective boundary symmetry $\widetilde{U}(1)$. Following the endpoint fusion and associator calculation of Sec.~\ref{sec:Z8F_SPT}, from Eq.~\eqref{eq:Z8F_boundary_effective_symmetry} to Eq.~\eqref{eq:Z8f_edge_defect}, gives the associator phase}
\begin{align}
    {\exp\bigl(2\pi i\nu_3(g,h,k)\bigr)}&=\exp \left(\frac{\pi\,i}{2} \, \alpha^{-1}g \left\lfloor\frac{ h+k}{2^{n-1}}\right\rfloor \right)  \\
    n_2(g,h)&=0 
\end{align}
where $g,h,k\in \ZZ_{2^{n-1}}$.
{This gives one explicit fSPT representative for each $n>3$.}

\section{Discussions}
\label{sec:discussions}

{
In this work, we introduced Majorana-Pauli stabilizer codes
and constructed lattice realizations of all fermionic twisted
quantum doubles with Abelian anyons, together with families
of fermionic symmetry-protected topological (SPT) phases
whose fully gauged theories are Abelian.
Our central example is the fermionic toric
code~\cite{GuWangWen14}, an intrinsically fermionic twisted
version of the toric code~\cite{Kitaev2006anyon}.
By combining $\ZZ_8$ generalized Pauli operators with
Majorana operators, we implement anyon condensation directly
at the stabilizer level on the square lattice
(Sec.~\ref{sec:fermionic_toric_code}).
This gives a simpler realization than the earlier string-net
construction~\cite{GuWangWen14}, with an explicit stabilizer
structure.
To accommodate more general fermionic symmetries, we
introduce fermionic clock and shift operators that extend
the Majorana algebra (Sec.~\ref{sec:fermionicclock}).
These operators enable a stabilizer realization of the
nontrivial $\ZZ_8^F$ SPT in Sec.~\ref{sec:Z8F_SPT},
a phase with no free-fermion realization.
}

{
These models are connected to bosonic topological orders
and other SPT and SET phases through the condensation
and gauging operations summarized in
Figs.~\ref{fig:Z_8_commutative_diagram}
and~\ref{fig:Z_16_commutative_diagram}.
The resulting duality webs retain a common stabilizer
description throughout. Their arrows relate distinct phases,
while the consistency of different routes to the same phase
is checked by comparing the resulting topological and
symmetry data.
Beyond these examples, Sec.~\ref{sec:general_constructions}
formulates the general condensation procedure and specifies
the SPT data realized by the split-symmetry and non-split
bosonic-shadow constructions, including explicit cyclic
lattice families.
Together, these results provide a basis for several
further developments.
}

{
A natural next step is to incorporate Kitaev-chain decorations
and thereby extend the SPT construction beyond supercohomology.
Our present models have vanishing decoration $n_1=0$.
When $n_1(g)=1$, the corresponding symmetry domain wall
carries a Kitaev chain, whose endpoints host Majorana zero
modes~\cite{WangGu2020Construction,Barkeshli2022Classification}.
In the nonchiral setting $c_-=0$, gauging fermion parity
produces a bosonic shadow with Abelian toric-code topological
order, in accordance with Kitaev's
16-fold way~\cite{Kitaev2006anyon}.
However, symmetry elements with $n_1(g)=1$ exchange the
$e$ and $m$ anyons and support non-Abelian symmetry defects.
Gauging the remaining symmetry promotes these defects to
deconfined anyons, yielding non-Abelian topological
order~\cite{Barkeshli2022Classification}.
The resulting fully gauged theories therefore fall outside
the class of Abelian parent topological orders used in our
present construction. Although commuting-projector
realizations of phases beyond supercohomology
exist~\cite{TarantinoFidkowski2016}, a corresponding
Majorana-Pauli stabilizer construction remains to be developed.
}

Although we have focused on stabilizer codes, the same condensation construction can be directly carried over to subsystem codes. We note that the construction of subsystem codes for bosonic Abelian anyon theories (more precisely, pointed braided fusion categories) is discussed extensively in Ref.~\cite{Ellison2023subsystem}. The fermionic case is similar. Recall that a stabilizer code is defined by an Abelian group of commuting checks $\mathcal{S}$. By contrast, a subsystem code is defined by a non-commuting group of checks $\mathcal{G}$. The stabilizer group is the center of the check group,
\begin{equation}
    \mathcal{S}_\text{subsystem} =\mathcal{Z}_\mathcal{G}(\mathcal{G} ) \, ,
\end{equation}
while the non-central elements of $\mathcal{G}$ act only on gauge degrees of freedom.

To construct a subsystem code corresponding to a fermionic anyon theory, potentially with degenerate braiding, we add a subset of short string operators to the check group. The logical strings that survive are those that still commute with the full check group, while not being generated by it. The subsystem codes could allow for constructions of Majorana-Pauli subsystem codes that correspond to chiral Abelian fermionic phases. Interestingly, on the bosonic side, recent work has also related such subsystem codes to mixed-state topological order, by viewing the non-commuting checks as a local decoherence channel \cite{Ellison25, Sohal25}. Thus, in such a construction, it is possible to decohere anyons that do not have bosonic statistics, and at maximal decoherence arrive at a mixed-state topological order corresponding to chiral fermion theories, such as invertible phases within the 16-fold way~\cite{Kitaev2006anyon}, or other fermionic braided fusion categories. We leave the detailed analysis to future work.

In Sec.~\ref{sec:general_constructions}, we described several generalizations of the Majorana-Pauli stabilizer constructions to gapped Abelian fermionic topological phases. In particular, we obtained stabilizer realizations of all Abelian fermionic twisted gauge theories. However, a systematic stabilizer construction for fermionic symmetry-enriched topological phases remains open. Developing a more complete stabilizer framework for fermionic phases, together with their associated condensation, gauging, and ungauging duality webs, is a natural direction for future work.

Another useful direction is to develop a polynomial representation for translation-invariant Majorana-Pauli stabilizer codes. For ordinary Pauli stabilizer codes, Laurent-polynomial methods provide a compact algebraic description of stabilizer generators, commutation relations, logical operators, and boundaries~\cite{haah_module_13, haah2016algebraic, liang2023extracting}. A similar polynomial formalism exists for translation invariant Majorana fermion codes~\cite{VijayHaahFu2015}. In the present hybrid setting, the commutation form should combine the symplectic pairing of the Pauli/qudit sector with the Euclidean pairing that records the fermionic signs from Majorana operators. Such a mixed symplectic--Euclidean polynomial formalism could provide a systematic language for classifying Majorana-Pauli stabilizer models, analyzing their excitations and logical operators, and implementing condensation, gauging, and ungauging directly at the algebraic level, generalizing Ref.~\cite{Tantivasadakarn20}.

Our stabilizer construction also has a natural interpretation in terms of fermionic tensor networks. Since all stabilizer terms in our Majorana-Pauli Hamiltonians are fermion-parity even, each local projector $\Pi_\alpha$ for $\alpha \in \mathcal{I}$, e.g. Eq.~\eqref{eq:projector_Hamiltonian_fTC}, defines an even fermionic tensor. The product of these local projectors therefore gives a fermionic tensor network representation of the groundspace projector, and, after fixing boundary conditions or applying the projector to a simple reference state, of a ground-state wavefunction. It would be interesting to explicitly write down tensor network representations of these wavefunctions. One way to do so would be to use the tensor-network description for bosonization to map a bosonic wavefunction into a fermionic one~\cite{Shukla20,Tantivasadakarn24,Obrien25}. Also, a recently developed graphical calculus, extending the ZX-calculus to fermionic modes \cite{ren2025graphicalcalculusfermionictensors} could be helpful. Extending this calculus to the Majorana-Pauli operators we developed, especially the fermionic clock operators would be an interesting direction.

Throughout this paper, although we worked with the square lattice, it is clear how to generalize this to an arbitrary surface beyond the torus with arbitrary triangulations. This can be done by using the general bosonization procedure of Ref.~\cite{CKR18} employing the cup product formalism and explicitly keeping track of the choice of spin structure. {For fermionic clock models with symmetry $G_f$, identifying the corresponding generalized spin structure remains an interesting direction for future work.}

Finally, since our models are generalizations of both Pauli stabilizer codes and Majorana quantum codes, it would be interesting to study their implementation for quantum simulation of fermionic phases of matter, as well as their practical usage as building blocks for quantum error correction and quantum computation in systems with native physical fermions.

\begin{acknowledgments}
We are especially grateful to Tyler Ellison for helpful discussions during the early stages of this work. Z.W. and {Y.-A.C.} are supported by the National Natural Science Foundation of China (Grant No.~12474491) and the Central University Fundamental Research Funds (Peking University).
\end{acknowledgments}

\appendix

\section{Fermionic \texorpdfstring{$K$}{K}-matrices}
\label{sec:K_matrix_fermionic_condition}

In this appendix, we explain the condition for a $K$-matrix to describe a fermionic order. In an Abelian $K$-matrix theory, an integer vector $\ell$ labels an excitation, with topological spin
\begin{equation}
    h_{\ell}
    =
    \frac12 \ell^T K^{-1}\ell
    \in \RR/\ZZ .
\end{equation}
Two vectors that differ by a vector of the form $K\Lambda$, with
$\Lambda\in \ZZ^r$, describe excitations that differ by a topologically trivial excitation. Indeed, for any excitation $\ell$,
\begin{equation}
    (K\Lambda)^T K^{-1}\ell
    =
    \Lambda^T \ell
    \in \ZZ .
\end{equation}
Therefore, $K\Lambda$ has trivial mutual braiding with every excitation.

However, a topologically trivial excitation does not have to be bosonic in a fermionic theory. Its spin is
\begin{equation}
    h_{K\Lambda}
    =
    \frac12 (K\Lambda)^T K^{-1}(K\Lambda)
    =
    \frac12 \Lambda^T K\Lambda .
    \label{eq:local_vector_spin}
\end{equation}
Thus, the question is whether $\Lambda^T K\Lambda$ can be odd. Expanding this
quadratic form gives
\begin{equation}
    \Lambda^T K\Lambda
    =
    \sum_a K_{aa}\Lambda_a^2
    +
    2\sum_{a<b} K_{ab}\Lambda_a\Lambda_b .
\end{equation}
The off-diagonal terms are always even because of the explicit factor of $2$. Hence, they cannot change whether $\Lambda^T K\Lambda$ is even or odd. The parity is controlled only by the diagonal entries of $K$
\begin{equation}
    \Lambda^T K\Lambda
    =
    \sum_a K_{aa}\Lambda_a
    \quad \text{mod } 2 .
\end{equation}

It follows that if every diagonal entry of $K$ is even, then $\Lambda^T K\Lambda$ is even for every integer vector $\Lambda$. In that case every topologically trivial excitation has integer spin and is bosonic. By contrast, if some diagonal entry $K_{aa}$ is odd, then choosing $\Lambda=r_a$, the $a$-th standard basis vector, gives a topologically trivial excitation $Kr_a$ with
\begin{equation}
    h_{Kr_a}
    =
    \frac{1}{2} r_a^T K \, r_a
    =
    \frac{1}{2} K_{aa}
    =
    \frac{1}{2} \quad \text{mod } 1 .
\end{equation}
Thus, $K r_a$ is a transparent fermion: it has trivial mutual braiding with all excitations but has fermionic spin.

Therefore, an integral Abelian $K$-matrix is fermionic precisely when at least one diagonal entry is odd. That is, a $K$-matrix with an odd diagonal defines a spin Chern--Simons theory. The presence of a transparent fermionic local excitation means that the theory must be defined with a choice of spin structure.

\section{Assumptions on the fermionic twisted quantum double and the fermionic symmetry group}\label{app: Assumptions on the fermionic twisted quantum double}
{In Sec.~\ref{sec:general_fermionic_topological_orders}, we used a
particular normal form for the matrices $N$ and $S_{\mathcal I}$ that
define a fermionic twisted quantum double.  This Appendix proves the
normal-form statement carefully.  There are three logically distinct
steps. The first and most basic argument is that every Abelian fermionic topological order with gapped boundary can be stably represented by the normal form. Then, we prove that after stacking with a trivial local-fermion block, odd
diagonal entries of $S_{\mathcal I}$ on odd-order cyclic factors can be
removed.  Finally, if any odd diagonal entries remain, an integral
congruence can combine them into one while leaving $N$ in its original
invariant-factor order.  We also explain why the surviving odd entry
cannot in general be required to occur on the first invariant factor.}

\subsection{{Boundary-adapted form for Abelian fermionic topological orders}}
\label{sec:boundary_adapted_normal_form}

{We now show that every Abelian fermionic topological order admitting a fully gapped boundary can, up to stable equivalence, be represented by a \(K\)-matrix of the form
\[
K_{\mathrm{fTQD}}
=
\begin{pmatrix}
0 & N\\
N & -S
\end{pmatrix},
\qquad
N=\operatorname{diag}(n_1,\ldots,n_M),
\]
where \(S\) is an integral symmetric matrix. 

We use additive notation for fusion. Let \(\mathcal A\) denote the finite Abelian group of topological sectors modulo local excitations, and let
\[
B:\mathcal A\times\mathcal A\longrightarrow \mathbb Q/\mathbb Z
\]
be the nondegenerate symmetric braiding pairing. A gapped boundary determines a Lagrangian subgroup
\[
\mathcal L\subset \mathcal A,
\qquad
\mathcal L=\mathcal L^\perp ,
\]
whose elements can be represented by mutually local bosons and condensed at the boundary. Choose the invariant-factor decomposition of this finite Abelian subgroup, with $n_1\mid\cdots\mid n_M$,
\[
\mathcal L
\simeq
\bigoplus_{i=1}^{M}\mathbb Z_{n_i},
\qquad
\mathcal L=\langle c_1,\ldots,c_M\rangle ,
\qquad
n_i c_i=0 .
\]

Because the braiding pairing is nondegenerate, it defines a canonical homomorphism
\[
\mathcal A/\mathcal L\longrightarrow \widehat{\mathcal L},
\qquad
[a]\longmapsto
\bigl(c\mapsto B(c,a)\bigr).
\]
Its kernel is \(\mathcal L^\perp/\mathcal L=0\), and \(|\mathcal A|=|\mathcal L|^2\) because \(\mathcal L\) is Lagrangian. Hence the map is an isomorphism,
\[
\mathcal A/\mathcal L\simeq\widehat{\mathcal L}
\simeq
\bigoplus_{i=1}^{M}\mathbb Z_{n_i}.
\]
We may therefore choose lifts \(\phi_i\in\mathcal A\) of the standard dual generators such that
\[
B(c_i,c_j)=0,
\qquad
B(c_i,\phi_j)=\frac{\delta_{ij}}{n_i}
\quad (\mathrm{mod}\ 1).
\]
The elements \(c_i\) and \(\phi_i\) together generate \(\mathcal A\).

Since the image of \(\phi_i\) in \(\mathcal A/\mathcal L\) has order \(n_i\), we have \(n_i\phi_i\in\mathcal L\) modulo a local excitation. Isotropy of \(\mathcal L\) then gives \(n_i n_j B(\phi_i,\phi_j)=0\pmod 1\), so there exist integers \(S_{ij}\) such that
\[
B(\phi_i,\phi_j)
=
\frac{S_{ij}}{n_i n_j}
\quad(\mathrm{mod}\ 1).
\]
Symmetry of \(B\) allows us to choose \(S_{ij}=S_{ji}\), so \(S=(S_{ij})\) is an integral symmetric matrix. We also match the spins of chosen representatives of the fluxes. Write \(\theta_{\phi_i}=e^{2\pi i h_{\phi_i}}\). Since \(n_i\phi_i\) is a boundary charge up to a local fermion, \(2n_i^2 h_{\phi_i}\) is integral. We can choose the diagonal representatives so that
\[
S_{ii}=2n_i^2h_{\phi_i}\pmod{2n_i^2}.
\]
This choice reproduces both the self-braiding and the spin of \(\phi_i\); attaching the physical fermion changes the spin by a sign, as expected. Together with the bosonic representatives of the \(c_i\), these choices fix the quadratic data of all generated excitations.

Consider now
\[
d_i
=
n_i\phi_i-\sum_{j=1}^{M}S_{ji}c_j .
\]
Its braiding with every \(c_k\) is trivial:
\[
B(d_i,c_k)
=
n_i B(\phi_i,c_k)
=
\delta_{ik}
=
0
\quad(\mathrm{mod}\ 1),
\]
while its braiding with every \(\phi_k\) is
\[
\begin{aligned}
B(d_i,\phi_k)
&=
n_i B(\phi_i,\phi_k)
-\sum_j S_{ji}B(c_j,\phi_k)
\\
&=
\frac{S_{ik}}{n_k}
-\frac{S_{ki}}{n_k}
=0
\quad(\mathrm{mod}\ 1).
\end{aligned}
\]
Since the \(c_k\) and \(\phi_k\) generate \(\mathcal A\) and \(B\) is nondegenerate, \(d_i\) is a local excitation. In the fermionic theory the only relevant distinction among local excitations is their fermion parity, so we can write
\[
n_i\phi_i-\sum_j S_{ji}c_j
=
\epsilon_i\psi ,
\qquad
\epsilon_i\in\mathbb Z_2,
\]
where \(\psi\) denotes the physical local fermion. Since the charge combination has trivial spin and its braiding with \(n_i\phi_i\) is integral, the spin of this local relation is
\[
h_{d_i}=n_i^2h_{\phi_i}=\frac{S_{ii}}2\pmod 1,
\qquad \epsilon_i=S_{ii}\pmod 2.
\]
Thus the parity of the \(i\)-th local fusion relation is precisely encoded by the diagonal entry \(S_{ii}\).

These fusion relations are exactly those generated by the columns of
\[
K_{\mathrm{fTQD}}
=
\begin{pmatrix}
0 & N\\
N & -S
\end{pmatrix}.
\]
Indeed, ordering the generators as
\[
(\phi_1,\ldots,\phi_M,c_1,\ldots,c_M),
\]
the first \(M\) columns impose
\[
n_i c_i=0,
\]
while the remaining \(M\) columns impose
\[
n_i\phi_i-\sum_j S_{ji}c_j
=\text{local}.
\]

The proposed matrix is nondegenerate and has
\[
|\det K_{\mathrm{fTQD}}|=(\det N)^2
=|\mathcal L|^2=|\mathcal A|.
\]
The map from its generators to the chosen charges and fluxes is surjective and respects the displayed local relations. Equality of orders makes it an isomorphism, so no additional topological sectors are introduced. The spins and mutual braiding agree by construction. Finally, the gapped-boundary assumption excludes a residual chiral invertible phase; the block matrix also has signature zero. This establishes the boundary-adapted representative at the level of the fermionic topological phase, up to trivial local-fermion blocks. The explicit integral congruences needed to reduce this representative further are given below.}

\subsection{{Invariant-factor and diagonal-presentation conventions}}
\label{sec:invariant_factor_conventions}

{It is useful first to distinguish two notions that are sometimes both
called a cyclic decomposition.  A \emph{diagonal cyclic presentation}
has
\begin{equation}
    N=\operatorname{diag}(n_1,\ldots,n_M),
\end{equation}
but places no divisibility condition on the $n_i$.  An
\emph{invariant-factor presentation} is the more restrictive Smith
normal form
\begin{equation}
    N=\operatorname{diag}(n_1,\ldots,n_M),
    \qquad
    1\leq n_1\mid n_2\mid\cdots\mid n_M .
    \label{eq:appendix_invariant_factor_N}
\end{equation}
Permuting unequal diagonal entries preserves a diagonal cyclic
presentation, but generally destroys the displayed invariant-factor
ordering.  Throughout this Appendix, $N$ means the ordered matrix in
Eq.~\eqref{eq:appendix_invariant_factor_N}.

For compactness, define
\begin{equation}
    K(N,S)
    =
    \begin{pmatrix}
        0_{M\times M} & N\\
        N & -S
    \end{pmatrix},
    \qquad S=S^T\in\operatorname{Mat}_M(\ZZ).
    \label{eq:KNS_notation}
\end{equation}
For $U,V\in GL_M(\ZZ)$, let
\begin{equation}
    W(U,V)
    =
    \begin{pmatrix}
        U^T & 0\\
        0 & V
    \end{pmatrix}.
\end{equation}
A direct block multiplication gives
\begin{equation}
    W(U,V)^T K(N,S)W(U,V)
    =
    \begin{pmatrix}
        0 & UNV\\
        V^TNU^T & -V^TSV
    \end{pmatrix}.
    \label{eq:general_UV_congruence}
\end{equation}
In particular, whenever $UNV=N$, this congruence changes $S$ to
$V^TSV$ while leaving the invariant-factor matrix $N$ exactly
unchanged.

The invariant-factor convention also explains the integrality
condition used in the lattice construction.  If $i>j$, then
$n_j\mid n_i$, and consequently
\begin{equation}
    \frac{2n_i s_{ij}}{n_j}
    =
    2\left(\frac{n_i}{n_j}\right)s_{ij}
    \in\ZZ .
    \label{eq:appendix_integrality_from_IF}
\end{equation}
Thus this condition is automatic in the invariant-factor
presentation; it is not an additional restriction on $S$.}

{
\subsection{{Removing odd diagonal entries on odd-order factors}}
\label{sec:noodd}

We next show that an odd diagonal entry $s_i$ associated with an odd
$n_i$ can be made even by a stable integral congruence.  Consider
\begin{equation}
    K
    =
    K(N,S_{\mathcal I})\oplus K_f,
    \qquad
    K_f
    =
    \begin{pmatrix}
        1 & 0\\
        0 & -1
    \end{pmatrix},
\end{equation}
The unimodular, zero-signature block $K_f$ describes a trivial pair of
local fermionic invertible phases.  Stacking with this block and
performing an integral congruence is the stable equivalence used below.

Let
\begin{equation}
    t_i=(S_{\mathcal I})_{ii}\pmod 2,
    \qquad
    t=(t_1,\ldots,t_M)^T,
\end{equation}
where each $t_i$ is represented by either $0$ or $1$.  To modify only
the unwanted odd-order entries, define
\begin{equation}
    \epsilon_i
    =
    \begin{cases}
        t_i, & n_i\text{ odd},\\
        0, & n_i\text{ even},
    \end{cases}
    \qquad i=1,\ldots,M .
\end{equation}
Collect these entries into the column vector $\epsilon$ and define the
row vector $\eta$ by
\begin{align}
    \epsilon&=(\epsilon_1,\ldots,\epsilon_M)^T,
    \nonumber\\
    \eta&=\epsilon^T N
    =(\epsilon_1n_1,\ldots,\epsilon_Mn_M).
    \label{eq:epsilon_eta_stable}
\end{align}\noindent
Consider the integer matrix
\begin{equation}
    W
    =
    \begin{pmatrix}
        \mathbb{I}_M & 0_{M\times M} & 0_{M\times 1} & \epsilon\\
        0_{M\times M} & \mathbb{I}_M & 0_{M\times 1} & 0_{M\times 1}\\
        0_{1\times M} & 0_{1\times M} & 1 & 0\\
        0_{1\times M} & \eta & 0 & 1
    \end{pmatrix}.
    \label{eq:stable_W_odd_ni}
\end{equation}
Subtracting from the last row the linear combination $\eta$ of the
second block of rows turns $W$ into an upper-triangular matrix with
unit diagonal.  Therefore, $\det W=1$ and
$W\in GL_{2M+2}(\ZZ)$.  A direct block multiplication gives
\begin{equation}
    W^T K W
    =
    \begin{pmatrix}
        0_{M\times M} & N\\
        N & -S'_{\mathcal{I}}
    \end{pmatrix}
    \oplus K_f,
    \label{eq:stable_K_transform}
\end{equation}
where
\begin{equation}
    S'_{\mathcal{I}}
    =
    S_{\mathcal{I}}+\eta^T\eta .
\end{equation}
The potentially new mixed block vanishes because
\begin{equation}
    N\epsilon-\eta^T=0,
\end{equation}
which is precisely the reason for the definition
$\eta=\epsilon^T N$.
In components,
\begin{equation}(S'_{\mathcal{I}})_{ii}
    =
    s_i+\epsilon_i n_i^2,
    \label{eq:stable_S_diag}
\end{equation}
and
\begin{equation}
    (S'_{\mathcal{I}})_{ij}
    =
    s_{ij}+\epsilon_i\epsilon_j n_i n_j .
    \label{eq:stable_S_offdiag}
\end{equation}

Eq.~\eqref{eq:stable_S_diag} gives the parity update
\begin{equation}
    (S'_{\mathcal I})_{ii}
    \equiv
    \begin{cases}
        t_i+t_i n_i^2\equiv0, & n_i\text{ odd},\\
        t_i, & n_i\text{ even},
    \end{cases}
    \pmod 2 .
    \label{eq:stable_parity_update}
\end{equation}
Thus every odd diagonal entry on an odd-order factor is removed, while
the parity of every even-order factor is left unchanged.  Notice also
that the matrix $N$ itself has not changed at all, so its
invariant-factor ordering is preserved.

There are now two cases.  If every diagonal entry of
$S'_{\mathcal I}$ is even, then the non-unimodular gauge block is a
bosonic twisted quantum double; the stacked $K_f$ still supplies the
transparent local fermion.  This branch can be realized by the bosonic
construction of Ref.~\cite{Ellison22} together with decoupled local
fermions.  Otherwise, the remaining parity vector is nonzero, and
every index on which it is supported has even $n_i$.  The next
subsection treats precisely this intrinsically fermionic gauge-block
branch.
}

\subsection{{One-odd-diagonal normal form at fixed invariant factors}}
\label{sec:single_odd_diagonal_fTQD}

{After the stable step above, relabel the resulting symmetric matrix as
$S_{\mathcal I}$ and define its diagonal-parity vector by
\begin{equation}
    t_i=(S_{\mathcal I})_{ii}\pmod2,
    \qquad
    t=(t_1,\ldots,t_M)^T .
    \label{eq:parity_vector_after_stable}
\end{equation}
When more than one symmetric matrix is under discussion, we write this
vector as $t(S_{\mathcal I})$.
In the branch considered here,
\begin{equation}
    t\neq0,
    \qquad
    t_i=1\ \Longrightarrow\ n_i\text{ is even}.
    \label{eq:odd_s_even_n}
\end{equation}

We will prove that there are $U,V\in GL_M(\ZZ)$ satisfying
\begin{equation}
    UNV=N
    \label{eq:fixed_N_UV}
\end{equation}
such that $S'_{\mathcal I}=V^TS_{\mathcal I}V$ has exactly one odd
diagonal entry.  The equality in Eq.~\eqref{eq:fixed_N_UV} is stronger
than merely preserving the Smith invariants: it keeps the displayed
invariant-factor matrix, including its order, exactly fixed.}

\subsubsection{{Transformation law for diagonal parity}}
\label{sec:diagonal_parity_transform}

{Let $V\in GL_M(\ZZ)$ and set
\begin{equation}
    S'_{\mathcal I}=V^TS_{\mathcal I}V .
\end{equation}
If $v_a$ denotes the $a$th column of $V$, then symmetry of
$S_{\mathcal I}$ gives
\begin{align}
    (S'_{\mathcal I})_{aa}
    &=
    v_a^TS_{\mathcal I}v_a
    \nonumber\\
    &=
    \sum_k (S_{\mathcal I})_{kk}(v_a)_k^2
    +
    2\sum_{k<\ell}
    (S_{\mathcal I})_{k\ell}(v_a)_k(v_a)_\ell
    \nonumber\\
    &\equiv
    \sum_k t_k(v_a)_k
    =
    (V^Tt)_a
    \pmod2 .
    \label{eq:diagonal_parity_transform}
\end{align}
In the last line, we used $x^2\equiv x\pmod2$ for every integer $x$.
Therefore,
\begin{align}
    t(S'_{\mathcal I})
    &=V^Tt(S_{\mathcal I}),
    \nonumber\\
    t(S'_{\mathcal I})^T
    &=t(S_{\mathcal I})^TV .
    \label{eq:parity_covector_transform}
\end{align}}

\subsubsection{{Pairwise reduction at fixed \texorpdfstring{$N$}{N}}}
\label{sec:Single_odd_diagonal_U_V_existence}

{Suppose the current parity vector has two nonzero components
$t_i=t_j=1$, with $i<j$.  Since $N$ is in invariant-factor order,
$n_i\mid n_j$, so there is a positive integer
\begin{equation}
    q=\frac{n_j}{n_i}.
    \label{eq:q_ratio_pair}
\end{equation}
We construct a pair of unimodular matrices acting only on coordinates
$i$ and $j$.  All unshown entries agree with the identity matrix.

If $q$ is odd, take the following $2\times2$ blocks:
\begin{equation}
    U_{\mathrm{o}}
    =
    \begin{pmatrix}
        1&1\\
        0&1
    \end{pmatrix},
    \qquad
    V_{\mathrm{o}}
    =
    \begin{pmatrix}
        1&-q\\
        0&1
    \end{pmatrix}.
    \label{eq:odd_q_pair_matrices}
\end{equation}
Both determinants are $1$.  Using $n_j=qn_i$, one obtains
\begin{equation}
    \begin{pmatrix}
        n_i&0\\
        0&n_j
    \end{pmatrix}
    V_{\mathrm{o}}
    =
    \begin{pmatrix}
        n_i&-n_j\\
        0&n_j
    \end{pmatrix}
\end{equation}
and hence
\begin{equation}
    U_{\mathrm{o}}
    \begin{pmatrix}
        n_i&0\\
        0&n_j
    \end{pmatrix}
    V_{\mathrm{o}}
    =
    \begin{pmatrix}
        n_i&0\\
        0&n_j
    \end{pmatrix}.
    \label{eq:odd_q_preserves_N}
\end{equation}
On the parity pair,
\begin{equation}
    (1,1)V_{\mathrm{o}}
    =
    (1,1-q)
    \equiv(1,0)
    \pmod2 ,
    \label{eq:odd_q_reduces_t}
\end{equation}
because $q$ is odd.

If $q$ is even, instead take
\begin{equation}
    U_{\mathrm{e}}
    =
    \begin{pmatrix}
        1&1\\
        q&q+1
    \end{pmatrix},
    \qquad
    V_{\mathrm{e}}
    =
    \begin{pmatrix}
        q+1&-q\\
        -1&1
    \end{pmatrix}
    .
    \label{eq:even_q_pair_matrices}
\end{equation}
Again, both determinants are $1$.  This time
\begin{equation}
    \begin{pmatrix}
        n_i&0\\
        0&n_j
    \end{pmatrix}
    V_{\mathrm{e}}
    =
    \begin{pmatrix}
        n_i+n_j&-n_j\\
        -n_j&n_j
    \end{pmatrix},
\end{equation}
so
\begin{equation}
    U_{\mathrm{e}}
    \begin{pmatrix}
        n_i&0\\
        0&n_j
    \end{pmatrix}
    V_{\mathrm{e}}
    =
    \begin{pmatrix}
        n_i&0\\
        0&n_j
    \end{pmatrix},
    \label{eq:even_q_preserves_N}
\end{equation}
where the lower-left entry vanishes because $qn_i=n_j$.  The parity
pair transforms as
\begin{equation}
    (1,1)V_{\mathrm{e}}
    =
    (q,1-q)
    \equiv(0,1)
    \pmod2 ,
    \label{eq:even_q_reduces_t}
\end{equation}
because $q$ is even.

In either parity case, the move replaces two nonzero components of
$t$ by one, leaves every other component unchanged, and leaves $N$
itself unchanged.  Both active factors have even order by
Eq.~\eqref{eq:odd_s_even_n}, so the surviving nonzero component is
still supported on an even-order factor.  Repeating the move reduces
the number of nonzero components by one at every step and therefore
terminates after finitely many steps.  The products of the embedded
pair matrices give $U,V\in GL_M(\ZZ)$ such that
\begin{equation}
    UNV=N,
    \qquad
    t^TV=r_{i_\star}^T
    \pmod2
    \label{eq:fixed_N_single_parity}
\end{equation}
for some distinguished index $i_\star$, where $r_{i_\star}$ is the
$i_\star$th standard basis vector and $n_{i_\star}$ is even.

Combining Eqs.~\eqref{eq:general_UV_congruence},
\eqref{eq:parity_covector_transform}, and
\eqref{eq:fixed_N_single_parity}, we obtain
\begin{align}
    W(U,V)^T K(N,S_{\mathcal I})W(U,V)
    &=K(N,S'_{\mathcal I}),
    \nonumber\\
    S'_{\mathcal I}&=V^TS_{\mathcal I}V,
    \label{eq:single_odd_transformed_K}
\end{align}}\noindent
where
{\begin{equation}
    (S'_{\mathcal I})_{i_\star i_\star}
    \text{ is odd},
    \qquad
    (S'_{\mathcal I})_{ii}
    \text{ is even for }i\neq i_\star .
    \label{eq:single_odd_final}
\end{equation}
This proves that the one-odd-diagonal normal form is compatible with
the invariant-factor decomposition.}

\subsubsection{{Final normal-form statement}}
\label{sec:final_fTQD_normal_form}

{Combining the stable transformation in Appendix~\ref{sec:noodd} with
the fixed-$N$ reduction above gives the following dichotomy:
\begin{equation}
    K(N,S_{\mathcal I})\oplus K_f
    \cong
    K(N,S_{\mathcal I}^{\star})\oplus K_f ,
    \label{eq:final_stable_normal_form}
\end{equation}
where $N$ remains in the original invariant-factor order and either
\begin{enumerate}
    \item every diagonal entry of $S_{\mathcal I}^{\star}$ is even; or
    \item there is a unique $i_\star$ such that
    $(S_{\mathcal I}^{\star})_{i_\star i_\star}$ is odd, all other
    diagonal entries are even, and $n_{i_\star}$ is even.
\end{enumerate}
The first case is the bosonic twisted gauge block stacked with local
fermions.  The intrinsically fermionic construction in
Sec.~\ref{sec:general_fermionic_topological_orders} uses the second
case.  Since $N$ has never been permuted, its invariant-factor
ordering and the automatic integrality property {in Eq.}~
\eqref{eq:appendix_integrality_from_IF} remain valid.}

\subsection{{Reduction of an Abelian fermion-parity extension}}
\label{sec:sufficiency_non_split_ZF2n}

{This subsection shows that the fermion-parity extension of a finite Abelian fermionic symmetry can be supported on at most one cyclic factor. This is a statement about the symmetry group, not about completeness of the cyclic fSPT construction in Sec.~\ref{sec:general_fermionic_SPTs}.} We first take the central extension for a fermionic symmetry ${G_f}$
\begin{eqs}
    1 \rightarrow \ZZ_2^F
    \rightarrow {G_f}
    \rightarrow G_b
    \rightarrow 1.
\end{eqs}
Since ${G_f}$ is assumed to be Abelian, the corresponding extension class is an element of
\begin{eqs}
    \operatorname{Ext}_{\ZZ}^1(G_b,\ZZ_2^F).
\end{eqs}
Using the primary decomposition theorem for finite Abelian groups, we may write
\begin{eqs}
    G_b
    \cong
    \GG_{\mathrm{odd}}
    \oplus
    \bigoplus_{\alpha=1}^{r} \ZZ_{2^{q_\alpha}},
\end{eqs}
where
\begin{eqs}
    \GG_{\mathrm{odd}}
    =
    \bigoplus_{p \ {\rm odd}}
    \bigoplus_j \ZZ_{p^{s_{p,j}}}
\end{eqs}
contains all odd-primary cyclic factors. By additivity of the $\operatorname{Ext}$ functor in its first argument, the extension class decomposes as
\begin{equation}
\begin{aligned}
    w_2 \in \, &\operatorname{Ext}_{\ZZ}^1(G_b,\ZZ_2^F) \\
    &\cong
    \operatorname{Ext}_{\ZZ}^1(\GG_{\mathrm{odd}},\ZZ_2^F)
    \oplus
    \bigoplus_{\alpha=1}^{r} \operatorname{Ext}_{\ZZ}^1 \bigl(\ZZ_{2^{q_\alpha}},\ZZ_2^F\bigr).
\end{aligned}
\end{equation}
For a cyclic factor, one has
\begin{eqs}
    \operatorname{Ext}_{\ZZ}^1(\ZZ_M,\ZZ_2^F)
    \cong
    \ZZ_{\gcd(M,2)}.
\end{eqs}
Consequently,
\begin{eqs}
    \operatorname{Ext}_{\ZZ}^1 \bigl(\ZZ_{p^s},\ZZ_2^F\bigr) = 0
    \quad \text{for } p \ \text{ odd} \, ,
\end{eqs}
whereas
\begin{eqs}
    \operatorname{Ext}_{\ZZ}^1
    \bigl(\ZZ_{2^q},\ZZ_2^F\bigr)
    \cong
    \ZZ_2 \, .
\end{eqs}
It then follows that the restriction of the fermionic extension to every odd-primary cyclic factor is necessarily split. Therefore, the odd-primary part of $G_b$ does not contribute to the nontrivial embedding of fermion parity into the full symmetry group.

It remains to consider the possibility that the extension class is non-trivial on more than one $2$-primary cyclic factor. Suppose that $\widetilde g_\alpha$ and $\widetilde g_\beta$ lift generators of $\ZZ_{2^{q_\alpha}}$ and $\ZZ_{2^{q_\beta}}$, respectively, and that
\begin{eqs}
    \widetilde g_\alpha^{\,2^{q_\alpha}}
    =
    \widetilde g_\beta^{\,2^{q_\beta}}
    =
    (-1)^F.
\end{eqs}
Assume without loss of generality that $q_\alpha\geq q_\beta$. We may redefine the second lifted generator by
\begin{eqs}
    \widetilde g_\beta'
    =
    \widetilde g_\beta
    \widetilde g_\alpha^{-2^{q_\alpha-q_\beta}}.
\end{eqs}
Since ${G_f}$ is Abelian, this new generator satisfies
\begin{eqs}
    \bigl(\widetilde g_\beta'\bigr)^{2^{q_\beta}}
    =
    \widetilde g_\beta^{\,2^{q_\beta}}
    \widetilde g_\alpha^{-2^{q_\alpha}}
    =
    1.
\end{eqs}
Thus, after an appropriate change of cyclic generators, the extension can be made nontrivial on at most one $2$-primary cyclic factor.

Therefore, if the extension is non-split, there exists a choice of generators such that
\begin{eqs}
    {G_f}
    \cong
    \ZZ_{2^n}^F
    \oplus
    \GG_{\mathrm{split}},
\end{eqs}
where $\GG_{\mathrm{split}}$ consists of ordinary bosonic symmetry factors whose
lifts commute with and are independent of fermion parity. Equivalently, the
bosonic quotient takes the form
\begin{eqs}
    G_b
    \cong
    \ZZ_{2^{n-1}}
    \oplus
    \GG_{\mathrm{split}}.
\end{eqs}
{Hence one non-split cyclic factor suffices to represent the fermion-parity extension.}

We emphasize that this statement concerns the structure of the fermion-parity extension itself. {The remaining split factors can still support fermion decorations in $n_2$, and the quotient can support homogeneous bosonic responses involving one or several factors. Nonzero self-decoration data additionally require the term in Eq.~\eqref{eq:nonsplit_cross_term}. The group-theoretic reduction alone does not realize these responses. Their bosonic-shadow construction, including the onsite symmetry and defect-associator data, is given in Sec.~\ref{sec:nonsplit_spt_scope} and Appendix~\ref{app:nonsplit_bosonic_shadow}.}

\Needspace{8\baselineskip}
\section{{Additional anyon-level checks for the \texorpdfstring{$\ZZ_8$}{Z8} duality web}}
\label{app:Z8_duality_condensation_checks}

{This appendix supplies the anyon-level checks for the two bosonic ungauging routes summarized in Sec.~\ref{sec:Z4F_duality_web}, together with the change of basis used for the column-vector labels in Fig.~\ref{fig:Z_8_commutative_diagram}.

We first consider the route from $U(1)_4\times U(1)_{-4}$ to the $\ZZ_2$ symmetry-enriched $\ZZ_2$ toric code. This point of view is also discussed in detail in Ref.~\cite{LuVishwanath16}. The lattice realization and its stable equivalence to the fermion-parity-gauging construction are proved in Appendix~\ref{app:stable_equivalence_Z2_SET}.
Abstractly, this is obtained by ungauging the order-two charge subgroup generated by the boson
\begin{equation}
    q^2=x^2y^2 =
    \begin{pmatrix}
        2\\2
    \end{pmatrix}_{\rm diag}.
\end{equation}
Here,
\begin{equation}
    q=xy=
    \begin{pmatrix}
        1\\1
    \end{pmatrix}_{\rm diag}
\end{equation}
is the parent $\ZZ_4$ gauge charge. Ungauging only the subgroup generated by $q^2$ is a partial ungauging of the original $\ZZ_4$ gauge theory: it removes the order-two subgroup generated by $q^2$, but leaves a residual $\ZZ_2$ gauge theory together with a dual global $\ZZ_2$ symmetry. In the original $\ZZ_8$ toric-code notation
\begin{equation}
    x^2y^2=[e^4]=[m^4] \, ,
\end{equation}
where the equality uses the previous condensation of $\beta=e^4m^4$.

An anyon $x^a y^b=(a,b)_{\rm diag}$ remains deconfined after this partial ungauging precisely when it has trivial mutual braiding with the charge $x^2y^2$
\begin{equation}
    M_{(a,b)_{\rm diag},(2,2)_{\rm diag}} = (-1)^{a-b} = 1 .
\end{equation}
Thus, the deconfined anyons satisfy
\begin{equation}
    a-b=0 \pmod 2 .
\end{equation}
In the intrinsic topological sector, we must also identify anyons that differ by fusion with the ungauged charge $x^2y^2$. A convenient set of representatives for the remaining deconfined anyons is
\begin{align}
    1
    &=
    \begin{pmatrix}0\\0\end{pmatrix}_{\rm diag},
    \nonumber\\
    e_{\rm TC}
    &=
    xy
    =
    \begin{pmatrix}1\\1\end{pmatrix}_{\rm diag},
    \nonumber\\
    m_{\rm TC}
    &=
    xy^{-1}
    =
    \begin{pmatrix}1\\-1\end{pmatrix}_{\rm diag},
    \nonumber\\
    f_{\rm TC}
    &=
    x^2
    =
    \begin{pmatrix}2\\0\end{pmatrix}_{\rm diag}
    \sim
    y^2
    =
    \begin{pmatrix}0\\2\end{pmatrix}_{\rm diag}.
\end{align}
Their spins and mutual braiding are
\begin{equation}
    \theta_{e_{\rm TC}}
    =
    \theta_{m_{\rm TC}}
    =
    1,
    \qquad
    \theta_{f_{\rm TC}}=-1,
\end{equation}
and
\begin{equation}
    M_{e_{\rm TC},m_{\rm TC}}=-1 .
\end{equation}
Therefore, the intrinsic topological order left after this partial ungauging is the ordinary $\ZZ_2$ toric code.

The remaining symmetry enrichment can also be read off from the parent $\ZZ_4$ gauge theory. Since this partial ungauging uses only the order-two subgroup generated by the $\ZZ_4$ gauge charge $q^2=x^2y^2$, a global $\ZZ_2$ symmetry remains. A unit $\ZZ_4$ flux may be chosen as
\begin{equation}
    v=x=\begin{pmatrix}1\\0\end{pmatrix}_{\rm diag},
\end{equation}
because it has charge-flux braiding with $q$
\begin{equation}
    M_{v,q}=i .
\end{equation}
After ungauging the charge subgroup generated by $q^2=x^2y^2$, the flux $v$ is confined, because it braids nontrivially with this charge
\begin{equation}
    M_{v,q^2}=-1.
\end{equation}
This confinement has a natural interpretation from the ungauging viewpoint. Gauging a global symmetry turns symmetry defects into dynamical gauge fluxes; conversely, ungauging turns the corresponding confined gauge fluxes back into extrinsic symmetry defects. Thus, in the partially ungauged theory, the confined parent flux $v$ becomes the defect of the remaining global $\ZZ_2$ symmetry. The fusion of two such defects is inherited from the parent anyon theory
\begin{equation}
    v^2=x^2=f_{\rm TC}.
\end{equation}
Hence, two $\ZZ_2$ symmetry defects fuse to the toric-code fermion, and the symmetry fractionalization class \cite{Barkeshli14} is
\begin{equation}
    t(1,1)=f_{\rm TC}.
\end{equation}
Equivalently, the local $\ZZ_2$ symmetry action on a deconfined anyon $\alpha$ squares to the mutual braiding phase with $f_{\rm TC}$
\begin{equation}
    \eta_\alpha(1,1)=M_{\alpha,f_{\rm TC}},
    \qquad
    \alpha\in\{e_{\rm TC},m_{\rm TC},f_{\rm TC}\}.
\end{equation}
Using the toric code braiding data, we find
\begin{equation}
    \eta_{e_{\rm TC}}(1,1)=-1,~
    \eta_{m_{\rm TC}}(1,1)=-1,~
    \eta_{f_{\rm TC}}(1,1)=+1 .
\end{equation}
Thus, both $e_{\rm TC}$ and $m_{\rm TC}$ carry nontrivial projective representations of the remaining global $\ZZ_2$ symmetry, while their bound state $f_{\rm TC}=e_{\rm TC}\times m_{\rm TC}$ transforms linearly. This identifies the resulting phase as the $\ZZ_2$ symmetry-enriched $\ZZ_2$ toric code discussed in Sec.~\ref{sec:Z2_enriched_TC}.

We next consider the route from $U(1)_4\times U(1)_{-4}$ to the $\ZZ_4$ SPT. This is obtained by ungauging the full $\ZZ_4$ charge subgroup generated by the order-four boson
\begin{equation}
    q=xy
    =
    \begin{pmatrix}
        1\\1
    \end{pmatrix}_{\rm diag}.
\end{equation}
In the original $\ZZ_8$ toric-code notation,
\begin{equation}
    xy=[em]\,[em^{-1}]=[e^2].
\end{equation}
The full charge algebra used in this ungauging is
\begin{equation}
    1\oplus xy\oplus (xy)^2\oplus (xy)^3
\end{equation}
This fully ungauges the $\ZZ_4$ gauge theory. Indeed, an anyon $x^a y^b$ braids trivially with $xy$ only if
\begin{equation}
    a-b=0\mod 4.
\end{equation}
Thus, the only deconfined anyons are powers of $xy$, and these become local symmetry charges after ungauging the full charge subgroup and hence belong to the intrinsic topological vacuum sector. Hence, no intrinsic topological order remains. Keeping the dual zero-form $\ZZ_4$ symmetry gives a $\ZZ_4$ SPT phase.

The SPT label is most transparent after translating to the standard twisted-quantum-double basis. Let
\begin{equation}
    K_{\rm std}
    =
    \begin{pmatrix}
        0&4\\
        4&-4
    \end{pmatrix}.
\end{equation}
The diagonal and standard bases are related by
\begin{equation}
    K_{\rm diag}
    =
    W^T K_{\rm std} W,
    \qquad
    W=
    \begin{pmatrix}
        1&0\\
        1&1
    \end{pmatrix}.
\end{equation}
Therefore, an anyon vector transforms as
\begin{equation}
    \ell_{\rm std}=W^{-T}\ell_{\rm diag}.
\end{equation}
Explicitly,
\begin{equation}
    \begin{pmatrix}
        a\\ b
    \end{pmatrix}_{\rm diag}
    \longmapsto
    \begin{pmatrix}
        a-b\\ b
    \end{pmatrix}_{\rm std}.
\end{equation}
Thus, the three anyons relevant for the condensation and ungauging arrows in Fig.~\ref{fig:Z_8_commutative_diagram} transform as
\begin{align}
    y^2
    =
    \begin{pmatrix}0\\2\end{pmatrix}_{\rm diag}
    &\longmapsto
    \begin{pmatrix}-2\\2\end{pmatrix}_{\rm std}
    \sim
    \begin{pmatrix}2\\2\end{pmatrix}_{\rm std}, \\
    x^2y^2 =
    \begin{pmatrix}2\\2\end{pmatrix}_{\rm diag}
    &\longmapsto
    \begin{pmatrix}0\\2\end{pmatrix}_{\rm std}, \\
    xy =
    \begin{pmatrix}1\\1\end{pmatrix}_{\rm diag}
    &\longmapsto
    \begin{pmatrix}0\\1\end{pmatrix}_{\rm std}.
\end{align}
In the standard basis, the fermionic toric code arrow condenses the emergent fermion
\begin{equation}
    \begin{pmatrix}2\\2\end{pmatrix}_{\rm std}
\end{equation}
after binding it to the physical fermion $\psi$. The $\ZZ_2$-enriched toric-code arrow ungauges the order-two subgroup generated by the gauge charge
\begin{equation}
    \begin{pmatrix}0\\2\end{pmatrix}_{\rm std} \, ,
\end{equation}
and the $\ZZ_4$ SPT arrow ungauges the full $\ZZ_4$ charge subgroup generated by
\begin{equation}
    \begin{pmatrix}0\\1\end{pmatrix}_{\rm std}.
\end{equation}
In the standard convention, the gauged $\ZZ_4$ SPT labelled by $p\in\ZZ_4$ has
\begin{equation}
    K_p=
    \begin{pmatrix}
        0&4\\
        4&-2p
    \end{pmatrix}.
\end{equation}
Since
\begin{equation}
    K_{\rm std}
    =
    \begin{pmatrix}
        0&4\\
        4&-4
    \end{pmatrix}
    =
    K_{p=2},
\end{equation}
the final ungauging of the full charge subgroup generated by $xy$ turns the $p=2$ twisted $\ZZ_4$ gauge theory into the $p=2$ $\ZZ_4$ SPT.

In summary, the three processes associated with
\begin{equation}
    y^2\psi,\qquad x^2y^2,\qquad xy
\end{equation}
are, respectively, fermion condensation, partial ungauging, and full ungauging. They lead respectively to the fermionic toric code, the $\ZZ_2$ symmetry-enriched $\ZZ_2$ toric code, and the $p=2$ $\ZZ_4$ SPT phase. This also explains the column-vector labels in Fig.~\ref{fig:Z_8_commutative_diagram}. They are written in the diagonal basis $K_{\rm diag}=\mathrm{diag}(4,-4)$, while the standard twisted-quantum-double basis identifies them as
\begin{equation}
    \psi\times\begin{pmatrix}2\\2\end{pmatrix}_{\rm std}~,\quad
    \begin{pmatrix}0\\2\end{pmatrix}_{\rm std}~,\quad
    \begin{pmatrix}0\\1\end{pmatrix}_{\rm std}
\end{equation}
respectively.}

\section{{Stable equivalence of the two \texorpdfstring{$\ZZ_2$}{Z2}-enriched toric-code constructions}}\label{app:stable_equivalence_Z2_SET}

{We now explain how the $\ZZ_2$-enriched toric-code construction obtained by gauging fermion parity in Sec.~\ref{sec:Z2_enriched_TC} is related to the purely bosonic route through \(U(1)_4\times U(1)_{-4}\) via ungauging the charge subgroup generated by the $\ZZ_2$ boson shown in Fig.~\ref{fig:Z_8_commutative_diagram}. At the microscopic level, the two constructions are not written on the same Hilbert space. The fermion-parity-gauging construction contains additional bosonization gauge qubits \((\widetilde X_e,\widetilde Z_e)\) on the edges, while the direct bosonic construction does not. These extra variables should be viewed as auxiliary gauge degrees of freedom of the bosonization map.

It is useful to first explain why these auxiliary variables can be removed. The important observation is that a local \(C'_e\)-stabilizer violation is topologically trivial. Indeed, \(C'_e\) contains \(\widetilde X_e\), and hence the local operator \(\widetilde Z_e\) creates a single \(C'_e\)-term violation. Since this excitation is created by an operator with bounded support, it carries the vacuum topological charge. In particular, it is a trivial boson in the topological order. Condensing this local trivial boson should therefore not change the SET phase; it only removes a locally redundant presentation of the same phase.

In the stabilizer language, condensing this local trivial boson is implemented by proliferating the commuting local operators \(\widetilde Z_e\), or equivalently by choosing the gauge slice
\begin{equation}
    \widetilde Z_e=1,\qquad \forall e .
\end{equation}
We denote the corresponding gauge-fixing group by
\begin{equation}
    \mathcal F
    =
    \big\langle \{\widetilde Z_e\}\big\rangle .
\end{equation}
The gauge-fixed stabilizer group is obtained in the usual stabilizer way. First keep only those stabilizers that commute with the condensate \(\mathcal F\), and then set \(\widetilde Z_e=1\)
\begin{equation}
    \mathcal S_{\rm bos}
    =
    \left.
    \operatorname{Cent}_{\mathcal S_{\ZZ_2\text{-SET}}}
    (\mathcal F)
    \right|_{\widetilde Z_e=1}.
    \label{eq:Z2SET_gauge_fixed_group}
\end{equation}
Here, \(\operatorname{Cent}_{\mathcal S}(\mathcal F)\) denotes the subgroup of \(\mathcal S\) in Eq.~\eqref{eq:Z4F_SET_stabilizers} commuting with every element of \(\mathcal F\).

Let us now evaluate this centralizer. From Eq.~\eqref{eq:Z4F_SET_stabilizers}, the stabilizers \(A'_v\) and \(D_e\) contain no \(\widetilde X\)'s, while \(B'_p\) contains only \(\widetilde Z\)'s in the auxiliary sector. Therefore, they survive the gauge fixing. The plaquette term becomes
\begin{equation}
    B'_p
    \longmapsto
    \overline B_p
    :=
    B'_p\big|_{\widetilde Z=1}.
\end{equation}
By contrast, each $C'_e$ contains $\widetilde X_e$, so it anticommutes with the corresponding $\widetilde Z_e$. Hence, $C'_e$ itself is not in the centralizer of $\mathcal F$.
However, its square
\begin{equation}
    K_e := (C'_e)^2
\end{equation}
does commute with all \(\widetilde Z\)'s and therefore survives. We thus obtain
\begin{equation}
    \mathcal S_{\rm bos}
    =
    \big\langle
        \{A'_v\},\{\overline B_p\},\{K_e=(C'_e)^2\},\{D_e\}
    \big\rangle,
\end{equation}
pictorially,
\begin{eqs}
A'_v &=
\begin{tikzpicture}[baseline=-0.6ex, line width=0.5pt]
  \def\L{1.2}
  \draw (-\L,0) -- (\L,0);
  \draw (0,0) -- (0,-\L);
  \draw (0,0) -- (0,\L);
  \draw (0,\L) -- (\L,\L);
  \draw (\L,\L) -- (\L,0);
  \node[op] at (0,0) {$v$};
  \node[op] at (-0.5*\L,0) {$\textcolor{blue}{X^\dagger}$};
  \node[op] at (0,-0.5*\L) {$\textcolor{blue}{X^\dagger}$};
  \node[op] at (0,0.5*\L) {$\textcolor{blue}{XZ}$};
  \node[op] at (0.5*\L,0) {$\textcolor{blue}{XZ^\dagger}$};
  \node[op] at (0.5*\L,\L) {$\textcolor{blue}{Z}$};
  \node[op] at (\L,0.5*\L) {$\textcolor{blue}{Z^\dagger}$};
  \node[op] at (0.5*\L,0.5*\L) {$\textcolor{orange}{X}$};
\end{tikzpicture}
,\quad
\overline B_p =
\raisebox{-0.6cm}{\begin{tikzpicture}[baseline=-0.6ex, line width=0.5pt]
  \def\L{1.2}
  \draw (0,0) rectangle (\L,\L);
  \node[op] at (0.58*\L,1.2*\L) {$\textcolor{blue}{Z^{-2}}$};
  \node[op] at (-0.1*\L,0.5*\L) {$\textcolor{blue}{Z^{-2}}$};
  \node[op] at (1.15*\L,0.5*\L) {$\textcolor{blue}{Z^2}$};
  \node[op] at (0.5*\L,-0.17*\L) {$\textcolor{blue}{Z^2}$};
  \node[op] at (0.5*\L,0.5*\L) {${p}$};
\end{tikzpicture}}
, \\
(C'_e)^2 &= ~~
\raisebox{-0.6cm}{\begin{tikzpicture}[baseline=-0.6ex, line width=0.5pt]
  \def\L{1.2}
  \draw (-\L,0) -- (0,0);
  \draw (0,0) -- (0,\L);
  \node[op] at (-0.5*\L,0.05*\L) {$\textcolor{blue}{Z^4}$};
  \node[op] at (0.15*\L,0.5*\L) {$\textcolor{blue}{X_e^{4}}$};
\end{tikzpicture}}
,\qquad 
\raisebox{0.6cm}{\begin{tikzpicture}[baseline=-0.6ex, line width=0.5pt]
  \def\L{1.2}
  \draw (0,0) -- (\L,0);
  \draw (0,0) -- (0,-\L);
  \node[op] at (0.5*\L,0) {$\textcolor{blue}{X_e^4}$};
  \node[op] at (0,-0.5*\L) {$\textcolor{blue}{Z^4}$};
\end{tikzpicture}}
\quad ,\\
D_e &= ~~
\tikz[baseline=-0.6ex]{
  \def\L{\Ledge}
  \draw (0,-0.5*\L) -- (0,0.5*\L);
  \node[op] at (-0.5*\L,0) {$\color{orange}Z$};
  \node[op] at (0,0) {$\color{blue}X_e^4$};
  \node[op] at (0.5*\L,0) {$\color{orange}Z$};
}
\;,\quad
\tikz[baseline=-0.6ex]{
  \def\L{\Ledge}
  \draw (-0.5*\L,0) -- (0.5*\L,0);
  \node[op] at (0,0.5*\L) {$\color{orange}Z$};
  \node[op] at (0,0) {$\color{blue}X_e^4$};
  \node[op] at (0,-0.5*\L) {$\color{orange}Z$};
}\;,
\label{eq:Z2SET_reduced_stabilizer_group}
\end{eqs}
where the \(\textcolor{blue}{X}\) and \(\textcolor{blue}{Z}\) operators denote \(\ZZ_8\) qudit Pauli operators, while \(\textcolor{orange}{X}\) and \(\textcolor{orange}{Z}\) denote ordinary \(\ZZ_2\) Pauli operators. This gives the desired Hilbert-space reduction. If we further drop the \(D_e\) terms and remove the \(\color{orange}X\) operator in \(A'_v\), the remaining stabilizer model reduces to the \(U(1)_4\times U(1)_{-4}\) twisted quantum double constructed in Ref.~\cite{Ellison22}. Keeping the \(D_e\) terms implements the additional ungauging of the charge subgroup generated by the order-two boson
\begin{equation}
    q^2=
    \begin{pmatrix}
        2\\
        2
    \end{pmatrix}_{\rm diag}
\end{equation}
in the $U(1)_4\times U(1)_{-4}$ theory. We discuss this bosonic ungauging route in more detail in Sec.~\ref{sec:U1_4_U1_minus4_TQD}. Consequently, the Hamiltonian defined by Eq.~\eqref{eq:Z2SET_reduced_stabilizer_group} gives a lattice realization of the $\ZZ_2$-enriched $\ZZ_2$ toric code obtained from this route.

The operator \(K_e=(C'_e)^2\) is the local hopping operator for the bosonic condensate
\begin{equation}
    (e^2m^6\psi)^2=e^4m^4
\end{equation}
in the original \(\ZZ_8\) toric code. Thus, the fermionic route contains a local square root \(C'_e\) of the bosonic \(e^4m^4\) hopping operator. This square root becomes local only after introducing the auxiliary bosonization gauge field. After condensing the local trivial \(C'_e\)-defect, or equivalently after fixing \(\widetilde Z_e=1\), only the gauge-invariant square \(K_e=(C'_e)^2\) remains.

The same statement can also be expressed at the level of local projectors. Let
\begin{equation}
    P_{\widetilde Z}
    =
    \prod_e \frac{1+\widetilde Z_e}{2}
\end{equation}
be the projector onto the gauge slice. Since \(C'_e\) flips \(\widetilde Z_e\), while \(K_e=(C'_e)^2\) preserves the gauge slice, we have
\begin{equation}
    P_{\widetilde Z} C'_e P_{\widetilde Z}=0,
    \qquad
    P_{\widetilde Z} (C'_e)^3 P_{\widetilde Z}=0 .
\end{equation}
For the order-four stabilizer projector
\begin{equation}
    \Pi_{C'_e}
    =
    \frac{1}{4}
    \left(
        1+C'_e+(C'_e)^2+(C'_e)^3
    \right),
\end{equation}
we therefore find
\begin{equation}
\begin{split}
    P_{\widetilde Z}\Pi_{C'_e}P_{\widetilde Z}
    &=
    \frac{1}{4}
    P_{\widetilde Z}
    \left(
        1+(C'_e)^2
    \right)
    P_{\widetilde Z}  \\
    &\propto
    P_{\widetilde Z}\Pi_{K_e}P_{\widetilde Z},
    \qquad
    \Pi_{K_e}:=\frac{1+K_e}{2}.
\end{split}
\label{eq:Ce_projector_reduces_to_Ce_square}
\end{equation}
Thus projection to the \(\widetilde Z_e=1\) condensate removes the linear square-root constraint \(C'_e=1\), and leaves exactly the bosonic constraint \(K_e=(C'_e)^2=1\).

Although the above comparison is naturally phrased as a gauge fixing, the relation can also be made fully unitary. We now give a finite-depth local unitary construction which restores the auxiliary bosonization gauge qubits starting from the reduced bosonic model. In this form, the projection $P_{\widetilde Z}$ should be viewed as the nonunitary shadow of a unitary encoding map.

Start from the reduced bosonic stabilizer group of Eq.~\eqref{eq:Z2SET_reduced_stabilizer_group}. For each edge \(e\), add two auxiliary qubits. The first is the bosonization gauge qubit \(\widetilde q_e\), with Pauli operators \(\widetilde X_e,\widetilde Z_e\), initialized in
\begin{equation}
    |0\rangle_{\widetilde Z_e},
    \qquad
    \widetilde Z_e |0\rangle_{\widetilde Z_e}
    =
    |0\rangle_{\widetilde Z_e}.
\end{equation}
The second is a work qubit \(\tau_e\), initialized in
\begin{equation}
    |+\rangle_{\tau^x_e}
    =
    \frac{
        |0\rangle_{\tau^z_e}
        +
        |1\rangle_{\tau^z_e}
    }{\sqrt 2}.
\end{equation}

For each edge, define the controlled-\(C'_e\) unitary
\begin{equation}
    \operatorname{Ctrl}_{\tau_e}(C'_e)
    =
    |0\rangle\langle 0|_{\tau^z_e}\otimes I
    +
    |1\rangle\langle 1|_{\tau^z_e}\otimes C'_e .
\end{equation}
Then set
\begin{equation}
    U_e
    =
    \operatorname{CNOT}_{\widetilde q_e\rightarrow \tau_e}
    \,
    \operatorname{Ctrl}_{\tau_e}(C'_e),
    \label{eq:edge_encoding_unitary}
\end{equation}
where the CNOT is taken in the
\(\widetilde Z_e\)- and \(\tau^z_e\)-bases:
\[
    |a\rangle_{\widetilde q_e}|b\rangle_{\tau_e}
    \longmapsto
    |a\rangle_{\widetilde q_e}|a+b\rangle_{\tau_e},
    \qquad a,b\in\ZZ_2 .
\]
The operator \(U_e\) is unitary on the full microscopic Hilbert space.
{The full encoding circuit is
\begin{equation}
    U_{\rm enc}=\prod_e U_e .
    \label{eq:encoding_circuit}
\end{equation}
The edge unitaries commute, and their bounded supports allow them to be scheduled in a number of disjoint-support layers independent of system size. Thus $U_{\rm enc}$ has finite depth. The added qubits transform trivially under the global $\ZZ_2$ symmetry, and every $C'_e$ commutes with $\prod_p\textcolor{orange}{X}_p$, so the circuit preserves this symmetry.}

{We now show that the local unitaries $U_e$ identify the ground spaces of the two stabilizer models after adjoining and removing product-state ancillas. Throughout the following conjugation calculation, we restrict to the common subspace satisfying $K_e=(C'_e)^2=1$ for every edge. Each $U_e$ preserves these constraints, and $C'_e$ is Hermitian on this subspace.

Consider the enlarged stabilizer group $\mathcal S=\braket{\mathcal S_{\rm bos},\widetilde Z_e,\tau_e^x}$. Conjugation by $\operatorname{Ctrl}_{\tau_e}(C'_e)$ sends $\tau_e^x$ to $\tau_e^x C'_e$. Any stabilizer that anticommutes with $C'_e$ acquires a factor of $\tau_e^z$: in particular, $\widetilde Z_e$ becomes $\tau_e^z\widetilde Z_e$, and each plaquette stabilizer $\overline B_p$ acquires this factor on every edge $e\in\partial p$.

Next, conjugation by $\operatorname{CNOT}_{\widetilde q_e\rightarrow\tau_e}$ sends
\begin{equation}
    \tau_e^z\mapsto\tau_e^z\widetilde Z_e,
    \qquad
    \widetilde X_e\mapsto\tau_e^x\widetilde X_e .
\end{equation}
Since $C'_e$ contains one factor of $\widetilde X_e$ and no $\widetilde Z_e$, the two factors of $\tau_e^x$ cancel: $\tau_e^x C'_e$ becomes $C'_e$. Meanwhile, $\tau_e^z\widetilde Z_e$ becomes $\tau_e^z$. The work qubits therefore decouple in the state $\tau_e^z=+1$. Removing them leaves the ground space of $\mathcal S_{\ZZ_2\text{-SET}}$. The graphical sequence below uses the same restriction $(C'_e)^2=1$.}
\begin{widetext}
{
\newcommand{\Dstage}[1]{\makebox[3.65cm][c]{$\displaystyle #1$}}
\newcommand{\Darrow}[1]{\makebox[2.25cm][c]{$\displaystyle\xrightarrow{#1}$}}
\newcommand{\Dedgepair}[2]{\vcenter{\hbox{%
\begin{tikzpicture}[line width=.5pt]
  \path[use as bounding box] (-.65,0) rectangle (2.65,1.2);
  \draw (0,0)--(0,1.2);
  \draw (1.2,.6)--(2.4,.6);
  \node[op] at (0,.6) {$#1$};
  \node[op] at (1.8,.6) {$#2$};
  \node[inner sep=0pt] at (.65,.6) {$,$};
\end{tikzpicture}}}}
\newcommand{\Dplaquette}[1]{\vcenter{\hbox{%
\begin{tikzpicture}[line width=.5pt]
  \def\L{1.35}
  \path[use as bounding box] (-.8*\L,-.32*\L) rectangle (1.8*\L,1.32*\L);
  \draw (0,0) rectangle (\L,\L);
  \node[op] at (.58*\L,1.2*\L) {$#1\textcolor{blue}{Z^{-2}}$};
  \node[op] at (-.1*\L,.5*\L) {$#1\textcolor{blue}{Z^{-2}}$};
  \node[op] at (1.15*\L,.5*\L) {$#1\textcolor{blue}{Z^2}$};
  \node[op] at (.5*\L,-.17*\L) {$#1\textcolor{blue}{Z^2}$};
  \node[op] at (.5*\L,.5*\L) {$p$};
\end{tikzpicture}}}}
\begin{alignat}{3}
\Dstage{\Dedgepair{\textcolor{magenta}{\widetilde Z_e}}{\textcolor{magenta}{\widetilde Z_e}}}
&\Darrow{\mathrm{Ctrl}_{\tau_e}(C'_e)}
&\Dstage{\Dedgepair{\textcolor{magenta}{\widetilde Z_e}\tau_e^z}{\textcolor{magenta}{\widetilde Z_e}\tau^z_e}}
&\Darrow{\mathrm{CNOT}_{\widetilde q_e\rightarrow\tau_e}}
&\Dstage{\Dedgepair{\tau_e^z}{\tau^z_e}}\,,
\\[6pt]
\Dstage{\Dedgepair{\tau_e^x}{\tau^x_e}}
&\Darrow{\mathrm{Ctrl}_{\tau_e}(C'_e)}
&\Dstage{\Dedgepair{\tau_e^x C'_e}{\tau^x_e C'_e}}
&\Darrow{\mathrm{CNOT}_{\widetilde q_e\rightarrow\tau_e}}
&\Dstage{\Dedgepair{C'_e}{C'_e}}\,,
\\[6pt]
\Dstage{\Dplaquette{}}
&\Darrow{\mathrm{Ctrl}_{\tau_e}(C'_e)}
&\Dstage{\Dplaquette{\tau^z}}
&\Darrow{\mathrm{CNOT}_{\widetilde q_e\rightarrow\tau_e}}
&\Dstage{\Dplaquette{\tau^z\textcolor{magenta}{\widetilde Z}}}\,.
\end{alignat}
}
\end{widetext}
{Thus $U_{\rm enc}$ maps the enlarged ground space to that of $\mathcal{S}_{\ZZ_2\text{-SET}}$ in Eq.~\eqref{eq:Z4F_SET_stabilizers}, together with decoupled work qubits satisfying $\tau_e^z=+1$.}

In summary, the two microscopic stabilizer realizations are related in two complementary ways. In the gauge-fixed description, one condenses the strictly local and topologically trivial \(C'_e\)-defect created by \(\widetilde Z_e\), which reduces the stabilizer group to
\begin{equation}
    \big\langle
        A'_v,\overline B_p,(C'_e)^2,D_e
    \big\rangle .
\end{equation}
In the coherent description, this reduction is reversed by adding product-state ancillas and applying the finite-depth unitary \(U_{\rm enc}\), thereby restoring the local square-root constraint \(C'_e=1\). Thus, the auxiliary \(\widetilde X\) and \(\widetilde Z\) degrees of freedom do not change the SET phase; they only provide a local bosonization-gauge-field realization of the square root of the bosonic \(e^4m^4\) condensate.}

\addtocontents{toc}{\protect\newpage}
\section{{Topological data of the general Majorana-Pauli stabilizer construction}}
\label{app:general_lattice_topological_data}

{We derive the fusion, exchange, and mutual-braiding data of the strings in Sec.~\ref{sec:general_fermionic_topological_order_lattice_model}. We use the surface-orientation convention fixed in Sec.~\ref{sec:fermionic_toric_code}: clockwise exchange in the displayed coordinates is positive, and reversing the surface orientation complex conjugates the statistics. The Pauli algebra in layer $i$ is
\begin{equation}
Z_iX_i=\omega_iX_iZ_i,
\qquad
\omega_i=\exp\!\left(\frac{2\pi i}{2n_i^2}\right).
\label{eq:general_lattice_Pauli_algebra_appendix}
\end{equation}
Write the electric dressing as
\begin{equation}
\hat Z_i=\prod_j Z_j^{a_{ij}},\qquad
 a_{ij}=\begin{cases}
 s_i,&j=i,\\
 2n_js_{ij}/n_i,&j>i,\\
 0,&j<i.
 \end{cases}
\label{eq:hatZ_appendix}
\end{equation}
All exponents are integers by the invariant-factor argument. Closed strings commute with the stabilizers, and local stabilizer multiplication deforms a path without changing its topological process. We can therefore choose straight crossing segments and a three-arm junction for the calculations below.}

\paragraph{{Charge and flux braiding.}}
{Charge strings contain only $Z_i^{2n_i}$. They commute with one another, and a charge exchange has no nontrivial reordering:
\begin{equation}
M_{c_i,c_j}=1,\qquad\theta_{c_i}=1.
\label{eq:cc_statistics_appendix}
\end{equation}
At a positive crossing of a charge string with a flux string, the only possible overlap is $Z_i^{2n_i}$ with $X_j$. Thus
\begin{equation}
M_{c_i,\phi_j}=\omega_i^{2n_i\delta_{ij}}
=\exp\!\left(\frac{2\pi i\delta_{ij}}{n_i}\right).
\label{eq:cphi_statistics_appendix}
\end{equation}

To keep track of both contributions at a flux--flux crossing, let $h_{xy}$ and $v_{xy}$ denote the horizontal and vertical edges starting at vertex $(x,y)$. Choose a right-going flux string $H_i$ and a down-going flux string $D_j$ with segments
\begin{equation}
\begin{aligned}
H_i&=\prod_x X_{v_{x0},i}\hat Z_{h_{x1},i},\\
D_j&=\prod_y X_{h_{0y},j}\hat Z_{v_{1y},j}^{\dagger}.
\end{aligned}
\end{equation}
The products extend beyond the crossing, and all factors away from it commute. At the crossing, the dressing of $D_j$ meets the $X_i$ factor of $H_i$, and the dressing of $H_i$ meets the $X_j$ factor of $D_j$. Consequently,
\begin{equation}
H_iD_j=\omega_i^{a_{ji}}\omega_j^{a_{ij}}D_jH_i.
\label{eq:general_flux_crossing_word}
\end{equation}
For $i<j$, only $a_{ij}$ is nonzero, while for $i=j$ both overlaps contribute. In both cases,
\begin{equation}
M_{\phi_i,\phi_j}
=\exp\!\left(\frac{2\pi i(S_{\mathcal I})_{ij}}{n_in_j}\right).
\label{eq:phiphi_mutual_statistics_appendix}
\end{equation}
In particular, $M_{\phi_i,\phi_i}=\omega_i^{2s_i}$. This factor of two distinguishes a full braid from exchange.}

\paragraph{{Exchange at a T-junction.}}
{Fix a layer $i$ and a central plaquette $p$. The three diagrams below define hops into $p$ from the left ($L$), bottom ($B$), and right ($R$). Each diagram shows the starting plaquette and the destination $p$. Blue labels specify the edge operators, and dashed arrows indicate the motion of the flux $\phi_i$:
\begin{equation}
\begin{aligned}
W_L&=
\begin{tikzpicture}[baseline=(pL.base),x=1.6cm,y=1.6cm,
    hop/.style={magenta,dashed,-{Latex[length=1.6mm]},line width=.6pt},
    localop/.style={inner sep=1.5pt,fill=white,text=blue},
    target/.style={inner sep=2pt,text=black}]
  \path[use as bounding box] (-1.1,-.08) rectangle (2.1,1.18);
  \draw[black!45,line width=.4pt] (-1,0) rectangle (1,1);
  \draw[black!45,line width=.4pt] (0,0)--(0,1);
  \node[target] (pL) at (.5,.5) {$p$};
  \draw[hop,shorten >=2pt] (-.5,.5)--(pL.west);
  \node[localop] at (0,.5) {$X_i$};
  \node[localop] at (.5,1) {$\hat Z_i$};
\end{tikzpicture}
\\[4pt]
W_B&=\vcenter{\hbox{%
\begin{tikzpicture}[baseline=(pB.base),x=1.6cm,y=1.6cm,
    hop/.style={magenta,dashed,-{Latex[length=1.6mm]},line width=.6pt},
    localop/.style={inner sep=1.5pt,fill=white,text=blue},
    target/.style={inner sep=2pt,text=black}]
  \path[use as bounding box] (-1.1,-1.08) rectangle (2.1,1.08);
  \draw[black!45,line width=.4pt] (0,-1) rectangle (1,1);
  \draw[black!45,line width=.4pt] (0,0)--(1,0);
  \node[target] (pB) at (.5,.5) {$p$};
  \draw[hop,shorten >=2pt] (.5,-.5)--(pB.south);
  \node[localop] at (.5,0) {$X_i^\dagger$};
  \node[localop] at (1,.5) {$\hat Z_i$};
\end{tikzpicture}}}
\\[4pt]
W_R&=
\begin{tikzpicture}[baseline=(pR.base),x=1.6cm,y=1.6cm,
    hop/.style={magenta,dashed,-{Latex[length=1.6mm]},line width=.6pt},
    localop/.style={inner sep=1.5pt,fill=white,text=blue},
    target/.style={inner sep=2pt,text=black}]
  \path[use as bounding box] (-1.1,-.08) rectangle (2.1,1.18);
  \draw[black!45,line width=.4pt] (0,0) rectangle (2,1);
  \draw[black!45,line width=.4pt] (1,0)--(1,1);
  \node[target] (pR) at (.5,.5) {$p$};
  \draw[hop,shorten >=2pt] (1.5,.5)--(pR.east);
  \node[localop] at (1,.5) {$X_i^\dagger$};
  \node[localop] at (1.5,1) {$\hat Z_i^\dagger$};
\end{tikzpicture}
.
\end{aligned}
\label{eq:general_Tjunction_hops}
\end{equation}

These are the oriented string segments of Eq.~\eqref{eq:general_phi_string_operator}; taking an adjoint reverses the hop. The left hop $W_L$ commutes with both other hops. The only noncommuting overlap between $W_B^\dagger$ and $W_R$ is on the right edge of $p$, where they act with $\hat Z_i^\dagger$ and $X_i^\dagger$, respectively. Their layer-$i$ factors obey
\begin{equation}
Z_i^{-s_i}X_i^\dagger
=\omega_i^{s_i}X_i^\dagger Z_i^{-s_i}.
\end{equation}
The higher-layer dressing factors commute with the $X_i$ factors and contribute no exchange phase. Hence the two three-hop sequences satisfy
\begin{equation}
W_LW_B^\dagger W_R
=\omega_i^{s_i}W_RW_B^\dagger W_L.
\end{equation}
Equivalently, with particles initially on the left and bottom arms, the six-hop exchange word is
\begin{equation}
\begin{aligned}
\mathcal E_i
&=W_B^\dagger W_RW_L^\dagger W_BW_R^\dagger W_L\\
&=W_B^\dagger W_RW_BW_R^\dagger
=\omega_i^{s_i}.
\end{aligned}
\label{eq:general_Tjunction_exchange_word}
\end{equation}
Operators act from right to left. The trajectories exchange the particles clockwise in the displayed coordinates. Thus
\begin{equation}
\theta_{\phi_i}=\omega_i^{s_i}
=\exp\!\left(\frac{\pi i s_i}{n_i^2}\right).
\label{eq:phi_spin_appendix}
\end{equation}
Parity-even local endpoint dressings and stabilizer multiplication preserve these phases. Attaching the transparent physical fermion $\psi$ instead reverses the spin and leaves all mutual braiding unchanged.}

\paragraph{{Fusion and reconstruction of the target order.}}
{The same strings have parent-theory representatives
\begin{equation}
\begin{aligned}
c_i&=e_i^{2n_i},\\
\phi_i&=m_i e_i^{s_i}\prod_{j>i}e_j^{2n_js_{ij}/n_i}.
\end{aligned}
\end{equation}
Raising the second expression to $n_i$ and multiplying by the condensate $q_i$ in Eq.~\eqref{eq:lattice_condensed_anyon_bi} cancels its magnetic component. Using $q_i\sim1$ gives the local fusion relations
\begin{equation}
c_i^{n_i}=1,\qquad
\phi_i^{n_i}=\psi^{\bar s_i}\prod_j c_j^{(S_{\mathcal I})_{ji}}.
\label{eq:general_string_fusion_relations}
\end{equation}
These are precisely the relations of $K_{\mathrm{fTQD}}$, with the parity of a local relation recorded by its diagonal entry.

For a chosen representative $\ell=(a,b)^T\in\ZZ^{2M}$, where $a_i$ counts fluxes and $b_i$ counts charges, fusion of the generator statistics gives
\begin{equation}
\theta_\ell=\exp\!\left\{\pi i\left[
 a^TN^{-1}S_{\mathcal I}N^{-1}a+2a^TN^{-1}b
\right]\right\},
\label{eq:general_spin_appendix}
\end{equation}
and
\begin{equation}
\begin{aligned}
M_{\ell,\ell'}=\exp\!\bigl\{2\pi i[&a^TN^{-1}S_{\mathcal I}N^{-1}a'\\
&+a^TN^{-1}b'+b^TN^{-1}a']\bigr\}.
\end{aligned}
\label{eq:general_mutual_appendix}
\end{equation}
The bilinear form read directly from the lattice is therefore
\begin{equation}
\mathcal B_{\mathrm{lattice}}
=\begin{pmatrix}
N^{-1}S_{\mathcal I}N^{-1}&N^{-1}\\
N^{-1}&0
\end{pmatrix}
=K_{\mathrm{fTQD}}^{-1}.
\label{eq:reconstructed_inverse_K_appendix}
\end{equation}
Its inverse is $K_{\mathrm{fTQD}}=\left(\begin{smallmatrix}0&N\\N&-S_{\mathcal I}\end{smallmatrix}\right)$, as required. The decoupled local block $K_f=\operatorname{diag}(1,-1)$ supplies the transparent physical fermion without changing this intrinsic mutual-braiding form.}

\section{{Bosonic-shadow construction for non-split fSPTs}}
\label{app:nonsplit_bosonic_shadow}

{This appendix supplies the cochain and $K$-matrix calculations used in Sec.~\ref{sec:nonsplit_spt_scope}. All cochain equalities involving $\nu_3$ are in $\RR/\ZZ$, and $\alpha,\beta$ run from $0$ to $M$. We retain the carry cocycles $B_\alpha=\delta\widetilde a_\alpha/N_\alpha$ and the type-I decoration of Eq.~\eqref{eq:nonsplit_n2_choice}.}

\subsection{{Canonical solution and the homogeneous labels}}
\label{app:nonsplit_shadow_cochains}

{Put $A_\alpha=\widetilde a_\alpha/N_\alpha$, so $\delta A_\alpha=B_\alpha$. Define
\begin{equation}
q_{\boldsymbol v,\boldsymbol k}
=\frac12\sum_{\alpha,\beta}
v_\alpha k_\beta A_\alpha\cup B_\beta .
\label{eq:shadow_q_canonical}
\end{equation}
Its coboundary is directly
\begin{equation}
\begin{aligned}
\delta q_{\boldsymbol v,\boldsymbol k}
&=\frac12
\left(\sum_\alpha v_\alpha B_\alpha\right)
\cup\left(\sum_\beta k_\beta B_\beta\right)\\
&=\frac12(\omega_2+n_2)\cup n_2 .
\end{aligned}
\label{eq:shadow_q_coboundary}
\end{equation}
This is the non-split Gu--Wen equation~\eqref{eq:nonsplit_GW_scope}. To compare with the solution already given in Secs.~\ref{sec:split_spt_scope} and~\ref{sec:nonsplit_spt_scope}, use the type-II cocycle $T_{ij}$ of Eq.~\eqref{eq:split_typeII_generator}. Equations~\eqref{eq:Q-n2} and~\eqref{eq:nonsplit_cross_term} imply
\begin{equation}
q_{\boldsymbol v,\boldsymbol k}
=Q(n_2)+R_{\omega,n}
+\sum_{1\leq i<j\leq M}k_i k_j T_{ij}.
\label{eq:shadow_q_vs_existing_solution}
\end{equation}
Here the cup-one terms are evaluated using the integer carry lifts; after multiplication by $1/2$ only their reductions modulo two matter. Thus the ordered solution and the earlier solution differ by a specified homogeneous type-II cocycle. In particular, changing between them while retaining the same fSPT requires shifting the homogeneous labels.

Let $H$ be a symmetric integer matrix with even diagonal, and write the homogeneous response in the generator convention of Eq.~\eqref{eq:split_typeII_generator} as
\begin{equation}
\nu_H=\sum_\alpha\frac{H_{\alpha\alpha}}2
A_\alpha\cup B_\alpha
+\sum_{\alpha<\beta}H_{\alpha\beta}T_{\alpha\beta}.
\label{eq:shadow_homogeneous_response}
\end{equation}
If the desired response is instead specified relative to
$Q+R_{\omega,n}$, Eq.~\eqref{eq:shadow_q_vs_existing_solution} requires
$H_{ij}=p_{ij}-k_i k_j$ for $1\le i<j\le M$, with
$H_{ji}=H_{ij}$, $H_{0i}=H_{i0}=p_{0i}$, and unchanged diagonal
entries $H_{\alpha\alpha}=2p_\alpha$.

To match $\nu^{\rm can}_3=q_{\boldsymbol v,\boldsymbol k}+\nu_H$ with the convention $K=(0,N;N,-S)$ in Eq.~\eqref{eq:abelian-K-SPT}, choose the lower-right block of Eq.~\eqref{eq:nonsplit_shadow_KB} to be
\begin{equation}
P=-H-\boldsymbol v\boldsymbol k^T
-\boldsymbol k\boldsymbol v^T .
\label{eq:shadow_P_response_dictionary}
\end{equation}
This is even on the diagonal. The signs and the integer shift in this formula are part of the response convention: an arbitrary symmetric matrix $P$ with even diagonal still defines the construction, but its entries are not directly the homogeneous labels.

One can check the dictionary without integrating out a half-integral field. Make the integral shear of gauge fields
\begin{equation}
x_{\rm old}=x_{\rm new}+\boldsymbol k^T\boldsymbol a .
\end{equation}
It sends the $y$--$\boldsymbol a$ block to $-\boldsymbol k^T$, the lower-right block to $-H$, and the anyon $f$ to $x_{\rm new}+y$. The bosonic gauge group is the Abelian extension $\widehat G$ of $G_b$ by $\ZZ_2$ with multiplication
\begin{equation}
\begin{aligned}
(g,s)(h,t)=\biggl(gh,\,
s+t+\sum_\alpha v_\alpha B_\alpha(g,h)
\pmod2\biggr).
\end{aligned}
\label{eq:shadow_auxiliary_group}
\end{equation}
This auxiliary group is not the physical symmetry $G_f$ in general. On $\widehat G$, the $\RR/\ZZ$-valued cochain
\begin{equation}
Y(g,s)=\frac12\left(s+\sum_\alpha v_\alpha A_\alpha(g)\right)
\end{equation}
is a character: its coboundary is integral. Hence the bosonic three-cocycle
\begin{equation}
\begin{aligned}
\alpha_3
&=Y\cup\sum_\beta k_\beta B_\beta+\nu_H\\
&=\frac12s\cup n_2+\nu^{\rm can}_3
\end{aligned}
\label{eq:shadow_auxiliary_cocycle}
\end{equation}
is closed. Pullbacks from $G_b$ are implicit, and the last line is understood modulo one. The constraint $2Y=\sum_\alpha v_\alpha A_\alpha$ gives the $-2,\boldsymbol v$ blocks of the sheared $K$ matrix; the first term of $\alpha_3$ gives the $-\boldsymbol k$ coupling, and $\nu_H$ gives the $-H$ block. This is the usual type-I/type-II Abelian twisted-gauge presentation, with the same cochain refinement as Eq.~\eqref{eq:corrected-K-nu}. Gauging the central $\ZZ_2$ first gives the equivariant toric-code shadow. Its defect associator is $\nu^{\rm can}_3$, while the $s\cup n_2/2$ term supplies the fermion decoration. This verifies the response data as well as the intrinsic parent order~\cite{BGK17,Barkeshli2022Classification}.}

\subsection{{Inverse matrix and maximality of the charge subgroup}}

{Define
\begin{equation}
\mathsf T=P+\frac12\left(
\boldsymbol v\boldsymbol k^T+\boldsymbol k\boldsymbol v^T\right).
\end{equation}
Block inversion gives, in the original $(x,\boldsymbol b,y,\boldsymbol a)$ basis,
\begin{equation}
\setlength{\arraycolsep}{3pt}
K_B^{-1}=\begin{pmatrix}
0&\frac12\boldsymbol k^T\mathsf N^{-1}&-\frac12&0\\
\frac12\mathsf N^{-1}\boldsymbol k&-\mathsf N^{-1}\mathsf T\mathsf N^{-1}&\frac12\mathsf N^{-1}\boldsymbol v&\mathsf N^{-1}\\
-\frac12&\frac12\boldsymbol v^T\mathsf N^{-1}&0&0\\
0&\mathsf N^{-1}&0&0
\end{pmatrix}.
\label{eq:shadow_KB_inverse}
\end{equation}
For the integer vector $\ell_f=(1,\boldsymbol0,1,-\boldsymbol k)^T$,
\begin{equation}
K_B^{-1}\ell_f
=\begin{pmatrix}
-1/2\\ \mathsf N^{-1}\boldsymbol w/2\\-1/2\\\boldsymbol0
\end{pmatrix},\qquad
\ell_f^T K_B^{-1}\ell_f=-1 .
\label{eq:shadow_f_inverse_check}
\end{equation}
This proves $\theta_f=-1$. It also shows that $q\ell_f$ is local precisely when $q$ is divisible by $2N$, proving the claimed order. The vectors for $a_\alpha$ have zero pairings with each other and with $\ell_f$; their orders are exactly $N_\alpha$. Finally,
\begin{equation}
2\ell_f-\ell_{a_0}=-K_B(\ell_x+\ell_y),
\end{equation}
which verifies $2f=a_0$ as an anyon relation. If $qf+\sum_{i>0}m_i a_i$ is local, Eq.~\eqref{eq:shadow_f_inverse_check} forces $q=0\pmod{2N}$ and $m_i=0\pmod{N_i}$. There are therefore no additional relations among the displayed generators of $\mathcal L$. Its order is $2\prod_\alpha N_\alpha$, exactly the square root of the parent anyon number. Nondegeneracy of the parent pairing then gives $\mathcal L^\perp=\mathcal L$. After adjoining the local fermion and binding it to odd-parity elements, these dressed charges form a maximal isotropic subgroup; ungauging this subgroup leaves only local fermions deconfined.}

\subsection{{Integral twisted-quantum-double presentation}}

{Group the basis as $(x,\boldsymbol b\mid y,\boldsymbol a)$. Then
\begin{equation}
\begin{gathered}
K_B=\begin{pmatrix}0&R\\R^T&C\end{pmatrix},\\
R=\begin{pmatrix}-2&\boldsymbol v^T\\0&\mathsf N\end{pmatrix},\quad
C=\begin{pmatrix}0&\boldsymbol k^T\\\boldsymbol k&P\end{pmatrix}.
\end{gathered}
\label{eq:shadow_integral_blocks}
\end{equation}
Choose a Smith decomposition $U R V=\mathsf N'$ with $U,V$ unimodular and $\mathsf N'$ positive diagonal. The integral basis change $W=\operatorname{diag}(U^T,V)$ gives
\begin{equation}
W^T K_B W
=\begin{pmatrix}0&\mathsf N'\\\mathsf N'&S'\end{pmatrix},
\qquad S'=V^T C V .
\label{eq:shadow_Smith_congruence}
\end{equation}
The matrix $S'$ has even diagonal because $C$ does. This is a bosonic Abelian twisted quantum double, with sign convention $S_{\mathcal I}=-S'$ in Eq.~\eqref{eq:K_fTQD}. Unit entries of $\mathsf N'$ give trivial unimodular sectors that may be removed after integral shears. The transformation of an anyon vector is $\ell'=W^T\ell$; retaining the vectors for $f$ and $a_i$ preserves the intended charge labels for ungauging. Since $R$ is invertible, a real congruence removes $C$ and exhibits equal positive and negative inertia. }

\bibliography{bibliography}

\end{document}